\documentclass[12pt]{aastex}
\usepackage{graphics,natbib,amsmath}
\usepackage[numberedappendix,onecolumn]{emulateapj5}
\newcommand{\Mspyr}{\hbox{$\,$M$_\odot\ {\rm yr}^{-1}$}}
\newcommand{\vterm}{\hbox{$\,v_\infty$}}
\newcommand{\mdot}{\hbox{$\,\dot{M}$}}
 
\newcommand{\ha}{H$\alpha$}
\newcommand{\hb}{H$\beta$} 
\newcommand{\hg}{H$\gamma$}
\newcommand{\hd}{H$\delta$}
\newcommand{\brg}{Br$\gamma$}

\newcommand{\kms}{\hbox{$\,$km$\,$s$^{-1}$}}
\newcommand{\Teff}{\hbox{$\,T_{\rm eff}$}}

\newcommand{\ADLD}{\hbox{$\,\theta_{\rm LD}$}}
\newcommand{\ADUD}{\hbox{$\,\theta_{\rm UD}$}}

\newcommand{\Logg}{\hbox{$\logg $}} 
\newcommand{\Loggeff}{\hbox{$\log g_{\rm eff}$}} 
\newcommand{\Loggrstar}{\hbox{$\log g(R_\star)$}} 
 
\newcommand{\logg}{\log g} 
\newcommand{\Msun}{\hbox{$M_\odot$}}
\newcommand{\Rsun}{\hbox{$R_\odot$}}
\newcommand{\Lsun}{\hbox{$L_\odot$}} 
\newcommand{\phoenix}{{\tt PHOENIX}} 
\newcommand{\oaotwo}{{\it OAO-2} } 
\newcommand{\iue}{{\it IUE} }

\def\myion#1#2{\hbox{{#1}\,{\sc {#2}}}}
\newcommand{\pder}[2]{{\frac{\partial#1}{\partial #2}}}
\def\g{\gamma} 
\def\b{\beta}

\shorttitle{The A-type Supergiant Deneb}
\shortauthors{Aufdenberg et al.}

\begin{document}

\bibliographystyle{apj}

\title{The Spectral Energy Distribution and Mass-loss Rate of the
A-Type Supergiant Deneb}

\author{J. P.~ Aufdenberg}
\affil{Solar and Stellar Physics Division, Harvard-Smithsonian Center
for Astrophysics, 60 Garden Street, Mail Stop 15, Cambridge, MA 02138} 
\email{jaufdenberg@cfa.harvard.edu}

\author{P. H.~ Hauschildt} \affil{Department of Physics and Astronomy
\& Center for Simulational Physics, University of Georgia, Athens, GA
30602-2451} \email{yeti@hal.physast.uga.edu}

\author{E.~ Baron}
\affil{Department of Physics and Astronomy, University of Oklahoma,
Norman, OK 73019-0260}
\email{baron@nhn.ou.edu}

\author{T. E.~ Nordgren}
\affil{Department of Physics, University of Redlands, 1200 East Colton
Ave, Redlands, CA 92373}
\email{tyler\_nordgren@redlands.edu}

\author{I. D.~ Howarth, A. W.~ Burnley} 
\affil{Department of Physics and Astronomy, University College London, 
Gower Street, London WC1E 6BT, UK}
\email{idh@star.ucl.ac.uk, awxb@star.ucl.ac.uk} 

\and

\author{K. D.~ Gordon, J. A.~ Stansberry} 
\affil{Steward Observatory, 933 N Cherry Ave, University of Arizona, Tucson, 
AZ 85721}         
\email{kgordon@as.arizona.edu,stansber@as.arizona.edu}

\begin{abstract}
A stellar wind module has been developed for the \phoenix\ stellar
atmosphere code for the purpose of computing non-LTE, line-blanketed,
expanding atmospheric structures and detailed synthetic spectra of hot
luminous stars with winds.  We apply the code to observations of
Deneb, for which we report the first positive detections of mm and cm
emission (obtained using the SCUBA and the VLA), as well a strong
upper limit on the 850$\mu$m flux (using the HHT).  The slope of the
radio spectrum shows that the stellar wind is partially ionized.  We
report a uniform-disk angular diameter measurement,
$\overline{\theta_{\rm UD}} = 2.40 \pm 0.06\ {\rm mas}$, from the Navy
Prototype Optical Interferometer (NPOI).  The measured bolometric flux
and corrected NPOI angular diameter yield an effective temperature of
$8600 \pm 500\ {\rm K}$.  Least-squares comparisons of synthetic
spectral energy distributions from 1220 \AA\ to 3.6 cm with the
observations provide estimates for the effective temperature and the
mass-loss rate of $\simeq 8400 \pm 100$ K and $8 \pm 3 \times 10^{-7}$
\Mspyr, respectively. This range of mass-loss rates is consistent with
that derived from high dispersion UV spectra when non-LTE metal-line
blanketing is considered.  We are unable achieve a reasonable fit to a
typical \ha\ P-Cygni profile with any model parameters over a
reasonable range.  This is troubling because the \ha\ profile is the
observational basis for Wind Momentum-Luminosity Relationship.
\end{abstract}


%\newpage

\section{Introduction}
\label{section_a1}

A-type supergiants are the brightest stars at visual wavelengths (up
to $M_V \simeq -9$) and are therefore among the brightest single
stars visible in galaxies.  For this reason, these stars have been of
increasing interest in extragalactic astronomy where they
show potential as independent distance indicators \cite[][hereafter
K99]{bresolin2001,kud99}.  This potential lies in the use of the Wind
Momentum-Luminosity Relationship (WMLR) which is derived from Sobolev
radiation-driven stellar wind theory (K99 and references therein).
Testing this relationship and the theory of radiation driven winds
most critically requires an accurate determination of the stellar
mass-loss rates.  The WMLR states that $\dot{M}v_{\infty} \propto
L^{1/\alpha}/R_\star^{0.5}$, where \mdot\ is the mass-loss rate,
\vterm\ is the terminal velocity of the wind, $R_\star$ is a
photospheric reference radius, and $\alpha$ is a parameter related to
the distribution of atomic line strengths for the spectral lines which
drive the wind.

The application of the WMLR to nearby galaxies requires a calibration
of the relationship using similar stars in the Milky Way.  To determine
mass-loss rates for A-type supergiants, K99 modeled the hydrogen
Balmer lines \ha\ and \hg\ for six stars in the Galaxy and M31 using a
unified stellar atmosphere model. 
In their analysis, K99 synthesized the observed line profiles
by adjusting model values for the mass-loss rate, the velocity law
exponent $\beta$, and a line-broadening parameter, $v_t$, after first
adopting parameters for the effective temperature, \Teff, the radius,
$R_\star$, the gravity, \Logg, and the projected stellar
rotational velocity.

The use of the \ha\ profile as the sole diagnostic of A-type
supergiant wind momenta is problematic because \ha\ emission observed
in stars of this type is known to be strongly variable
\cite[]{rose72,kaufer96}.  Nevertheless, the accuracy of mass-loss
rate determinations for A-type supergiants in M31 is claimed to be at
the level of 15\% \cite[]{mccarthy97}.  However, there has been no
consensus on the mass-loss rate for Deneb ($\alpha$~Cyg, HD~197345,
spectral type A2Ia according to \cite{mk73}), the brighest, nearest,
and best-studied A-type supergiant, which we might expect to serve as
an archetype and benchmark.  Published estimates of its mass-loss rate
range over more than three orders of magnitude, from methods which
include modeling the line profiles \ion{Mg}{2} $\lambda$2802
\cite[][$3.1\times10^{-9} - 1.5\times10^{-7}$ \Mspyr]{kp81} and \ha\
\cite[][$1.7\pm0.4\times10^{-7}$ \Mspyr; $3.7\pm0.8\times10^{-6}$
\Mspyr]{km82,scuderi92}, analysing low-excitation \ion{Fe}{2} lines in
the ultraviolet \cite[][$1-5\times10^{-9}$ \Mspyr]{hens82}, modeling
the IR excess \cite[][$6\times10^{-7}$ \Mspyr]{bc77}, and radio-flux
limits \cite[][$\leq 2\times10^{-7}$ \Mspyr]{atw84}.

It is not unreasonable to suggest that the degree to which
we understand A-type supergiants in general may be tested 
by Deneb.  Therefore, it is troubling that the published
mass-loss estimates for Deneb vary so widely and this fact casts doubt
on the what we can infer about the properties of more distant A-type
supergiants. The proximity of Deneb allows one to more rigorously
constrain its physical parameters and apply several different
techniques to estimate its mass-loss rate. 

For these reasons, in this paper we investiage Deneb's fundamental
properties, and in particular its mass-loss rate.  To do this we
employ a new stellar wind-model atmosphere package developed by
\cite{jpaphd} for use with the {\tt PHOENIX}\footnote{Also see
http://phoenix.physast.uga.edu} general-purpose stellar and planetary
atmosphere code \cite[][and references therein]{hb99} and bring
together new observations of Deneb at visible, infrared, millimeter,
and radio wavelengths.

Our principal motivation for the constructing unified wind models has
been that most analyses of early-type supergiants still suffer from
severe limitations present in the model atmospheres which are
employed.  Recent systematic analyses \cite[]{vtg99} of the
atmospheric parameters of A-type supergiants have employed LTE,
line-blanketed, plane-parallel, static model atmospheres.
These models do not include the effects (on the temperature
structure and the synthetic spectrum) of metal line-blanketing 
and a spherically extended expanding atmosphere.
Furthermore, most analyses do not allow for departures from LTE,
particularly for species heavier than H and He, in computation of the
opacity and line formation, which are especially important in
low-density extended atmospheres.  \cite{aph97} have developed models,
employed by K99, which address many of these limitations including the
solution of the spherical expanding transfer equation in the co-moving
frame and the non-LTE treatment of hydrogen and helium.  Despite these
important improvements, these models have had limited success in
fitting spectral energy distributions (SEDs) due to the lack of metal
line-blanketing in both the atmospheric structure and the synthetic
spectrum.  Fundamental stellar parameters derived from line-profile
fitting of hydrogen and helium lines alone may not be consistent with
the SED.  Therefore, we believe that fitting both the line spectrum
and the SED simultaneously should lead to more robust atmospheric
parameters for A-type supergiants.

In \S 2 we describe the spectrophotometric, spectroscopic, and
interferometic observations used here. In \S3 limb-darkening models
and corrections to the angular diamter are discussed.  The effective
temperature, reddening and other parameters constrained by the
observations are discussed in \S 4.  In \S 5 we compare the
spectrophotometic data from the UV to the radio with our synthetic
spectral energy distributions.  The comparison of our models to
portions of Deneb's line spectrum is discussed in \S 6.  We summarize
our results and conclusions in \S 7.  Details on the model atmosphere
construction and computation are left for Appendix A.

\section{Observations}

Previous radio observations of Deneb \cite[][and references
therein]{dl89} have yielded only upper limits.  We report
positive detections of radiation from Deneb at 1.35 mm with the
Submillimetre Common-User Bolometer Array (SCUBA) located on the James
Clerk Maxwell Telescope (JCMT) and at 3.6 cm with the Very Large Array
(VLA).
%by (IDH and A. Brown?).
In addition, two of us (KDG \& JAS)
observed Deneb at 870 \micron\ with the Heinrich Hertz Telescope (HHT)
and we report an upper limit on the flux density at this wavelength.
Futhermore, one of us (TEN) measured Deneb's angular diameter, previously
uncertain at the 30\% level, with the Navy Prototype Optical
Interferometer (NPOI) and we present an angular diameter measurement
accurate to 3\%.  In addition to these new data, we have brought
together observations of Deneb from Infrared Processing and Analysis
Center (IPAC) data archive.  The IR, millimeter and radio data are
summarized in Table \ref{tab:irdata}.

\begin{deluxetable}{rrl}
\tablecolumns{3} 
\tabletypesize{\small}
\tablecaption{Infrared, Millimeter, and Radio Photometry of Deneb} 
\tablewidth{0pt}
\tablehead{
\colhead{Wavelength} & \colhead{Flux Density} & \colhead{Observation}}
\startdata
1.25  \micron  &$609 \pm 19$ Jy     & COBE/DIRBE band 1A  \tablenotemark{a} \\
2.2   \micron  &$268 \pm  8$ Jy     & COBE/DIRBE band 2A  \tablenotemark{a} \\
2.2   \micron  &$282 \pm 28$ Jy     & IRTF $K$ band       \tablenotemark{b} \\
3.5   \micron  &$125 \pm  4$ Jy     & COBE/DIRBE band 3A  \tablenotemark{a} \\
3.5   \micron  &$122 \pm 12$ Jy     & IRTF $L$ band       \tablenotemark{b} \\
4.29  \micron  &$89.5\pm  9.5$ Jy   & MSX band B1         \tablenotemark{c} \\
4.35  \micron  &$84.5\pm  8.0$ Jy   & MSX band B2         \tablenotemark{c} \\
4.8   \micron  &$81  \pm   8$   Jy  & IRTF $M$ band       \tablenotemark{b} \\
4.9   \micron  &$74.5 \pm 2.2$Jy    & COBE/DIRBE band 4   \tablenotemark{a} \\
7.76  \micron  &$33.6\pm 1.7$ Jy    & MSX band A          \tablenotemark{c} \\
10.2  \micron  &$22 \pm    2$ Jy    & IRTF $N$ band       \tablenotemark{b} \\
11.99 \micron  &$15.6\pm 0.5$ Jy    & MSX band B          \tablenotemark{c} \\
14.55 \micron  &$11.2\pm 0.5$ Jy    & MSX band C          \tablenotemark{c} \\
20.0  \micron  &$6.7 \pm 0.7$ Jy    & IRTF $Q$ band       \tablenotemark{b} \\
20.68 \micron  &$5.38\pm 0.45$ Jy   & MSX band D          \tablenotemark{c} \\
60    \micron  &$957 \pm 287$ mJy   & ISOPHOT             \tablenotemark{d} \\
870   \micron  &$<24$ mJy           & HHT                 \tablenotemark{d} \\
1.35  mm       &$7.8 \pm  1.9$ mJy  & SCUBA               \tablenotemark{d} \\
2     cm       &$<1.0$ mJy          & VLA                 \tablenotemark{e} \\
3.6   cm       &$0.23 \pm 0.05$ mJy & VLA                 \tablenotemark{d} \\
6     cm       &$<0.15$ mJy         & VLA                 \tablenotemark{e} \\ 
\enddata
\tablenotetext{a}{http://cobe.gsfc.nasa.gov/cio/}
\tablenotetext{b}{\cite{atw84}}
\tablenotetext{c}{\cite{msxdata,msxguide}}
\tablenotetext{d}{This paper}
\tablenotetext{e}{\cite{dl89}}
\label{tab:irdata}
\end{deluxetable}

\subsection{SCUBA Observation}
SCUBA was used to observe Deneb on 4 May 1998.  Four runs were made in
the photometery mode using the 1350 $\micron$ filter. Each run
consisted of 50 integrations of 18 sec each.  Each integration
includes both the on-source and off-source time, the on-source time is
half the value.  Deneb was positively detected at 1.35 mm with flux
density of 7.8$\pm$ 1.9 mJy.  

\subsection{VLA Observation}
Deneb was observed by 22 antennas of the VLA in the
D$\longrightarrow$A array configuration on 29 January 1990.  The
central frequency of the observation was 8414 MHz with bandwidth of 50
MHz.  The total observation time was 4800 sec, with 3450 sec on
source.  Deneb was positively detected at 3.6 cm with a flux density of
0.23$\pm$0.05 mJy.

\subsection{HHT Observation}
Deneb was observed using the 10-m Heinrich Hertz Telescope 
\cite[e.g.,][]{baars99} on Mt. Graham, Arizona on 22 December 2000.  We
used the MPIfR 19-element bolometer array which is sensitive at
$870~\micron = 345$~GHz.  The observations were done during good weather
conditions ($\tau_{225GHz} < 0.06$).  We used source-sky offsets of
$100\arcsec$ compared to the telescope beam of $22.4\arcsec$.  The
total on source integration time was 1600 seconds with individual
integration times of 20 seconds.  Mars was observed to provide
calibration to physical units.  While we did not detect Deneb, we
were able to put a $3\sigma$ upper limit on this star of 24.0~mJy at
870~$\micron$.

\subsection{ISO Data Reduction, DIRBE and MSX Data}
Infrared Space Observatory (ISO) observed Deneb at 60 $\micron$ with
the ISOPHOT instrument \cite[]{lemke96}.  This data set was downloaded
from the ISO archive and reduced using the PIA package
\cite[]{gabriel97}.  The observations consist of three pointings
bracketed by observations of the FCS1 calibration source.  Each
pointing was offset by almost the full array width from the last
pointing and Deneb was observed in the middle pointing.  Deneb was
centered on the middle pixel of the array.  The sky level at the
position of Deneb was taken as the mean of the sky levels on the
middle pixel in the two adjacent observations.  The flux of Deneb was
the difference between the observed flux and the sky level so
computed.  Including the correction for the flux falling outside the
central pixel, the $60\ \micron$ flux of Deneb was 957~mJy.  The
uncertainty of this measurement was estimated at 30\%
\cite[]{schulz99}.

Infrared photometric data between 1 \micron\ to 5 \micron\ are from
the Diffuse Infrared Background Experiment (DIRBE) aboard the Cosmic
Background Explorer (COBE). These data were obtained from the
Astrophysical Data Facility using the the DIRBE Point Source
Photometry Research Tool.  Additional infrared photometric data
between 4 \micron\ and 20 \micron\ from the Midcourse Space
Experiment (MSX) \cite[]{msxdata,msxguide} were obtained through IPAC.

\subsection{Ultraviolet Data Sources}
\begin{figure}
\includegraphics[scale=0.7,angle=90]{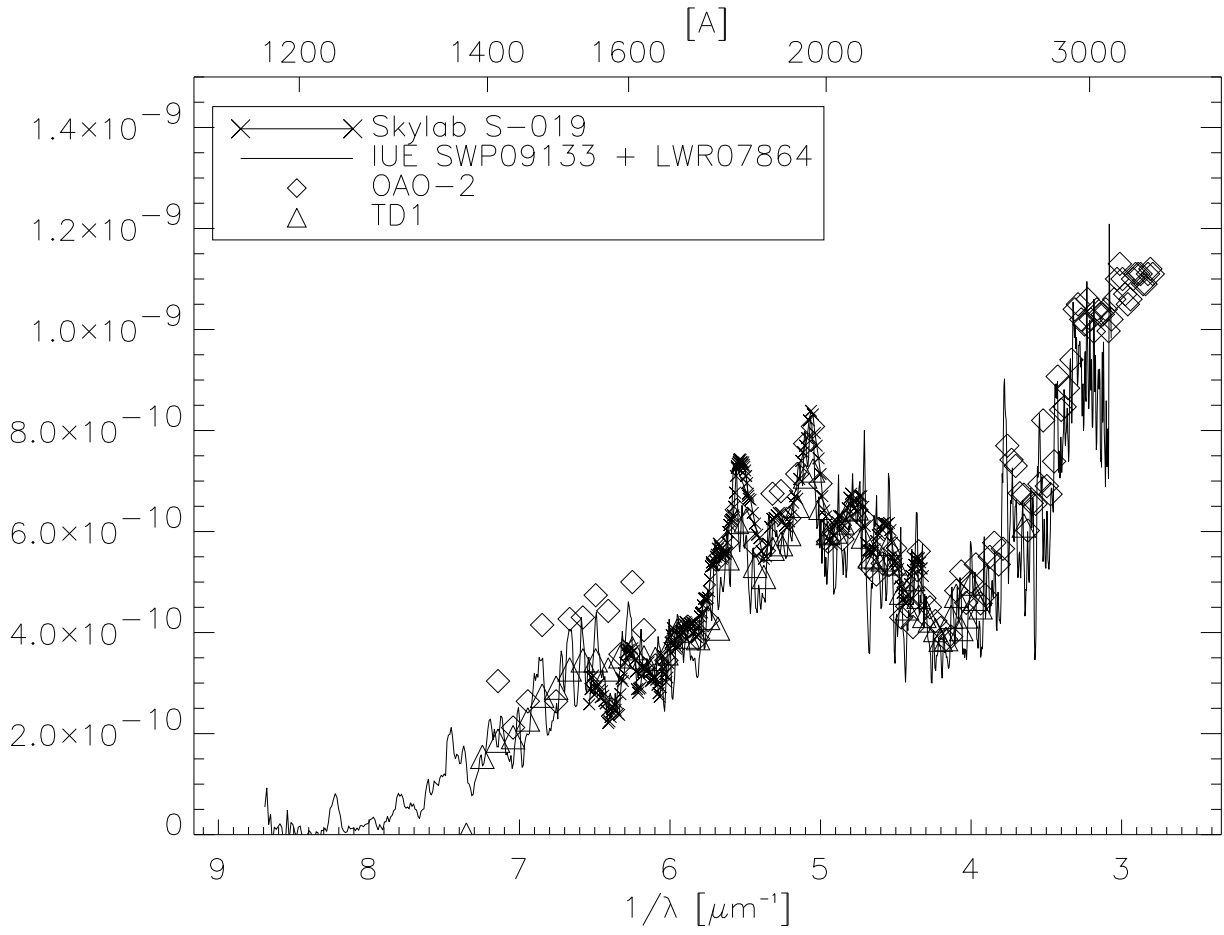}
\caption{
Comparison of ultraviolet absolute spectrophotometry
of Deneb from {\it Skylab}, {\it IUE}, {\it OAO-2}, and 
{\it TD1}. See text for details.
}
\label{uvdata}
\end{figure}

Spectrophotometric data of Deneb from the {\it Skylab} S-0109
Far-UV Objective Prism Spectrophotometer \cite[]{skylab}, the {\it
Orbiting Astronomical Observatory-2 (OAO-2)} spectrometers
\cite[]{oao2}, the S2/68 Ultraviolet Sky Survey Telescope aboard {\it
TD1} \cite[]{td1}, and the {\it International Ultraviolet Explorer
(IUE)} are all in close agreement regarding the shape and absolute
flux level despite being obtained at different epochs (see Table
\ref{tab:uvdata}).  
Figure \ref{uvdata} shows these data 
over their common interval from 7 ${\micron}^{-1}$ to 4
${\micron}^{-1}$ (1400 \AA\ to 2500 \AA). 
The \oaotwo data do show systematically higher fluxes relative to the
other data sets between 7 ${\micron}^{-1}$ and 6 ${\micron}^{-1}$.
This \oaotwo excess has been noted in studies on the UV
spectrophotometry of B-type stars \cite[]{boholm,betacma} and is
expected.  There is good agreement between \iue and \oaotwo longward
of 2500 \AA\ until the end of the LWR camera beyond 3300 \AA.  The only
absolute spectrophotometric data shortward of 1400 \AA\ are from {\it IUE}.

\begin{deluxetable}{lccc}
\tablecolumns{3} 
\tabletypesize{\small}
\tablecaption{Low-Resolution UV Spectrophotometry of Deneb} 
\tablewidth{0pt}
\tablehead{
\colhead{Observatory} &Wavelength Range (\AA) & \colhead{Dates} &\colhead{\# of Scans}}
\startdata
{\it OAO-2}\tablenotemark{a}  &1840--3600   &1969 June 19 &1 \\
{\it OAO-2} &1400--1800   &1970 May 11     &1\\
            &             &1970 June 21    &4 \\
{\it TD1}\tablenotemark{b}   &1360--2740   &1972 March 19   &3\\
            &             &1972 October 31 &-- \\
{\it Skylab}&1347--2290   &1973 August 10  &3 \\
{\it IUE}   &1150--2000   &1980 May 26     &1 \\
{\it IUE}   &1850--3300   &1980 May 26     &1 \\
\enddata
\tablenotetext{a}{{\it OAO-2} observation dates from \cite{meade2001}.}
\tablenotetext{b}{No observation dates available. Dates shown are those of the {\it TD1} mission.}
\label{tab:uvdata}
\end{deluxetable}

%%%%%%%%%%%%%%%%%%%%%%%%%%%%
%% FIGURE 4 
%%%%%%%%%%%%%%%%%%%%%%%%%%%%

Only two large aperture, low dispersion spectra of Deneb were obtained
with {\it IUE}, one each in the Short Wavelength Primary (SWP) and
Long Wavelength Redundant (LWR) cameras, exposures SWP 09133 and LWR
07864, respectively.  We use the \iue data set exclusively in our SED
analysis because it has the greatest wavelength coverage, is
consistent with the other UV data, has probably the best absolute (and
relative) flux calibration, and both the long and short wavelength
spectra were taken on the same date.  For the high-dispersion
analysis, we have chosen the well-exposed, large-aperture spectra SWP
10375 and LWR 07760. The \iue\ data were obtained from the ESA ``IUE
Newly Extracted Spectra'' (INES) system \cite[]{ines99,ines00}.

\subsection{Additional Spectroscopic and Spectrophotometric Data Sources}
Optical spectrophotometric data are from \cite{glush92} and are in
good agreement with data from the compilation by \cite{breger} except
at 3975 \AA, where the datum from \citeauthor{glush92} is $\simeq
50\%$ lower.  Two high-dispersion ($R\simeq 20000$) optical spectra with wavelength
coverage from 3500 \AA\ to 5400 \AA\ (``blue'') and 5775 \AA\ to 8630 \AA\
(``red'') were obtained on 1 August 1995 with the Heidelberg echelle spectrograph
HEROS \cite[]{kaufer99}.  Fourier Transform Spectroscopy ($R\simeq 3000$)
in the $K$-band is from \cite{wallhink97}.

\subsection{NPOI Observations}
\begin{figure}
\includegraphics[scale=0.9]{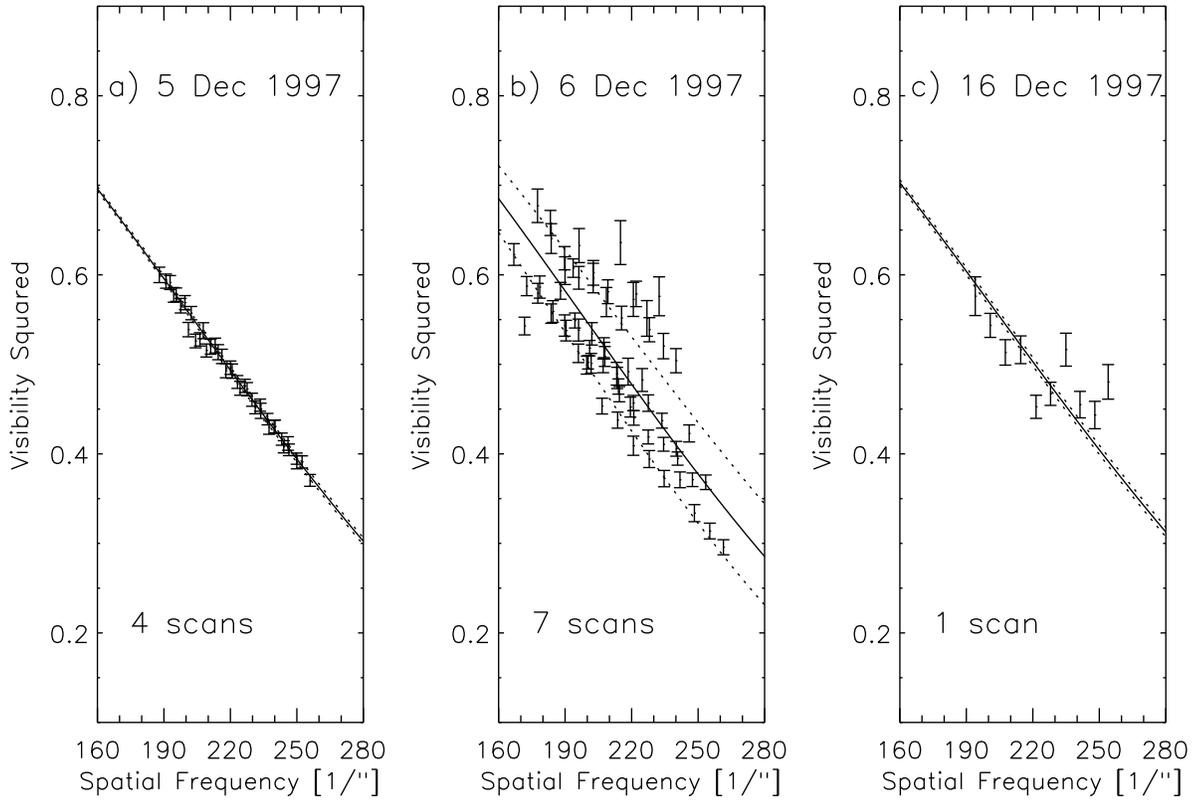}
\caption{NPOI visibility data of Deneb with uniform-disk (UD) fits.
For each date, a model visibility curve (solid line) for the mean UD
diameter, $\overline{\theta_{\rm UD}}$ (see Table
\ref{ud_fits_table}), is plotted along with the corresponding
visibility curves for the $\pm$ 1$\sigma$ error in the mean UD
diameter (dotted lines).  (a) 5 December 1997: $\overline{\theta_{\rm
UD}}$ = $2.36 \pm 0.01$ mas (4 scans); (b) 6 December 1997:
$\overline{\theta_{\rm UD}}$ = $2.41 \pm 0.17$ mas (7 scans); (c) 16
December 1997: \ADUD\ = $2.33\pm 0.02$ mas (1 scan).}
\label{ud_fits}
\end{figure}

Each interferometric observation consists of a scan made over ten
spectral channels spread in increments of equal wavenumber from 850 nm
to 650 nm with a mean wavelength of 740 nm.  Here we consider only the
visibility data from the 38-meter East-West baseline, considered the
most robust \cite[]{nordgren99b}.  Deneb was observed 74 times over 5
nights in 1997 December.  Unfortunately, 62 of these observations have
no suitable calibration data.  Of the remaining data, 5 scans from
1997 December 5, 7 scans from 1997 December 6, and 1 scan from 1997
December 16 have suitable calibration data taken within one hour of
each science scan. These data are listed in Table
\ref{npoi_data_table} and plotted in Figure \ref{ud_fits}.  The
calibration star for each night and its estimated angular size
\cite[]{nordgren99b} are noted in Table \ref{npoi_data_table}.  For a
detailed description of the calibration process used for these data
see \cite{nordgren99b}.

\begin{deluxetable}{cccllc}
\tabletypesize{\scriptsize}
\tablecaption{NPOI Deneb Observations} 
\tablewidth{0pt}
\tablecolumns{6}
\tablehead{
 \colhead{Observation Time} &\colhead{Cal. Time\tablenotemark{a}} 
&\colhead{Wavelength}       &\colhead{$V^2$} 
&\colhead{\ }                &\colhead{Spatial Frequency}\\
 \colhead{(UT)}             &\colhead{(hours)} 
&\colhead{(nm)}             &\colhead{ } 
&\colhead{\ }                &\colhead{(cycles/arcsecond)}
}
\startdata
\cutinhead{1997 December 5\tablenotemark{b}}
 01:40:05  &0.305 &849.4    &0.5778 $\pm$   &0.0075     &195.5441  \\
 01:40:05  &0.305 &820.9    &0.5572    &0.0074     &202.3308    \\
 01:40:05  &0.305 &793.9    &0.5155    &0.0073     &209.2131    \\
 01:40:05  &0.305 &768.3    &0.5089    &0.0080     &216.1828    \\
 01:40:05  &0.305 &744.2    &0.4803    &0.0072     &223.1807    \\
 01:40:05  &0.305 &722.9    &0.4605    &0.0073     &229.7603    \\
 01:40:05  &0.305 &701.5    &0.4388    &0.0064     &236.7691    \\
 01:40:05  &0.305 &683.1    &0.4165    &0.0068     &243.1457    \\
 01:40:05  &0.305 &664.6    &0.3947    &0.0063     &249.9149    \\
 01:40:05  &0.305 &648.9    &0.3704    &0.0066     &255.9606    \\
 01:52:05  &0.505 &849.4    &0.5914    &0.0078     &192.7878    \\
 01:52:05  &0.505 &820.9    &0.5689    &0.0079     &199.4806    \\
 01:52:05  &0.505 &793.9    &0.5279    &0.0074     &206.2675    \\
 01:52:05  &0.505 &768.3    &0.5210    &0.0083     &213.1379    \\
 01:52:05  &0.505 &744.2    &0.4962    &0.0075     &220.0392    \\
 01:52:05  &0.505 &722.9    &0.4759    &0.0074     &226.5235    \\
 01:52:05  &0.505 &701.5    &0.4544    &0.0068     &233.4355    \\
 01:52:05  &0.505 &683.1    &0.4302    &0.0076     &239.7208    \\
 01:52:05  &0.505 &664.6    &0.4045    &0.0065     &246.3961    \\
 01:52:05  &0.505 &648.9    &0.3910    &0.0068     &252.3545    \\
 01:59:49  &0.634 &849.4    &0.5929    &0.0079     &190.9612    \\
 01:59:49  &0.634 &820.9    &0.5655    &0.0081     &197.5905    \\
 01:59:49  &0.634 &793.9    &0.5261    &0.0076     &204.3120    \\
 01:59:49  &0.634 &768.3    &0.5197    &0.0084     &211.1178    \\
 01:59:49  &0.634 &744.2    &0.4928    &0.0078     &217.9575    \\
 01:59:49  &0.634 &722.9    &0.4733    &0.0078     &224.3784    \\
 01:59:49  &0.634 &701.5    &0.4519    &0.0071     &231.2249    \\
 01:59:49  &0.634 &683.1    &0.4297    &0.0078     &237.4519    \\
 01:59:49  &0.634 &664.6    &0.4079    &0.0068     &244.0599    \\
 01:59:49  &0.634 &648.9    &0.3907    &0.0074     &249.9678    \\
 02:12:32  &0.846 &849.4    &0.5999    &0.0086     &187.9221    \\
 02:12:32  &0.846 &820.9    &0.5776    &0.0082     &194.4449    \\
 02:12:32  &0.846 &793.9    &0.5388    &0.0080     &201.0607    \\
 02:12:32  &0.846 &768.3    &0.5376    &0.0089     &207.7591    \\
 02:12:32  &0.846 &744.2    &0.5136    &0.0084     &214.4886    \\
 02:12:32  &0.846 &722.9    &0.4924    &0.0080     &220.8083    \\
 02:12:32  &0.846 &701.5    &0.4722    &0.0074     &227.5415    \\
 02:12:32  &0.846 &683.1    &0.4481    &0.0085     &233.6719    \\
 02:12:32  &0.846 &664.6    &0.4306    &0.0071     &240.1741    \\
 02:12:32  &0.846 &648.9    &0.4116    &0.0076     &245.9880    \\
\cutinhead{1997 December 6\tablenotemark{c}}
 01:16:52  &0.087 &849.4    &0.4991 $\pm$   &0.0100    &199.7180  \\
 01:16:52  &0.087 &820.9    &0.4534    &0.0088    &206.6525  \\
 01:16:52  &0.087 &793.9    &0.4377    &0.0090    &213.6804  \\
 01:16:52  &0.087 &768.3    &0.4091    &0.0107    &220.8014  \\
 01:16:52  &0.087 &744.2    &0.3942    &0.0095    &227.9508  \\
 01:16:52  &0.087 &722.9    &0.3724    &0.0091    &234.6640  \\
 01:16:52  &0.087 &701.5    &0.3710    &0.0088    &241.8242  \\
 01:16:52  &0.087 &683.1    &0.3338    &0.0095    &248.3401  \\
 01:16:52  &0.087 &664.6    &0.3138    &0.0088    &255.2507  \\
 01:16:52  &0.087 &648.9    &0.2957    &0.0084    &261.4261  \\
 01:44:35  &0.074 &849.4    &0.6075    &0.0103    &193.6147  \\
 01:44:35  &0.074 &820.9    &0.5023    &0.0075    &200.3382  \\
 01:44:35  &0.074 &793.9    &0.4984    &0.0077    &207.1531  \\
 01:44:35  &0.074 &768.3    &0.4752    &0.0085    &214.0514  \\
 01:44:35  &0.074 &744.2    &0.4401    &0.0081    &220.9846  \\
 01:44:35  &0.074 &722.9    &0.4189    &0.0078    &227.4956  \\
 01:44:35  &0.074 &701.5    &0.4107    &0.0077    &234.4356  \\
 01:44:35  &0.074 &683.1    &0.3948    &0.0074    &240.7516  \\
 01:44:35  &0.074 &664.6    &0.3711    &0.0076    &247.4509  \\
 01:44:35  &0.074 &648.9    &0.3683    &0.0081    &253.4416  \\
 02:14:50  &0.048 &849.4    &0.5821    &0.0094    &187.9029  \\
 02:14:50  &0.048 &820.9    &0.5490    &0.0084    &194.4257  \\
 02:14:50  &0.048 &793.9    &0.5181    &0.0086    &201.0415  \\
 02:14:50  &0.048 &768.3    &0.5153    &0.0093    &207.7393  \\
 02:14:50  &0.048 &744.2    &0.4667    &0.0085    &214.4687  \\
 02:14:50  &0.048 &722.9    &0.4568    &0.0085    &220.7885  \\
 02:14:50  &0.048 &701.5    &0.4567    &0.0092    &227.5210  \\
 02:14:50  &0.048 &683.1    &0.4371    &0.0083    &233.6514  \\
 02:14:50  &0.048 &664.6    &0.4057    &0.0083    &240.1537  \\
 02:14:50  &0.048 &648.9    &0.4228    &0.0095    &245.9638  \\
 02:27:25  &0.054 &849.4    &0.6579    &0.0140    &183.3663  \\
 02:27:25  &0.054 &820.9    &0.6186    &0.0130    &189.7327  \\
 02:27:25  &0.054 &793.9    &0.5967    &0.0127    &196.1848  \\
 02:27:25  &0.054 &768.3    &0.6022    &0.0138    &202.7225  \\
 02:27:25  &0.054 &744.2    &0.5811    &0.0133    &209.2878  \\
 02:27:25  &0.054 &722.9    &0.5517    &0.0132    &215.4553  \\
 02:27:25  &0.054 &701.5    &0.5787    &0.0142    &222.0274  \\
 02:27:25  &0.054 &683.1    &0.5387    &0.0135    &228.0098  \\
 02:27:25  &0.054 &664.6    &0.5204    &0.0142    &234.3555  \\
 02:27:25  &0.054 &648.9    &0.5041    &0.0135    &240.0260  \\
 02:51:50  &0.060 &849.4    &0.6771    &0.0187    &177.5475  \\
 02:51:50  &0.060 &820.9    &0.6405    &0.0166    &183.7125  \\
 02:51:50  &0.060 &793.9    &0.6047    &0.0155    &189.9600  \\
 02:51:50  &0.060 &768.3    &0.6327    &0.0190    &196.2900  \\
 02:51:50  &0.060 &744.2    &0.5964    &0.0165    &202.6481  \\
 02:51:50  &0.060 &722.9    &0.5696    &0.0163    &208.6200  \\
 02:51:50  &0.060 &701.5    &0.6360    &0.0245    &214.9848  \\
 02:51:50  &0.060 &683.1    &0.5725    &0.0190    &220.7742  \\
 02:51:50  &0.060 &664.6    &0.5513    &0.0199    &226.9180  \\
 02:51:50  &0.060 &648.9    &0.5760    &0.0217    &232.4076  \\
 03:18:22  &0.065 &849.4    &0.5428    &0.0099    &171.7017  \\
 03:18:22  &0.065 &820.9    &0.5794    &0.0111    &177.6646  \\
 03:18:22  &0.065 &793.9    &0.5567    &0.0108    &183.7040  \\
 03:18:22  &0.065 &768.3    &0.5434    &0.0119    &189.8256  \\
 03:18:22  &0.065 &744.2    &0.5123    &0.0103    &195.9740  \\
 03:18:22  &0.065 &722.9    &0.5012    &0.0110    &201.7472  \\
 03:18:22  &0.065 &701.5    &0.5088    &0.0111    &207.9038  \\
 03:18:22  &0.065 &683.1    &0.4848    &0.0119    &213.5034  \\
 03:18:22  &0.065 &664.6    &0.4523    &0.0111    &219.4446  \\
 03:18:22  &0.065 &648.9    &0.4825    &0.0128    &224.7564  \\
 03:44:02  &0.095 &849.4    &0.6227    &0.0121    &166.8693  \\
 03:44:02  &0.095 &820.9    &0.5874    &0.0107    &172.6637  \\
 03:44:02  &0.095 &793.9    &0.5863    &0.0124    &178.5318  \\
 03:44:02  &0.095 &768.3    &0.5590    &0.0121    &184.4802  \\
 03:44:02  &0.095 &744.2    &0.5377    &0.0115    &190.4547  \\
 03:44:02  &0.095 &722.9    &0.5383    &0.0123    &196.0689  \\
 03:44:02  &0.095 &701.5    &0.5341    &0.0125    &202.0478  \\
 03:44:02  &0.095 &683.1    &0.5186    &0.0111    &207.4907  \\
 03:44:02  &0.095 &664.6    &0.4895    &0.0127    &213.2697  \\
 03:44:02  &0.095 &648.9    &0.4918    &0.0150    &218.4261  \\
\cutinhead{1997 December 16\tablenotemark{d}}
 01:03:29  &0.214 &849.4    &0.5761 $\pm$   &0.0216      &194.0322  \\
 01:03:29  &0.214 &820.9    &0.5435    &0.0134      &200.7689  \\
 01:03:29  &0.214 &793.9    &0.5133    &0.0142      &207.5969  \\
 01:03:29  &0.214 &768.3    &0.5163    &0.0155      &214.5110  \\
 01:03:29  &0.214 &744.2    &0.4526    &0.0129      &221.4574  \\
 01:03:29  &0.214 &722.9    &0.4671    &0.0129      &227.9843  \\
 01:03:29  &0.214 &701.5    &0.5164    &0.0184      &234.9374  \\
 01:03:29  &0.214 &683.1    &0.4550    &0.0148      &241.2693  \\
 01:03:29  &0.214 &664.6    &0.4434    &0.0153      &247.9844  \\
 01:03:29  &0.214 &648.9    &0.4804    &0.0193      &253.9841  \\
\enddata
\tablenotetext{a}{Time between the science source scan
and the calibration scan used to calibrate it.}
\tablenotetext{b}{Calibration: $\gamma$ Peg, $\theta_{\rm UD}$ = 0.421 mas}
\tablenotetext{c}{Calibration: $\alpha$ Lac, $\theta_{\rm UD}$ = 0.502 mas}
\tablenotetext{d}{Calibration: $\zeta$  Cas, $\theta_{\rm UD}$ = 0.264 mas}
\label{npoi_data_table}
\end{deluxetable}

%%%%%%%%%%%%%%%%%%%%%%%%%%%%
%% FIGURE 1 
%%%%%%%%%%%%%%%%%%%%%%%%%%%%

A uniform-disk (UD) angular diameter \ADUD\ was derived for each
calibrated scan by fitting an Airy pattern
\begin{equation}	
V^2(kw) = \Biggl[\frac{J_1(2kw\alpha_{\rm UD})}{kw\alpha_{\rm UD}}\Biggl]^2
\label{eqn:airy}
\end{equation}
to the visibility data, where $kw$ is the spatial frequency,
$k=2\pi/\lambda$ is the wavenumber and $w$ is the projected baseline,
$\alpha_{\rm UD}= \theta_{\rm UD}/2$ is the angular radius, and $J_1$ is the Bessel
function of first order.  Table \ref{ud_fits_table} lists the least squares
best fit value of \ADUD\ for each scan and the mean value of \ADUD\ for
each night.  The model visibility curves corresponding to the mean and
$\pm 1\sigma$ values of \ADUD\ for each night are shown in Figure \ref{ud_fits}.

\begin{deluxetable}{crc}
%\tabletypesize{\scriptsize}
\tablecaption{Uniform-Disk Angular Diameters} 
\tablewidth{0pt}
\tablecolumns{3}
\tablehead{
\colhead{Observation Time} &\colhead{   \ADUD} &\colhead{$\chi^2$} \\
\colhead{(UT)}             &\colhead{   (mas)} &\colhead{ }}
\startdata
\cutinhead{1997 December 5: $\overline{\theta_{\rm UD}}$ =  2.364 $\pm$ 0.011}
 01:40:05   & 2.364 $\pm$ 0.007  &4.7\\
 01:52:05  & 2.353 $\pm$ 0.007  &5.0\\
 01:59:49   & 2.379 $\pm$ 0.007  &4.9\\
 02:12:32   & 2.360 $\pm$ 0.008  &7.1\\
\cutinhead{1997 December 6:  $\overline{\theta_{\rm UD}}$  = 2.410 $\pm$ 0.168}
 01:16:52   & 2.578 $\pm$ 0.009  &17 \\
 01:44:35   & 2.470 $\pm$ 0.008  &57 \\
 02:14:50   & 2.432 $\pm$ 0.009  &38 \\
 02:27:25   & 2.180 $\pm$ 0.014  &20 \\
 02:51:50   & 2.169 $\pm$ 0.020  &30 \\
 03:18:22   & 2.541 $\pm$ 0.012  &90 \\
 03:44:02   & 2.501 $\pm$ 0.014  &32 \\
\cutinhead{1997 December 16:  $\theta_{\rm UD}$  = 2.325 $\pm$ 0.015}
 01:03:29    &2.325 $\pm$ 0.015 &61  \\
\enddata
\label{ud_fits_table}
\end{deluxetable}

\begin{figure}
\includegraphics[scale=0.7,angle=90]{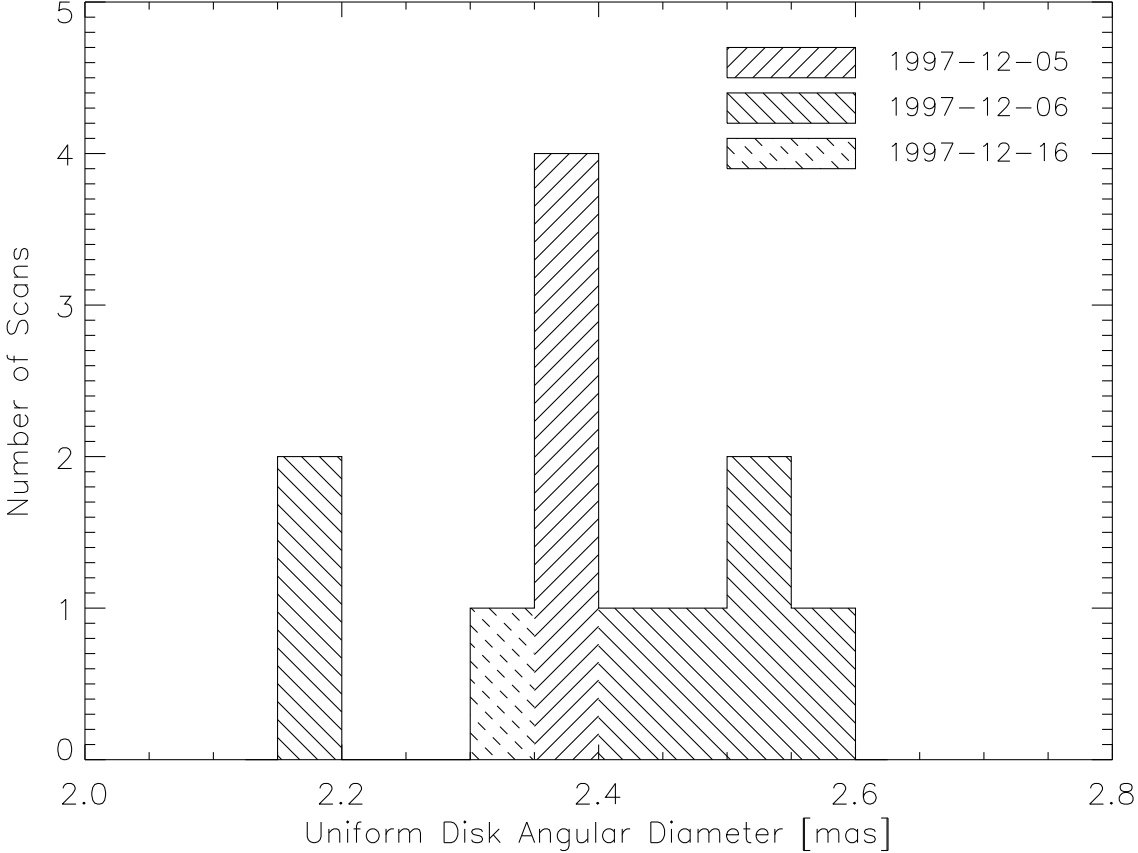}
\caption{Histogram of the uniform-disk angular diameter values derived
from each of the 12 scans of Deneb by the NPOI (see Table
\ref{ud_fits_table}).  The data are binned to 0.05 mas.  The different
cross-hatched regions specify the date of each scan.}
\label{histogram}
\end{figure}

The scans of December 5th are very consistent with each other and with
the single scan from December 16th which, while noisier, yields a
consistent angular diameter.  The scans from December 6th, however,
show considerable scatter and each scan is considerably noiser than
the December 5th data, yielding higher $\chi^2$ values for each of the
fits.  A histogram of all the derived \ADUD\ values is shown in Figure
\ref{histogram}.  Experience has shown that with well-observed stars 
the distribution of angular diameters follows a Gaussian
distribution in the limit of many ($\sim$100) scans.  
The weighted mean angular diameter for the 12 scans is
\begin{equation}	
\overline{\theta_{\rm UD}} = 2.40 \pm 0.06\ {\rm mas},
\label{eqn:udad}
\end{equation}
where the uncertainty represents our estimate for the error in the
mean.  This value is consistent with the published measurements by the
CERGA group: \ADUD = $2.04 \pm 0.45$ mas (in the spectral band $600
\pm 25$ nm) \cite[]{korab85} and \ADUD = $2.6 \pm 0.3$ mas (in the
spectral range 500-650 nm) \cite[]{bonneau81}.  Hanbury Brown's
interferometer at Narrabri could not observe Deneb due to its high
declination.

%%%%%%%%%%%%%%%%%%%%%%%%%%%%
%% 
%%%%%%%%%%%%%%%%%%%%%%%%%%%%

\section{Angular Diameter}
Our model atmospheres (described in detail in Appendix~A) predict
substantial limb darkening for the Deneb parameters determined in the
remainder of this paper.  This motivates further analysis of the NPOI
observations in order to obtain a more physically based estimate of
the angular diameter.  A limb-darkened angular diameter, \ADLD, is
normally employed in the calculation of the fundamental effective
temperature
\begin{equation}
{\rm T}_{\rm eff} = \Biggl[\frac{4\mathcal{F}_0}
{\sigma\theta_{\rm LD}^2}\Biggr]^{1/4}.
\label{eqn:fundteff}
\end{equation}
where $\mathcal{F}_0$ is the bolometric flux (corrected for interstellar
extinction) and $\sigma$ is the Stefan-Boltzmann constant.
Unfortunately, in extended atmospheres the effective temperature is
not well defined or is ambiguous because the atmosphere is not
sufficiently compact to have a unique radius \cite[]{baschek91}.
Hence, a value for \Teff\ cannot be separated from an adopted value
for the radius and the corresponding angular size.

For well resolved sources (e.g., Betelgeuse, Arcturus), observations
can directly yield limb-darkening information from the shape of the
visibility curve at spatial frequencies beyond the first null.  Deneb,
however, has yet to be observed at sufficiently long baselines ($>
100$ m) for limb-darkening information to be obtained in this
way. Therefore, the details of Deneb's center-to-limb variation (CLV)
must for now be supplied by theory. 

\begin{figure}
\includegraphics[scale=0.7,angle=90]{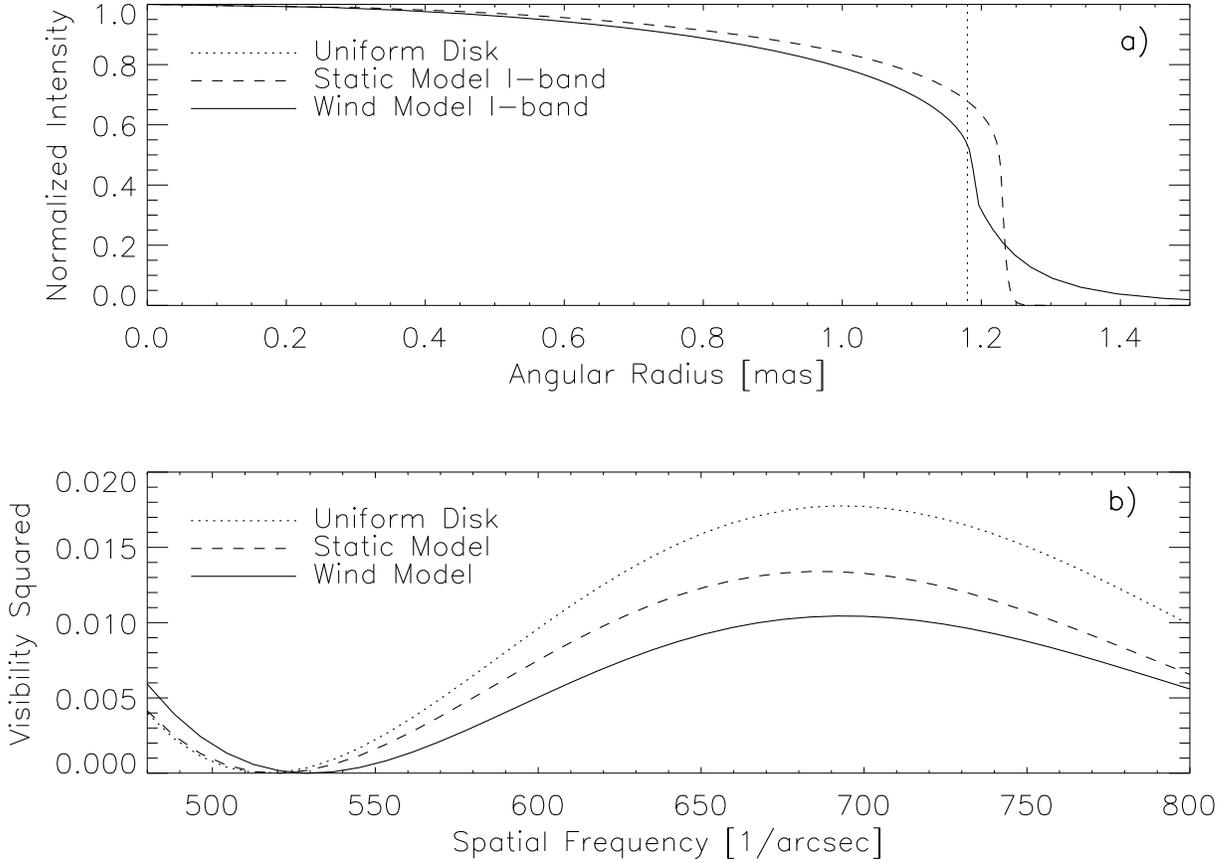}
\caption{(a) Comparison of the $I$-band center-to-limb variations
(CLVs) from two model atmospheres and a uniform-disk: uniform-disk
(dotted line), spherical, hydrostatic model (dashed line), and
spherical hydrostatic plus expanding wind model (solid line).  The
three CLVs have been scaled so that their visibility curves match the
NPOI data from 5 December 1997 at 220 arcsec$^{-1}$. (b) Different
CLVs can be most easily discriminated beyond the first null; the
uniform-disk squared visibilities are predicted to be 60\% larger than
the wind model squared visibilities at a spatial frequency of 700
arcsec$^{-1}$ in the $I$-band.  
The spherical, expanding atmosphere
model is parameterized by: $T_\star$ = 9000 K at $R_\star$ =
$1.2\times 10^{13}$ cm and a surface gravity, $\log g(R_\star)$ = 1.3,
equivalent to \Teff\ = 8543 K at $R(\tau_{\rm Ross} = 2/3)$ and $\log
g(R)$ = 1.21, see equation \ref{eqn:teffross}; a mass-loss rate,
\mdot\ = $5.0\times10^{-7}$ \Mspyr; a terminal wind velocity,\vterm\ =
225 \kms; and a velocity law exponent, $\beta$ = 3.0 (see \S A.2). The
spherical, hydrostatic model is parameterized by: $T_\star$ = 8500 K
at $R_\star$ = $1.2\times 10^{13}$ cm and $\log g(R_\star)$ = 1.3,
equivalent to \Teff\ = 8746 K at $R(\tau_{\rm Ross} = 2/3)$ and $\log
g(R)$ = 1.35.}
\label{clv_comp}
\end{figure}

Limb-darkening coefficients provided in the literature (e.g.,
\cite{claret2000}) are derived from plane-parallel models and are not
appropriate for Deneb.  For example, Figure {\ref{clv_comp}a} shows
that the model CLV from an expanding atmosphere or ``wind'' model
differ significantly from that of a hydrostatic model.  Unlike the
wind model CLV which falls very gradually from 40\% of the central
disk intensity to zero, the hydrostatic model CLV shows a fairly well
defined edge.  The unambiguous signature of the predicted wind model
CLV lies in the height of the visibility function beyond the first
null (see Figure {\ref{clv_comp}b}). Future measurements should be
able to test this prediction.

The gradual decline of the wind model CLV presents a problem for
deriving \ADLD.  As noted by \cite{baschek91}, a flat intensity
decrease toward the outer disk leaves an ``intensity radius'', defined
by an arbitrary cutoff of the CLV, e.g. at 1\%, poorly defined.  Such
a profile prohibits a simple relation between an ``optical depth
radius'', which corresponds to reference optical depth along a ray
normal to the atmosphere, and an ``intensity radius'', which
corresponds to the impact parameter of a ray at the angle of the
intensity cutoff.  We choose to define the limb-darkened edge to be
the angle beyond which the model intensity profile would have an
immeasurable effect on the observed visibilities.  This is defined
quantitatively by the following procedure.

The intensity of the stellar disk as a function of the angular radius,
$I(\alpha)$, is extracted from the model atmosphere.  The coordinate
system for the solution of the spherical model atmosphere problem is
specified by a set of parallel rays tangent to nested spherical shells
\cite[e.g., see Figure 1 in][]{hr71, mkh75}.  The rays are
parameterized by the perpendicular distance $p$ from the line of
symmetry though the center of the star.  A solution to the model
atmosphere problem provides the specific intensity $I_\lambda$ at
wavelength $\lambda$ at depths $z$ along each ray.  The observed
intensity for a given ray is taken to be the intensity at the
intersection of that ray with the outermost shell.  The relation
between the directional cosine of the emergent radiation,
$\mu=\cos\theta$, and the angular displacement on the sky, $\alpha$,
is
\begin{equation}	
\alpha = (1-\mu^2)^{1/2}.
\label{eqn:alpha}
\end{equation}

The spectral response of
the NPOI is approximated by the mean the $R$- and
$I$-bands \cite[]{nordgren99b}.  Hence, after calculating 
$I_\lambda(\alpha)$ on a fine wavelength grid, a passband-weighted 
CLV for each filter,
\begin{equation}	
\bar{I}_i(\alpha) = \frac{\int_{a_i}^{b_i}I_\lambda(\alpha)S_i(\lambda)\,d\lambda}
{\int_{a_i}^{b_i}S_i(\lambda)\,d\lambda}
\label{eqn:filter}
\end{equation}
is computed, where $S_i(\lambda)$ is the response function of the photometric
system for the filter $i$ with wavelength limits $a_i$ and $b_i$.
Response functions for the $V$-, $R$- and $I$-bands are taken from
\cite{bessell90}.  

\begin{figure}
\includegraphics[scale=0.7,angle=90]{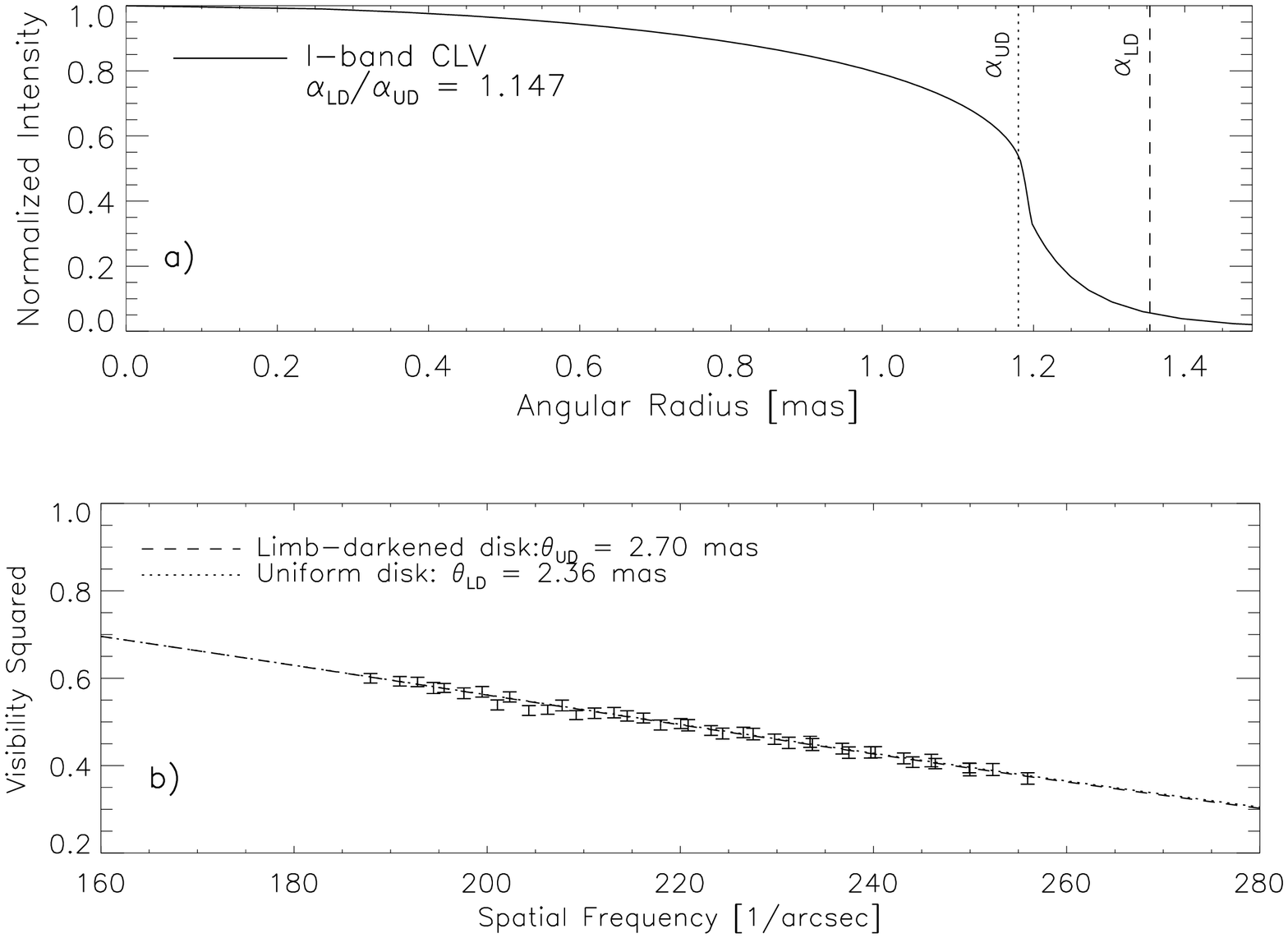}
\caption{(a) Model center-to-limb variation in the $I$-band for an
spherical, expanding atmosphere model with parameters: 
$T_\star$ = 9000 K at $R_\star$ =
$1.2\times 10^{13}$ cm and $\log g(R_\star)$ = 1.3,
equivalent to \Teff\ = 8490 K at $R(\tau_{\rm Ross} = 2/3)$ and $\log
g(R)$ = 1.19, see equation \ref{eqn:teffross}; \mdot\ = $1.0\times10^{-6}$ 
\Mspyr; \vterm\ = 225 \kms; and $\beta$ = 3.0.
The vertical dashed line, labeled $\alpha_{\rm LD}$, marks the angular
radius beyond which $\Delta V^2(220\ {\rm arcsec^{-1}}) < 1\%$ (see
text) .  The dotted line, labeled $\alpha_{\rm UD}$, marks the angular
radius of a uniform-disk with a visibility curve that matches the
limb-darkened visibility curve at $V^2 = 0.5$.  For this model the
limb-darkened radius is 15\% larger than the uniform-disk radius. (b)
Corresponding visibility curves for the model I-band CLVs (dashed,
limb-darkened disk; dotted, uniform-disk) over plotted on the
visibilities measured by NPOI on 1997 December 5.}
\label{ldcorr}
\end{figure}

Following \cite{quirrenbach96}, the model visibilities, $V_i(kw)$, for
each passband are computed by numerically integrating
\begin{equation}	
V^2_i(kw) = \Biggl[2\pi \int_0^{\alpha_{\rm limb}} 
J_0(kw\alpha)\bar{I}_i(\alpha)\alpha\,d\alpha \Biggl]^2
\label{eqn:vis}
\end{equation}
where $\alpha_{\rm limb} = \alpha_{\rm LD} = \theta_{\rm LD}/2$ is the
limb-darkened angular radius and $J_0$ is the Bessel function of
zeroth order.  The squared visibility will be insensitive to the
low-intensity tail of the CLV beyond some angle.  Therefore, we define
$\alpha_{\rm limb}$ to be the angle beyond which the CLV affects the
visibility at 220 arcsec$^{-1}$ by less than 1\%, approximately the
precision of the NPOI observations.  This corresponds to an intensity
cutoff at $\sim$5\% of the central disk intensity (see Figure
{\ref{ldcorr}a}).

Figure {\ref{ldcorr}b} shows that well inside the first null, the
shapes of UD and LD visibility curves are essentially identical and
both fit the NPOI data, but correspond to quantitatively different
diameters, \ADUD\ and \ADLD, respectively.  The ratio \ADLD/\ADUD =
$\alpha_{\rm LD}/\alpha_{\rm UD}$ is the limb-darkening correction.
The value of $\alpha_{\rm UD}$ parameterizes the UD visibility curve,
equation (\ref{eqn:airy}).  The average limb-darkening correction from
the $R$- \& $I$-bands is 15.0\%. Therefore, we adopt
\begin{equation}	
\theta_{\rm LD} = 1.15 \cdot \theta_{\rm UD} = 2.76\pm0.06\ {\rm mas}
\label{eqn:ad}
\end{equation}
as the limb-darkened angular diameter of Deneb.

%%%%%%%%%%%%%%%%%%%%%%%%%%%%%%%%%%%%%%%%%%%%%%%%%%%%%%%%%%%%%%%%%%%%%%%%
%
%FIGURE CLV - static-wind CLV comparison
%
%%%%%%%%%%%%%%%%%%%%%%%%%%%%%%%%%%%%%%%%%%%%%%%%%%%%%%%%%%%%%%%%%%%%%%%%
%%%%%%%%%%%%%%%%%%%%%%%%%%%%%%%%%%%%%%%%%%%%%%%%%%%%%%%%%%%%%%%%%%%%%%%%
%FIGURE CLV - wind model ld correction
%

%%%%%%%%%%%%%%%%%%%%%%%%%%%%%%%%%%%%%%%%%%%%%%%%%%%%%%%%%%%%%%%%%%%%%%%%

An accurate and precise limb-darkened angular diameter measurement may
yield an effective temperature, via equation (\ref{eqn:fundteff}),
which is not consistent with the visual color temperature or
previously determined values for \Teff\ from model atmosphere fitting.
A proper interpretation of this \Teff\ can be made only after an
appropriate CLV has been modeled (or measured) and the relationships
between the intensity and optical depth radii have been established.
Three different radii, and their corresponding angular diameters, are
useful for the comparison of our model atmospheres with the
observational data: (1) $R_\star$, the input radius to the model
atmosphere, $L = 4{\pi}R_{\star}^2{\sigma}T_{\star}^4$, which
also corresponds to the radius at the base of the wind
(see \S A.2.2); (2) $R(\tau_{\rm Ross} =
2/3)$, the optical depth radius corresponding a Rosseland optical
depth of 2/3; (3) $R_{\rm limb}$, the wavelength-dependent radius
corresponding to the limb-darkened edge of the atmosphere, the impact
parameter at the angle $\alpha_{\rm limb}$.

An example of the relationship between these radii is provided by one
of the best fitting models to Deneb's spectral energy distribution
(see Table 6, \S 5.2).  For this model the ratio $R_{\rm
limb}/R(\tau_{\rm Ross} = 2/3) = 1.19$.  This says that the intensity
radius derived from the CLV is 19\% larger than the optical depth
radius in the $V$-band.  Furthermore, this model is parameterized with
$R_\star$, where $R_\star/R(\tau_{\rm Ross} = 2/3) = 0.89$, and
$T_\star = 9000$ K.  Therefore, the effective temperature corresponding to
$R_{\rm limb}$ and the same luminosity is
\begin{equation}	
T_{\rm eff}^{\rm limb} = T_\star \sqrt{R_\star/R_{\rm limb}} = 7780 {\rm\ K}.
\label{eqn:tlimb}
\end{equation}
This model effective temperature can be directly compared to \Teff\
from equation \ref{eqn:fundteff}.  However, $T_{\rm eff}^{\rm limb}$
is not the most physically relevant temperature and it can be easily
shown that a plane-parallel, hydrostatic model with this effective
temperature is a very poor match to the spectral energy distribution.

The most useful effective temperature is one which corresponds to 
the radius at a Rosseland optical depth of $\tau_{\rm Ross} = 2/3$.
In this model, the $V$-, $R$- and $I$-band optical depth radii all differ by less
than 1\% from the Rosseland depth. The corresponding effective temperature is
\begin{equation}	
T_{\rm eff}^{\rm Ross} = T_\star \sqrt{R_\star/R(\tau_{\rm Ross} = 2/3)} \simeq 8490 {\rm\ K}.
\label{eqn:teffross}
\end{equation}
Interestingly, the ratio $R_{\rm limb}/R(\tau_{\rm Ross} = 2/3)$ is larger than the
limb-darkening correction, $\alpha_{\rm LD}/\alpha_{\rm UD}$. 
This means that the angular diameter
corresponding to $T_{\rm eff}^{\rm Ross}$ is {\it smaller} than the uniform-disk diameter.
\begin{equation}	
\theta_{\rm Ross} = \theta_{\rm LD} \frac{R(\tau_{\rm Ross} = 2/3)}{R_{\rm limb}} 
= 2.32 \pm 0.06 {\rm\ mas}
\label{eqn:thetaross}
\end{equation}
With more compact stars, the
limb-darkening correction is used to scale up the angular diameter
from the uniform-disk measurement ($\theta_{\rm Ross} > \theta_{\rm
UD}$) in order to bring the fundamental effective temperature in line
with the color temperature.  However, in the case of Deneb (and 
likely for other BA-supergiants) the angular diameter must be scaled down
from uniform-disk measurement ($\theta_{\rm Ross} < \theta_{\rm UD}$)
for this purpose.
In effect, the observed uniform-disk diameter is biased toward a
larger diameters than it otherwise would be in the absence of the 
gradual drop-off in the center-to-limb profile predicted by our wind
models.  Longer baseline interferometry of Deneb is needed to test
this prediction.

\section{Observational Constraints}
\subsection{Effective Temperature and Reddening}
The most fundamental atmospheric parameter of a stellar atmosphere is its
effective temperature. As shown above, \Teff\ must be carefully defined
in the case of an extended atmosphere.  Published values for Deneb's
\Teff\ range from 7635 K to 10080 K \cite[]{vtg99}.  Efforts to
establish an accurate and precise value of \Teff\ have been ongoing for
the last thirty years.  Constraints on Deneb's fundamental effective
temperature have been limited by the large uncertainties in its
angular diameter and in the foreground extinction which enters into
the calculation through the bolometric flux.  As a result, the value of \Teff\
determined from the bolometric flux and the angular diameter,
has been uncertain at a level of $\sim 30\%$.  We can now use the more
precise ($\sim 3$\%) angular diameter obtained by NPOI to better
constrain both \Teff\ and the foreground extinction.

The extinction curve toward Deneb is not well established due to the
lack of nearly unreddened A-type supergiants with which to perform a
pair method extinction curve analysis.  As a best guess, we have used an
extinction curve derived for the well-studied line-of-sight toward
the O9 V((f)) star HD 199579 \cite[]{fm90}.
Deneb and HD 199579 lie in the same extinction field, 
demarcated (84/0.0) by \cite{nk80}.
The correlation \cite[]{jg93} between the linear 
slope component \cite[$c_2$ = 0.898 for HD 199579, ][]{fm90} 
of the UV extinction curve and $R_V$, the ratio
of total to selective extinction, is
\begin{equation}	
R_V =(4.61\pm0.22) - (1.89\pm0.34)c_2 
\label{eqn:rv}
\end{equation}
which yields,
\begin{equation}	
R_V = 2.9\pm0.5.
\label{eqn:rv2}
\end{equation}

The color excess, $E(B-V)$, toward Deneb is not well constrained in
the literature.  Published values range from 0.04 \cite[]{wt95} to
0.10 \cite[]{takeda94} to 0.15 \cite[]{whk94,atw84}.  The NPOI angular
diameter, corrected to the $V$-band (equation \ref{eqn:thetaross}), allows
us to estimate $E(B-V)$ through the empirical surface brightness to
intrinsic $(V-R)$ color relation of \cite{BE2}.  This is done by
finding a simultaneous solution to the $V$-band surface brightness
\begin{equation}
F_V = 4.2207 - 0.1 V_0 - 0.5 \log \theta_{\rm Ross}
\label{eqn:ber1}
\end{equation}
and the Barnes-Evans relation 
in the color range appropriate for Deneb:
\begin{equation}
F_V = 3.977 - 1.390(V-R)_0, \qquad -0.17 \leq (V-R)_0 \leq 0.00
\label{eqn:ber2}
\end{equation}
These two equations can be expressed in terms of the observed
$V$ magnitude, the observed $(V-R)$ color, and $E(B-V)$:
\begin{equation}
F_V = 4.2207 - 0.1\Bigl[V - R_V\cdot E(B-V)\Bigr] - 0.5 \log \theta_{\rm Ross}
\label{eqn:ber3}
\end{equation}
\begin{equation}
F_V = 3.977 - 1.390\Bigl[(V-R)-0.82E(B-V)\Bigr] 
\label{eqn:ber4}
\end{equation}
using the relationship $E(V-R) =0.82E(B-V)$ from \cite{BE3}.
With the observed $V$ and $R$ magnitudes \cite[]{LPLmags} and
$R_V$ (equation \ref{eqn:rv2}) the solution equations
(\ref{eqn:ber3}) and (\ref{eqn:ber4}) yields
\begin{equation}	
E(B-V) = 0.09^{+0.04}_{-0.03} 
\label{eqn:ebv}
\end{equation}
\noindent The uncertainty in $E(B-V)$ is primarily due to the scatter
(0.025 mag) in the Barnes-Evans relation, but also to the
uncertainties in $\theta_{\rm Ross}$ and $R_V$.  This method of estimating the
reddening yields a best estimate for $E(B-V)$; however the range of
uncertainty still corresponds to the range of values in the
literature.

With best estimates for the angular diameter and reddening we proceed
calculate Deneb's effective temperature.  The UV extinction curve of
HD 199579 is used to deredden Deneb's UV spectrum, while the optical
data are dereddened using the extinction relations from \cite{ccm}.
The bolometric flux is calculated by integrating the dereddened UV and
optical spectrophotometric data from 1150 \AA\ to 10725 \AA.  Fluxes
at longer and shorter wavelengths provide a negligible contribution to the
bolometric flux.  Spectra SWP08133 and LWP07864 were used for the
range 1150 \AA\ $\leq \lambda < 3260$ \AA, and the optical and
near-infrared data from \cite{glush92} were used for the range 3260
\AA\ $\leq \lambda \leq 10750$ \AA.  The uncertainty in the bolometric
flux is primarily a combination of uncertainties in the extinction,
the monochromatic flux measurements, and the absolute flux
calibrations.  Absolute calibration uncertainties are taken to be 10\%
in the UV and 5\% in the optical, while uncertainties in the angular
diameter and extinction are taken from equations
(\ref{eqn:thetaross}), (\ref{eqn:rv2}), and (\ref{eqn:ebv})
respectively.  This yields a fundamental effective temperature for
Deneb of
\begin{equation}
T_{\rm eff}^{\rm Ross}  = 8600 \pm 500\ {\rm K}.
\label{eqn:teff}
\end{equation}
\subsection{Photospheric Radius and Gravity}
Taking the Rosseland mean angular diameter of Deneb from equation
(\ref{eqn:thetaross}) and the mean {\it Hipparcos} parallax, $\pi =
1.01\pm0.57$ mas \cite[]{hipparcos}, yields a radius of $R(\tau_{\rm
Ross} = 2/3) \simeq 1.7 \times 10^{13}$ cm = 240 \Rsun.  The large
uncertainty in the parallax causes the distance probability
distribution function to have a long tail extending to large
distances. If the most probable value from the distance probability
distribution function is adopted (685 pc, as opposed to the mean, 990
pc), the radius is $R(\tau_{\rm Ross} = 2/3) \simeq 1.2 \times
10^{13}$ cm = 172 \Rsun.

\cite{humphreys78} assigns Deneb to the Cyg OB7 association, with an
adopted true distance modulus of 9.5 (794 pc), which yields
$R(\tau_{\rm Ross} = 2/3) \simeq 1.38 \times 10^{13}$ cm = 196 \Rsun.
We adopt $R(\tau_{\rm Ross} = 2/3) = 180$ \Rsun.  This radius and
$T_{\rm eff}^{\rm Ross} = 8600$ K yield a luminosity $L = 1.6 \times
10^{5}$ \Lsun.  For this luminosity, evolutionary tracks of
\cite{schaller} and \cite[][considering the effects
of rotation and the composition gradient]{heger_langer2000} suggest Deneb's mass lies
in the range 20 - 25 \Msun.  Therefore, surface gravities should lie
in the range $\log g(R(\tau_{\rm Ross} = 2/3)) = 1.2 - 1.4$.  A
full list of the stellar parameters discussed in this paper are
compiled in Table \ref{tab:obsparams}.

\begin{deluxetable}{lcll}
\tablecolumns{4} 
\tabletypesize{\footnotesize}
\tablecaption{Stellar Parameters for Deneb} 
\tablewidth{0pt}
\tablehead{\colhead{Parameter} &\colhead{Symbol} & \colhead{Value} &\colhead{Reference}}
\startdata
\cutinhead{Measured values}
Uniform-disk angular diameter   &$\theta_{\rm UD}$ &$2.40 \pm 0.06\ {\rm mas}$                       &\S 2.7, eqn. \ref{eqn:udad} \\
Bolometric flux (reddened)      & $\mathcal{F}$ &$7.1 \pm 0.4 \times 10^{-6}$ erg cm$^{-2}$ s$^{-1}$ &\S 4.1 \\
Parallax                        &$\pi$ &$1.01\pm 0.57\ {\rm mas}$                                    &\cite{hipparcos} \\
\cutinhead{Derived values}
Limb-darkened angular diameter  &$\theta_{\rm LD}$     &$2.76 \pm 0.06\ {\rm mas}$  &\S 3, eqn. \ref{eqn:ad}  \\
Rosseland mean angular diameter &                            &                    & \\ 
\multicolumn{1}{r}{interferometric model}   &$\theta_{\rm Ross}$   &$2.32 \pm 0.06\ {\rm mas}$ &\S 3, eqn. \ref{eqn:thetaross} \\
\multicolumn{1}{r}{SED $\chi^2$ fit}  &$\theta_{\rm Ross}$   &$2.35 \pm 0.01\ {\rm mas}$ &\S 5.2.1 \\
Color Excess                    &                            &                    & \\
\multicolumn{1}{r}{Barnes-Evans relation} & $E(B-V)$             &$0.09^{+0.04}_{-0.03}$ & \S 4.1, eqn. \ref{eqn:ebv} \\
\multicolumn{1}{r}{SED $\chi^2$ fit} & $E(B-V)$                  & $0.06\pm0.01$         &\S 5.2.1 \\
Total/selective extinction      & $R_{V}$              &$2.9\pm 0.5$  &\S 4.1, eqn. \ref{eqn:rv2} \\ 
Bolometric flux (unreddened)    & $\mathcal{F}_0$      &$9.8 \pm 2.3 \times 10^{-6}$ erg cm$^{-2}$ s$^{-1}$ &\S 4.1 \\
Effective Temperature           &                            &                    & \\
\multicolumn{1}{r}{interferometry}   & $T_{\rm eff}^{\rm Ross}$   &$8600 \pm 500$ K    &\S4.1, eqn. \ref{eqn:teff} \\
\multicolumn{1}{r}{SED $\chi^2$ fit}  & $T_{\rm eff}^{\rm Ross}$   &$8420 \pm 100$K     &\S 5.2.1, eqn. \ref{eqn:teff_ls}, Fig. \ref{contour} \\
Radius                          & $R(\tau_{\rm Ross} = 2/3)$ & $\simeq$ 180 \Rsun &\S 4.2\\
Luminosity                      & $L$  &$\simeq 1.6\pm 0.4 \times 10^{5}$ \Lsun  &\S 4.2\\
Mass                            & $M$                   &$\sim$ 20 - 25 \Msun    &\S 4.2\\
Surface gravity                 &                            &                    & \\                 
\multicolumn{1}{r}{$R(\tau_{\rm Ross} = 2/3)$ and $M$}  & $\log g(R(\tau_{\rm Ross} = 2/3))$  &$\simeq$ 1.3 &\S 4.2\\
\multicolumn{1}{r}{stability, Balmer edge}  & $\log g(R(\tau_{\rm Ross} = 2/3))$  &1.1 - 1.6    &\S 5.2.2, Table \ref{tab:SEDmodels2} \\
Mass-loss Rate                  &                            &                    & \\                 
\multicolumn{1}{r}{SED $\chi^2$ fit}  & \mdot\                &$8\pm3\times 10^{-7}$ \Mspyr & \S 5.2.1 \\
\multicolumn{1}{r}{spectral lines}    & \mdot\                &$10^{-7} - 10^{-6}$ \Mspyr   &\S 6 \\
Terminal Velocity               & $v_\infty$            &$\simeq$ 225 \kms &\S 6.1, Fig. 19 \\
Velocity Law Exponent           & $\beta$               &$\simeq$ 3 & \S 5.2.2, eqn. \ref{betalaw}, Table \ref{tab:SEDmodels2} \\
Microturbulent velocity         & $\xi_t$               &$\simeq$ 15 \kms & \S 5.2.2, Table \ref{tab:SEDmodels2} \\
\enddata
\label{tab:obsparams}
\end{deluxetable}

\section{Deneb's Spectral Energy Distribution}

\subsection{Model Parameters}
The most important input parameters which characterize our model
atmospheres (for details see Appendix A) are: a model effective
temperature, $T_\star$, at a reference radius, $R_\star$, the gravity,
$g$, at $R_\star$, a terminal velocity, $v_\infty$, and beta-law
exponent, $\beta$, a mass-loss rate, $\dot{M}$, a microturbulent
velocity, $\xi_t$, assumed to be depth independent, and the chemical
composition.  Each of these parameters may be determined to varying
degrees of precision by the data.  In this section we therefore
conduct sensitivity tests in order to examine how well each parameter
is constrained by the observations.

As outlined by \cite{venn95_2,venn95_1}, a number of studies find that
metal abundances for Galactic, early A-type supergiants are near solar
($\pm 0.2$ dex).  Nitrogen abundances, while enhanced in A-type
supergiants relative to unevolved B-stars, are consistent within the
uncertainties with solar abundances, since the sun is metal-rich
relative to the nearby B-type stars \cite[see, e.g.][]{edvardsson93}.
A detailed abundance study of Deneb using LTE plane-parallel model
atmospheres was recently published by \cite{Albayrak00}. While we
would very much like to understand how the accuracy of these abundance
determinations is influenced by the lack of non-LTE, spherical, and expanding
atmosphere effects, this is beyond the present scope of this work.
In this paper we use solar abundances \cite[]{abund}
for all models.

\subsection{Least-squares Analysis}

To compare quantitatively a set of model spectra to the observed
spectrum of Deneb from the UV to the radio, both the model and observed spectra
have been binned to 20 \AA\ in the UV and the model spectra have been
binned to 50 \AA\ in the optical and near-IR to match the resolution
of the spectrophotometry in this region.  At longer wavelengths,
monochromatic fluxes from model spectra are interpolated at
wavelengths corresponding to DIRBE, MSX, ISO, SCUBA, and VLA
bands. 

The model and the observed SED are compared in each wavelength
bin:
\begin{equation}
{\chi}^2 = \sum_{i=1}^{265} \Biggl(\frac{f^0_{\oplus}(\lambda_i) -
 F_{\rm model}(\lambda_i)\frac{1}{4}\theta_\star^2}
{\sigma_{f^0_\oplus} (\lambda_i)}\Biggr)^2
\label{eqn:chisqr}
\end{equation}
\noindent where, 
\begin{equation}
f^0_{\oplus}(\lambda_i) = f_\oplus(\lambda_i)\cdot
10^{0.4\frac{A_{\lambda_i}}{A_V}R_VE(B-V)}
\end{equation}
\noindent is the unreddened flux measured at the earth in wavelength
bin $i$, $\sigma_{f^0_\oplus}(\lambda_i)$ is the 1$\sigma$
uncertainty in this value. 
In bin $i$, the interstellar extinction is $A_{\lambda_i}/{A_V}$
and the model flux is $F_{\rm model}(\lambda_i)$.
The angular
diameter, $\theta_\star$ in radians, scales the model surface flux 
(at $R_\star$) to the measured flux.
\begin{equation}
\theta_\star  = \theta_{\rm Ross} \frac{R_\star}{R_{\rm Ross}}  
\label{eqn:theta_star}
\end{equation}

\subsubsection{Constraining $T_\star$ and $\dot{M}$}

\begin{figure}
\includegraphics[scale=0.7,angle=90]{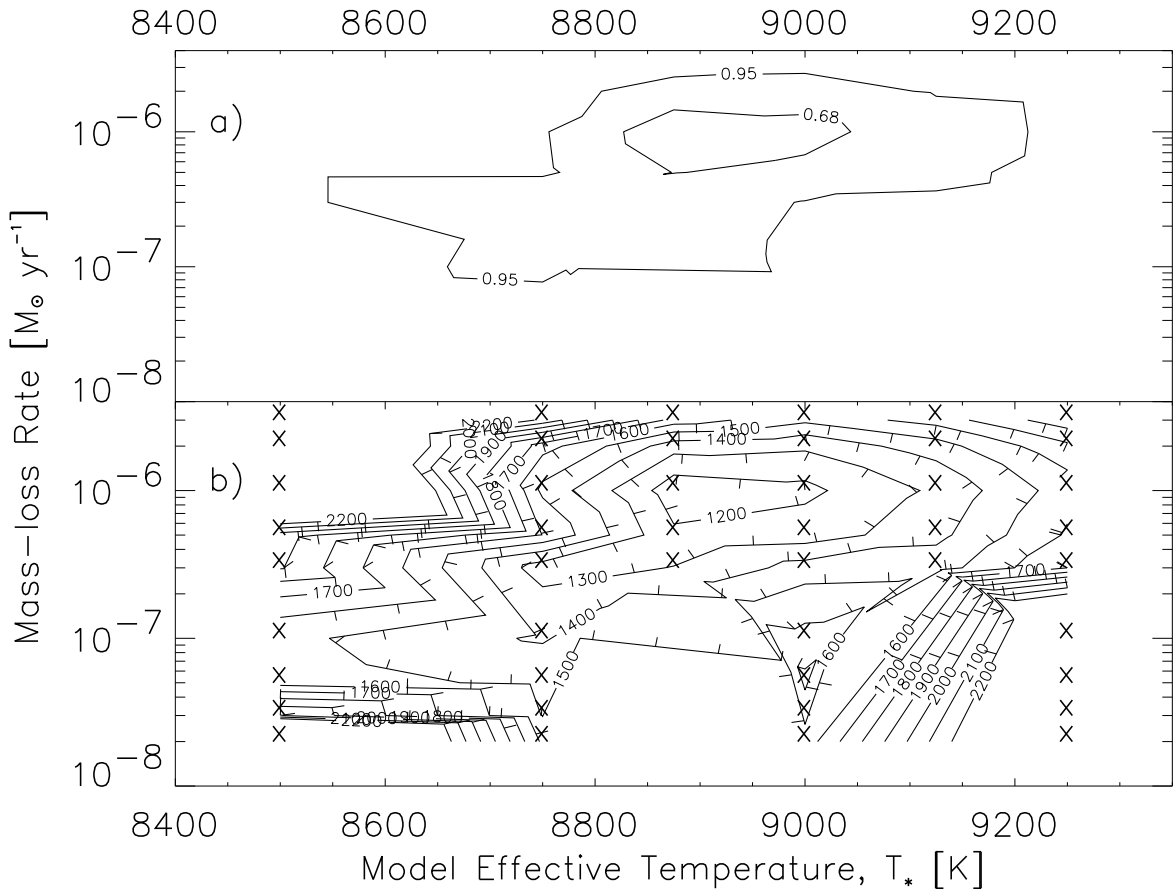}
\caption{(a) Regions (1$\sigma$ and 2$\sigma$) in the ($T_\star,
\dot{M}$) parameter plane containing the best fit model parameters to
the observed spectral energy distribution (SED) of Deneb.  The
probabilities are listed in Table \ref{tab:SEDmodels}.  (b) 
Contour map of ${\chi}^2$ in the ($T_\star, \dot{M}$) plane of the model
fits to the observed SED.  The grid of model locations in the
parameter plane is marked by $\mathsf{x}$.}
\label{contour}
\end{figure}

A least squares sum was computed for set of 45 models with a range of
effective temperatures and mass-loss rates (the $T_\star$, $\dot{M}$
grid is shown in Figure \ref{contour}b).  A combination of parameters
$\theta_\star$ (range: 1.5--3.5 mas) and $E(B-V)$ (range: 0.05--0.12
mag) was found which minimized the ${\chi}^2$ for each model.  The
models, sorted by ${\chi}^2$, are listed in Table \ref{tab:SEDmodels}
along with best fitting values for $\theta_\star$ and $E(B-V)$ at each
$T_\star$ and $\dot{M}$.  

\begin{figure}
\includegraphics[scale=0.7,angle=90]{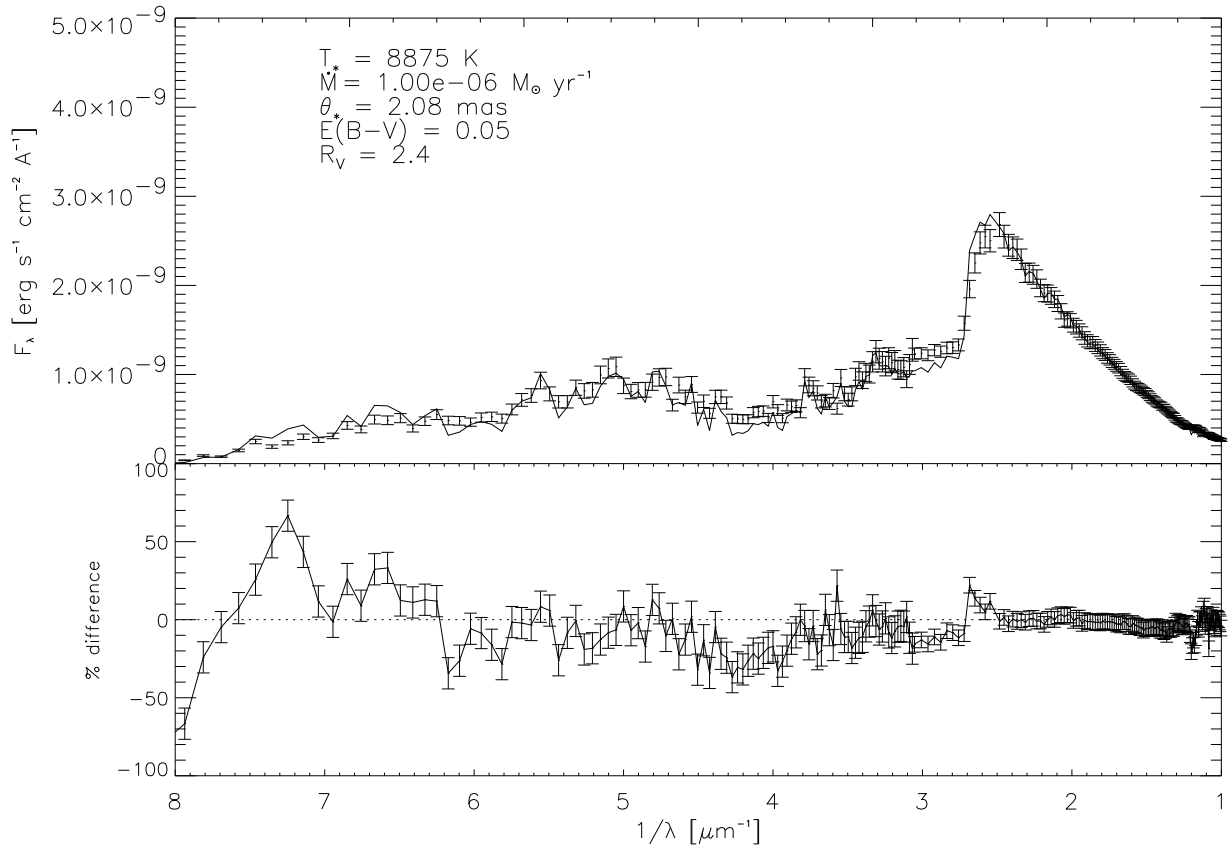}
\caption{Top Panel: Best least squares fit (solid line) to the UV
to radio spectrum compared to the UV-optical spectrum of Deneb (error
bars), which is binned to 20 \AA\ in the UV and to 50 \AA\ in the
optical.  The model has been scaled with the angular diameter
$\theta_\star$ (see text) and the data have been corrected for
interstellar extinction using the reddening parameters shown.
$T_\star = 8875$ K is equivalent to $T_{\rm eff}^{\rm Ross}  = 8370$ K.
Bottom Panel: Percentage difference, (model - data)/data, plotted as a
function of wavenumber.}
\label{figopt}
\end{figure}

\begin{figure}
\includegraphics[scale=0.7,angle=90]{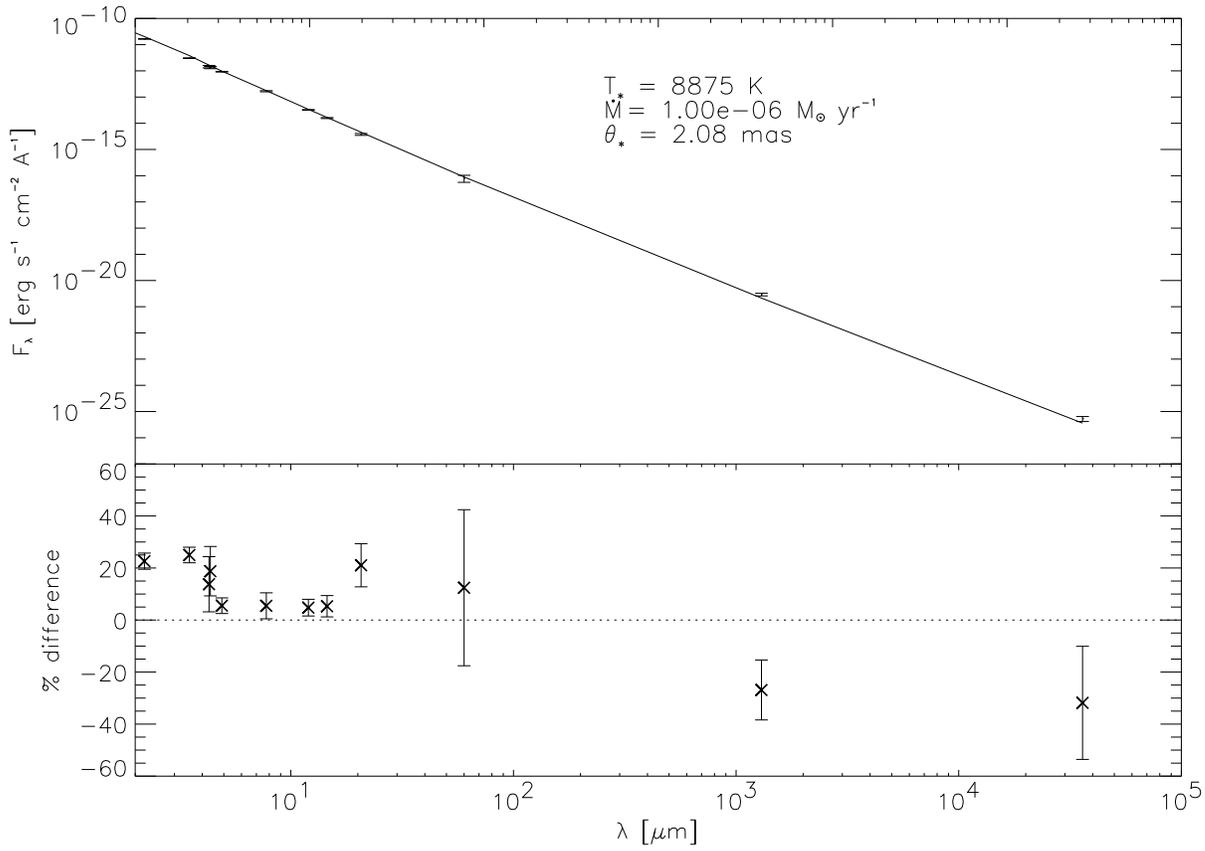}
\caption{Top Panel: Best least squares fit (solid line) to the UV
to radio spectrum compared to IR, mm, radio spectrum of Deneb (error
bars) from Table \ref{tab:irdata}.  The model has been scaled with the
angular diameter $\theta_\star$.  
$T_\star = 8875$ K is equivalent to $T_{\rm eff}^{\rm Ross}  = 8370$ K.
Bottom Panel: Percentage
difference, (model - data)/data, plotted as a function of wavelength.}
\label{figir}
\end{figure}

Although the best fit model is a good match to the spectrum (see
Figures \ref{figopt} and \ref{figir}), the reduced chi-square value,
$\tilde{\chi}^2 \simeq 4.4$, is large and the ${\chi}^2$ test yields a
very low probability of agreement between the observed and 
theoretical distributions indicative of a bad fit.  In cases like this
\cite{bevington} suggests the use of the $F$ test because:
``the statistic ${\chi}^2$ measures not only the discrepancy between
the estimated function [model spectrum] and the parent function
[actual spectrum], but also the deviations between the data [observed
spectrum] and the parent function [actual spectrum] simultaneously.''
The $F$ test concentrates on the discrepancy between the model
spectrum and actual spectrum.  The probability that model $n$ with $F
= {{\chi}_n^2}/{{\chi}_{\rm min}^2}$ is as good a fit to the data as
the best fit model is listed in Table \ref{tab:SEDmodels}.  From these
probabilities the $1\sigma$ and $2\sigma$ contours are plotted in
Figure \ref{contour}a.  The $1\sigma$ region encloses the parameter
space $T_\star \simeq 8925 \pm 100$ K and $\dot{M} \simeq 8 \pm 3 \times 10^{-7}$ \Mspyr. 
This effective temperature range is
equivalent to 
\begin{equation}
T_{\rm eff}^{\rm Ross} \simeq 8420 \pm 100\ {\rm K}
\label{eqn:teff_ls}
\end{equation}
after scaling from $R_\star$ to $R(\tau_{\rm Ross} = 2/3)$.

This effective temperature is in good agreement with the
interferometric Rosseland fundamental effective temperature from
equation \ref{eqn:teff}.  The least-squares determination of $T_{\rm
eff}^{\rm Ross}$ has a smaller random uncertainty because a correction
for the interstellar extinction is incorporated into the fitting procedure
(with the value for $R_V$ fixed), while the full uncertainties in both $R_V$
(equation \ref{eqn:rv}) and $E(B-V)$ (equation \ref{eqn:ebv}) from the
Barnes-Evans relation enter into the interferometric value of $T_{\rm
eff}^{\rm Ross}$. We note that the derived reddening
for the best fitting models (Table \ref{tab:SEDmodels}, column 6) yield 
values, $E(B-V) = 0.06\pm 0.01$, at the low end of the range
of equation \ref{eqn:ebv}.  This result is consistent with the
lower mean least-squares $T_{\rm eff}^{\rm Ross}$ relative to the mean
interferometric $T_{\rm eff}^{\rm Ross}$, since the derived bolometric
flux will be lower when a lower interstellar extinction is adopted.
In addition, the derived angular diameters for the best fitting models
(Table \ref{tab:SEDmodels}, column 5) correspond (via equation
\ref{eqn:theta_star}) to $\theta_{\rm Ross} = 2.35 \pm 0.01$ mas,
which is in good agreement with interferometric value of $\theta_{\rm
Ross}$ from equation \ref{eqn:thetaross}.  In short, the derived
interferometric and spectrophotometric values for $T_{\rm eff}^{\rm
Ross}$, $\theta_{\rm Ross}$, and $E(B-V)$ are in good agreement.

Figure \ref{contour}b shows the contours of the ${\chi}^2$ values from
Table \ref{tab:SEDmodels}.  Models in the region (8500 K, 2$\times
10^{-8}$ \Mspyr) poorly fit the observed spectrum because the model UV
fluxes are too small, particularly right below the Balmer jump and
below 1600 \AA, and the model millimeter and radio fluxes are too
small by roughly one order of magnitude.  Models with much higher
mass-loss rates and similar effective temperatures (8500 K, 2$\times
10^{-6}$ \Mspyr) provide a closer match to the millimeter and radio fluxes,
but still come up short, and the model UV fluxes are even smaller.
Models in the region (9250 K, 2$\times 10^{-8}$ \Mspyr) poorly fit the
observed spectrum because the model UV fluxes below 1600 \AA\ are too
large, and the predicted millimeter and radio fluxes are too small by
an order of magnitude.  For a given $T_\star$, larger $\dot{M}$ values
have two effects on the SED relative to lower $\dot{M}$ values:
(1) the UV continuum is suppressed, and (2) the radio continuum is
enhanced.  The hottest, highest mass-loss rate models explored (eg.,
9250 K, 3$\times 10^{-6}$ \Mspyr) produce millimeter and radio fluxes
in excess of those observed, and the model UV continuum, while
suppressed relative to lower $\dot{M}$ models, is still too large below
1600 \AA.  The best fitting models are slightly cooler, to match the
UV, and have slightly lower mass-loss rates, to match the millimeter
and radio fluxes.

\begin{deluxetable}{lllllll}
\tablecolumns{7} \tabletypesize{\scriptsize} \tablecaption{SED (UV to Radio) Fitting 
Results\tablenotemark{a}}
\tablewidth{0pt} \tablehead{ 
\multicolumn{2}{c}{Model} & & \multicolumn{4}{c}{Fit Results}\\ 
\cline{1-2} \cline{4-7} \\ 
\colhead{$T_\star$ (K)} & \colhead{$\dot{M}$ (\Mspyr)} &
& \colhead{${\chi}^2$} & \colhead{$P(F\ {\rm test})$} &
\colhead{$\theta_\star$ (mas)} & \colhead{$E(B-V)$}} 
\startdata 
8875 &1.00E-06 &&1.14E+03 &5.00E-01 &2.10 &0.05 \\ 
9000 &1.00E-06 &&1.17E+03 &4.13E-01 &2.08 &0.06 \\ 
8875 &5.00E-07 &&1.21E+03 &3.18E-01 &2.10 &0.05 \\ 
9000 &5.00E-07 &&1.23E+03 &2.57E-01 &2.08 &0.06 \\ 
8750 &3.00E-07 &&1.24E+03 &2.44E-01 &2.16 &0.05 \\ 
8875 &3.00E-07 &&1.30E+03 &1.44E-01 &2.16 &0.07 \\ 
9125 &1.00E-06 &&1.31E+03 &1.26E-01 &2.08 &0.08 \\ 
9000 &2.00E-06 &&1.32E+03 &1.21E-01 &2.06 &0.05 \\ 
8875 &2.00E-06 &&1.34E+03 &9.28E-02 &2.10 &0.05 \\ 
9125 &5.00E-07 &&1.37E+03 &7.08E-02 &2.10 &0.09 \\ 
8750 &1.00E-07 &&1.38E+03 &5.97E-02 &2.20 &0.06 \\ 
9000 &3.00E-07 &&1.43E+03 &3.22E-02 &2.16 &0.09 \\ 
9125 &3.00E-07 &&1.45E+03 &2.72E-02 &2.10 &0.09 \\ 
9125 &2.00E-06 &&1.45E+03 &2.50E-02 &2.06 &0.07 \\ 
8750 &1.00E-06 &&1.47E+03 &1.99E-02 &2.16 &0.05 \\ 
8750 &5.00E-08 &&1.47E+03 &1.98E-02 &2.26 &0.08 \\ 
8750 &3.00E-08 &&1.49E+03 &1.46E-02 &2.18 &0.05 \\ 
9000 &3.00E-06 &&1.52E+03 &9.91E-03 &2.04 &0.05 \\ 
8500 &1.00E-07 &&1.52E+03 &9.72E-03 &2.28 &0.05 \\ 
9000 &2.00E-08 &&1.54E+03 &7.58E-03 &2.18 &0.09 \\ 
9250 &1.00E-06 &&1.55E+03 &6.71E-03 &2.08 &0.10 \\ 
8500 &5.00E-08 &&1.55E+03 &6.38E-03 &2.28 &0.05 \\ 
8750 &2.00E-08 &&1.57E+03 &4.53E-03 &2.18 &0.05 \\ 
8750 &5.00E-07 &&1.59E+03 &3.53E-03 &2.16 &0.05 \\ 
9250 &5.00E-07 &&1.61E+03 &2.75E-03 &2.10 &0.11 \\ 
9000 &1.00E-07 &&1.62E+03 &2.23E-03 &2.20 &0.10 \\ 
9000 &3.00E-08 &&1.63E+03 &1.89E-03 &2.20 &0.10 \\ 
9125 &3.00E-06 &&1.64E+03 &1.65E-03 &2.02 &0.06 \\ 
8750 &2.00E-06 &&1.64E+03 &1.65E-03 &2.14 &0.05 \\ 
8875 &3.00E-06 &&1.65E+03 &1.32E-03 &2.10 &0.05 \\ 
9250 &2.00E-06 &&1.67E+03 &9.48E-04 &2.06 &0.09 \\ 
9250 &3.00E-07 &&1.70E+03 &6.61E-04 &2.10 &0.11 \\ 
9000 &5.00E-08 &&1.72E+03 &4.52E-04 &2.22 &0.11 \\ 
9250 &3.00E-06 &&1.86E+03 &3.70E-05 &2.00 &0.07 \\ 
8500 &3.00E-07 &&1.90E+03 &1.74E-05 &2.28 &0.05 \\ 
8500 &5.00E-07 &&1.95E+03 &8.16E-06 &2.26 &0.05 \\ 
8500 &3.00E-08 &&2.00E+03 &2.98E-06 &2.30 &0.05 \\ 
8750 &3.00E-06 &&2.17E+03 &1.19E-07 &2.12 &0.05 \\ 
9250 &1.00E-07 &&2.69E+03 &0.00     &2.14 &0.12 \\ 
9250 &2.00E-08 &&2.71E+03 &0.00     &2.14 &0.12 \\ 
9250 &3.00E-08 &&2.81E+03 &0.00     &2.14 &0.12 \\ 
8500 &3.00E-06 &&2.93E+03 &0.00     &2.22 &0.05 \\ 
8500 &2.00E-06 &&2.94E+03 &0.00     &2.22 &0.05 \\ 
9250 &5.00E-08 &&3.11E+03 &0.00     &2.12 &0.12 \\ 
8500 &1.00E-06 &&3.25E+03 &0.00     &2.24 &0.05 \\ 
8500 &2.00E-08 &&3.36E+03 &0.00     &2.28 &0.05 \\ 
\enddata
\tablenotetext{a}{Columns (1-2): With the exception of parameters
$T_\star$ and $\dot{M}$, all models listed share these parameters:
$R_\star$ = 1.2$\times 10^{13}$ cm, $\log g(R_\star)$ = 1.3, $\xi$ = 15 \kms,
\vterm\ = 225 \kms, $\beta$ = 3.0.  Column (3): ${\chi}^2$ values are
from equation (\ref{eqn:chisqr}) for 265 wavelength bins from 1220
\AA\ to 3.6 cm. 
Column (4): The probability from the $F$ test that the ${\chi}^2$ value
for each model is statistically equivalent to the best fit model.
Column (5): The best fit angular diameter value for each $T_\star$, \mdot\
combination.  Column (6): The best fit color excess 
for each $T_\star$, \mdot combination. $R_V$=2.9 is fixed.}
\label{tab:SEDmodels}
\end{deluxetable}

%%%%%%%%%%%%%%%%%%%%%%%%%%%%%%%%%%%%%%%%%%%%%%%%%%%%%%%%%%%%%%%%%%%%%%%%
%FIGURE Contour plot
%
%%%%%%%%%%%%%%%%%%%%%%%%%%%%%%%%%%%%%%%%%%%%%%%%%%%%%%%%%%%%%%%%%%%%%%%%

%%%%%%%%%%%%%%%%%%%%%%%%%%%%%%%%%%%%%%%%%%%%%%%%%%%%%%%%%%%%%%%%%%%%%%%%
%FIGURE Best Fit UV/Optical SED
%
%%%%%%%%%%%%%%%%%%%%%%%%%%%%%%%%%%%%%%%%%%%%%%%%%%%%%%%%%%%%%%%%%%%%%%%%

%%%%%%%%%%%%%%%%%%%%%%%%%%%%%%%%%%%%%%%%%%%%%%%%%%%%%%%%%%%%%%%%%%%%%%%%
%FIGURE Best Fit UV/Optical SED
%
%%%%%%%%%%%%%%%%%%%%%%%%%%%%%%%%%%%%%%%%%%%%%%%%%%%%%%%%%%%%%%%%%%%%%%%%

\subsubsection{Constraining Other Parameters}

\begin{figure}
\includegraphics[scale=0.7,angle=90]{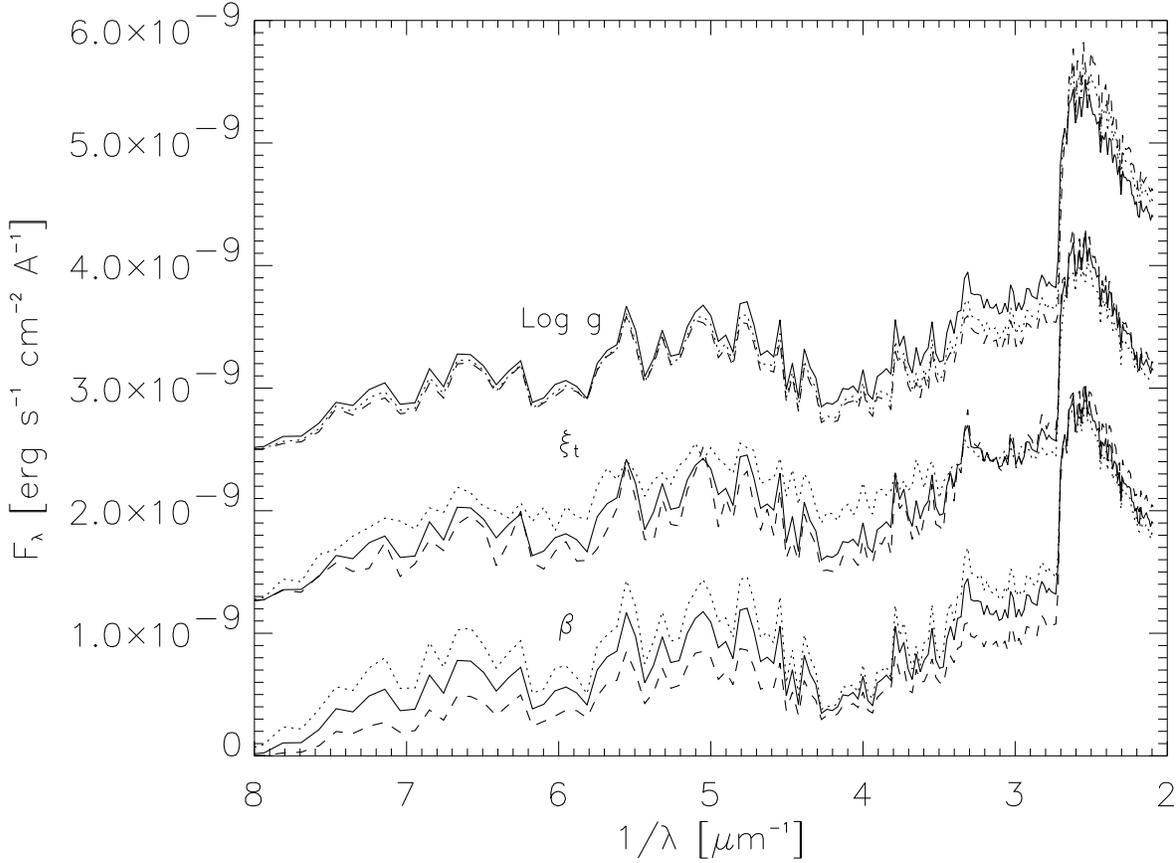}
\caption{Synthetic UV continua from three sets (from top to bottom)
of three models: surface gravity, \Logg\ = 1.3 (solid), 1.5 (dotted),
1.7 (dashed); the microturbulent velocity, $\xi_t$ = 5 \kms (dotted),
15 \kms (solid), 25 \kms (dashed); and the velocity law exponent,
$\beta$ = 2.0 (dotted), 3.0 (solid), 4.0 (dashed).  These synthetic
spectra have been binned to 20 \AA\ resolution to show the gross
features of the spectrum.  All models have the following parameters,
unless otherwise noted: $T_\star = 9000$ K , $\log g(R_\star) = 1.3$,
$\dot{M} = 10^{-6}$ \Mspyr, $\beta = 3.0$, $\xi_t$ = 15 \kms.  All
models have been scaled by $\theta_\star$ = 2.08 mas.  Constants of
$1.25\times 10^{-9}$ and $2.5 \times 10^{-9}$ erg s$^{-1}$ cm$^{-2}$
\AA$^{-1}$ have been added to the top two sets of models for clarity.
The effects of \Logg\ can be seen on both sides of the Balmer
discontinuity.  The effects of varying $\xi_t$ from 5 to 25 \kms\ are
apparent below 3000 \AA\ ($> 3.5\ \mu$m$^{-1}$) where the flux drops by
roughly a factor of two.  The effects of varying $\beta$ from 2.0 to
4.0 are significant below the Balmer jump where the flux drops by
roughly 50\%.}
\label{beta_xi}
\end{figure}

Our preliminary fits to Deneb's SED revealed that model parameters
$T_\star$ and $\dot{M}$ have the largest effects on the model SED.  As
a result, the parameters $R_\star$, $\log g$, $v_\infty$, $\beta$, and
$\xi_t$ were fixed in the analysis discussed above.  The most
straightforward parameter to constrain is the terminal velocity.  We
find \vterm\ $\simeq 225$ \kms\ by fitting the blue edge of the
\ion{Mg}{2} {\it h} resonance line profile (see Figure \ref{mg2_vterm}).
Parameters $\beta$, $\xi_t$, and \Logg\ can be constrained with
reference to least squares fits to the observed SED (see Table
\ref{tab:SEDmodels2}).  Parameters $\beta$ and $\xi_t$ are constrained
due to their significant effect on the ultraviolet SED (see Figure
\ref{beta_xi}). Despite the uncertainty in the reddening, $\beta$
values 2.0 and 4.0 can be rejected with confidence.  Similarly, a low
microturbulence, $\xi_t = 5$\kms, can be rejected.

We find that adjusting the model radius $R_\star$ by $\pm 25 \%$ has
no significant effect on the model SED. However, as shown below in \S
6.2, the radius can affect synthetic P-Cygni profiles because, via the
continuity equation (\ref{cont}), the parameter $R_\star$, for a fixed
mass-loss rate, sets the gas density at the base of the wind.

The model with \Loggrstar\ = 1.7 can be rejected at the $3\sigma$
level due to this parameter's strong effect on the Balmer jump.  From
the requirement of the hydrostatic equilibrium (equations
\ref{eqn:hydro}, \ref{geff}, and \ref{arad}) in the deepest layers
($\tau_{\rm Ross} \simeq 100$) of the model atmosphere we can assign a
lower limit of to the gravity of \Loggrstar\ $\geq$ 1.2.  Models with
lower values of \Loggrstar\ fail to have positive effective gravities
in the deepest layers of the atmosphere, thus making a hydrostatic
solution impossible.  The radiative acceleration is quite significant
in the hydrostatic layers.  For example, in the \Loggrstar\ = 1.7 model
the effective gravity, \Loggeff, drops as low as 1.43 in the
hydrostatic layers, while for the \Loggrstar\ = 1.3 model, the minimum
\Loggeff\ is 0.4.  
The surface gravity constraint is then 
$1.2 \leq$ \Loggrstar\ $\leq 1.7$. This range of values 
can be scaled to the reference depth 
$R(\tau_{\rm Ross} = 2/3)$ via 
\begin{equation}
\log g(R(\tau_{\rm Ross} = 2/3))  = \log g(R_\star) + 2\log{(R_\star/R(\tau_{\rm Ross} = 2/3))} 
\label{eqn:loggscale}
\end{equation}
yielding
$1.1 \leq \log g(R(\tau_{\rm Ross} = 2/3)) \leq 1.6$ for the same ratio of
of radii (0.89) as equation \ref{eqn:teffross}.

\begin{deluxetable}{llllcllll}
\tablecolumns{9}\tabletypesize{\scriptsize} 
\tablecaption{Least Squares Results for Parameters $R_\star$, $\log g$, $\beta$, $\xi_t$\tablenotemark{a}}
\tablewidth{0pt} \tablehead{ 
\multicolumn{4}{c}{Model} & &\multicolumn{4}{c}{Fit Results}\\ 
\cline{1-4} \cline{6-9} \\ 
\colhead{$\log g(R_\star)$}   &\colhead{$R_\star$} &\colhead{$\beta$} &\colhead{$\xi_t$} 
&\colhead{ } 
&\colhead{${\chi}^2$} &\colhead{$P(F\ {\rm test})$} &\colhead{$\theta_\star$} &\colhead{$E(B-V)$}}  
\startdata 
1.3   &1.50e+13  &3.0         &15 & &1.15e+03   &5.00e-01       &2.080     &0.060      \\ %radius
1.3   &1.20e+13  &3.0         &15 & &1.17e+03   &4.32e-01       &2.080     &0.060      \\ %base      
1.3   &9.00e+12  &3.0         &15 & &1.26e+03   &2.25e-01       &2.080     &0.060      \\ %radius
1.3   &1.20e+13  &3.0         &25 & &1.31e+03   &1.36e-01       &2.060     &0.050      \\ %xi
1.5   &1.20e+13  &3.0         &15 & &1.46e+03   &2.66e-02       &2.060     &0.050      \\ %logg
1.3   &1.20e+13  &2.0         &15 & &1.69e+03   &8.17e-04       &2.260     &0.120      \\ %beta
1.7   &1.20e+13  &3.0         &15 & &1.87e+03   &3.63e-05       &2.040     &0.050      \\ %logg
1.3   &1.20e+13  &3.0         &5  & &2.13e+03   &2.98e-07       &2.120     &0.090      \\ %xi
1.3   &1.20e+13  &4.0         &15 & &2.23e+03   &5.96e-08       &2.020     &0.050      \\ %beta
\enddata
\tablenotetext{a}{Columns (1-4): The models listed here have the fixed values
$T_\star$ = 9000 K and $\dot{M} = 10^{-6}$ \Mspyr. 
Columns (5-8) as described in Table \ref{tab:SEDmodels}.}
\label{tab:SEDmodels2}
\end{deluxetable}

%%%%%%%%%%%%%%%%%%%%%%%%%%%%%%%%%%%%%%%%%%%%%%%%%%%%%%%%%%%%%%%%%%%%%%%%
%FIGURE UV SED changes for beta and xi 
%
%%%%%%%%%%%%%%%%%%%%%%%%%%%%%%%%%%%%%%%%%%%%%%%%%%%%%%%%%%%%%%%%%%%%%%%%

\subsection{Deneb's Continuum Beyond 1 \micron}
%%%%%%%%%%%%%%%%%%%%%%%%%%%%%%%%%%%%%%%%%%%%%%%%%%%%%%%%%%%%%%%%%%%%%%%%
%FIGURE Radio Model-Data Comparison in detail
%
\begin{figure}
\includegraphics[scale=0.7,angle=90]{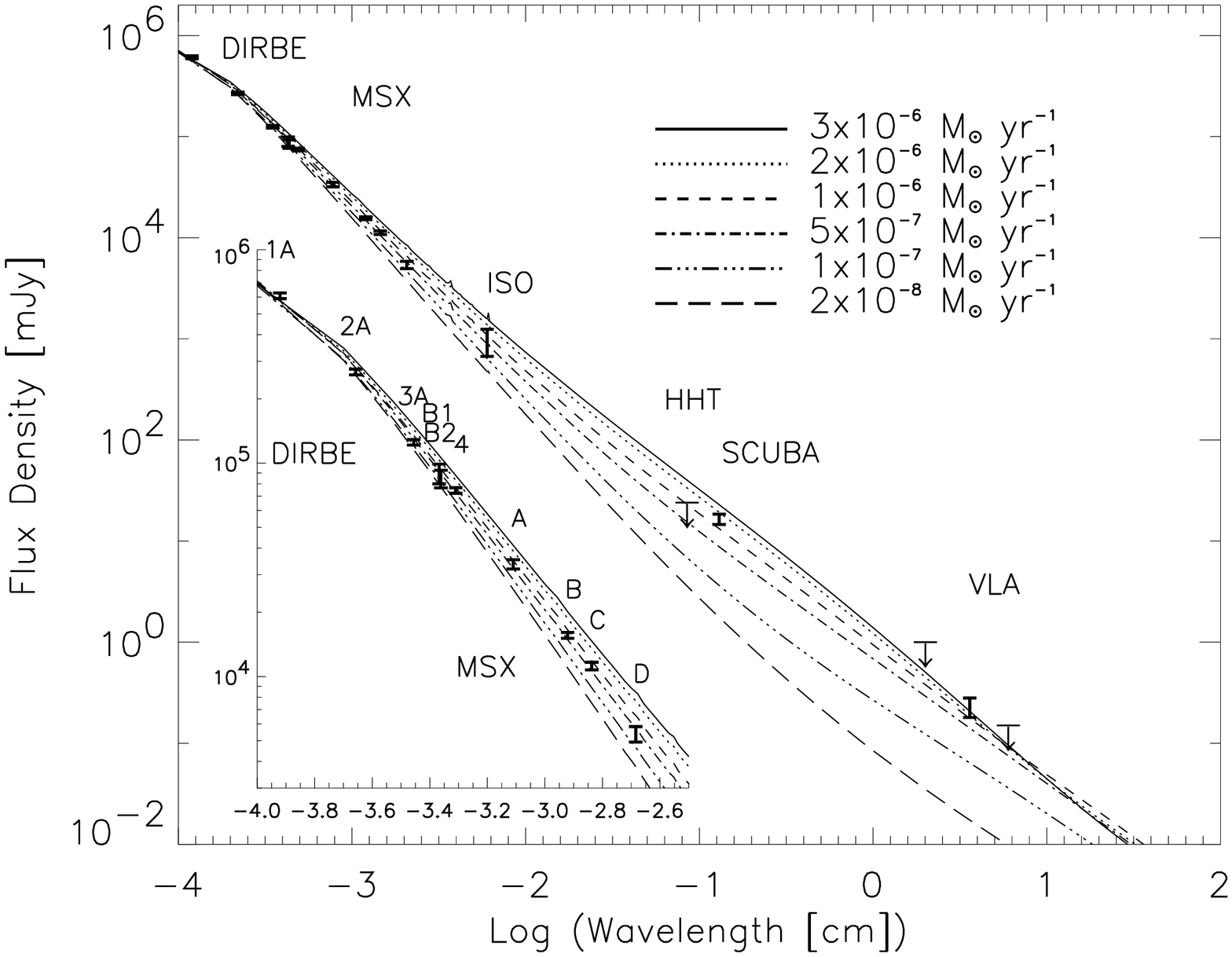}
\caption{Infrared, millimeter and radio photometry of Deneb from Table
\ref{tab:irdata}.  Synthetic continua for models listed in Table
\ref{tab:SEDmodels} with $T_\star$ = 9000 K and 6 different mass-loss
rates are shown for comparison.  The inset shows a close up of the
COBE/DIRBE and MSX photometry.}
\label{radio_data}
\end{figure}

%%%%%%%%%%%%%%%%%%%%%%%%%%%%%%%%%%%%%%%%%%%%%%%%%%%%%%%%%%%%%%%%%%%%%%%%
Figure \ref{radio_data} shows the IR, millimeter, and radio
observations from Table \ref{tab:irdata} compared with the synthetic
continua from a set of $T_\star$ = 9000 K models listed in
Table \ref{tab:SEDmodels}.  The synthetic continua are scaled with
$\theta_\star$ = 2.08 mas, consistent with the NPOI diameter.  Models
with mass-loss rates $\dot{M} \gtrsim 3\times 10^{-6}$ \Mspyr\ can be
ruled out by the 3$\sigma$ upper limit at 870 \micron\ from the HHT.
Except for the DIRBE band 1A, observations at shorter wavelengths are
consistent with this.  The detections at 1.35 mm and 3.6 cm, with
SCUBA and the VLA respectively, rule out models with mass-loss rates
\mdot\ $\lesssim 5 \times 10^{-7}$ \Mspyr.  These results are
consistent with the least squares results from the full UV to radio
spectrum.

%%%%%%%%%%%%%%%%%%%%%%%%%%%%%%%%%%%%%%%%%%%%%%%%%%%%%%%%%%%%%%%%%%%%%%%%
%FIGURE Radio Line-Blanketing Effects
%
\begin{figure}
\includegraphics[scale=0.7,angle=90]{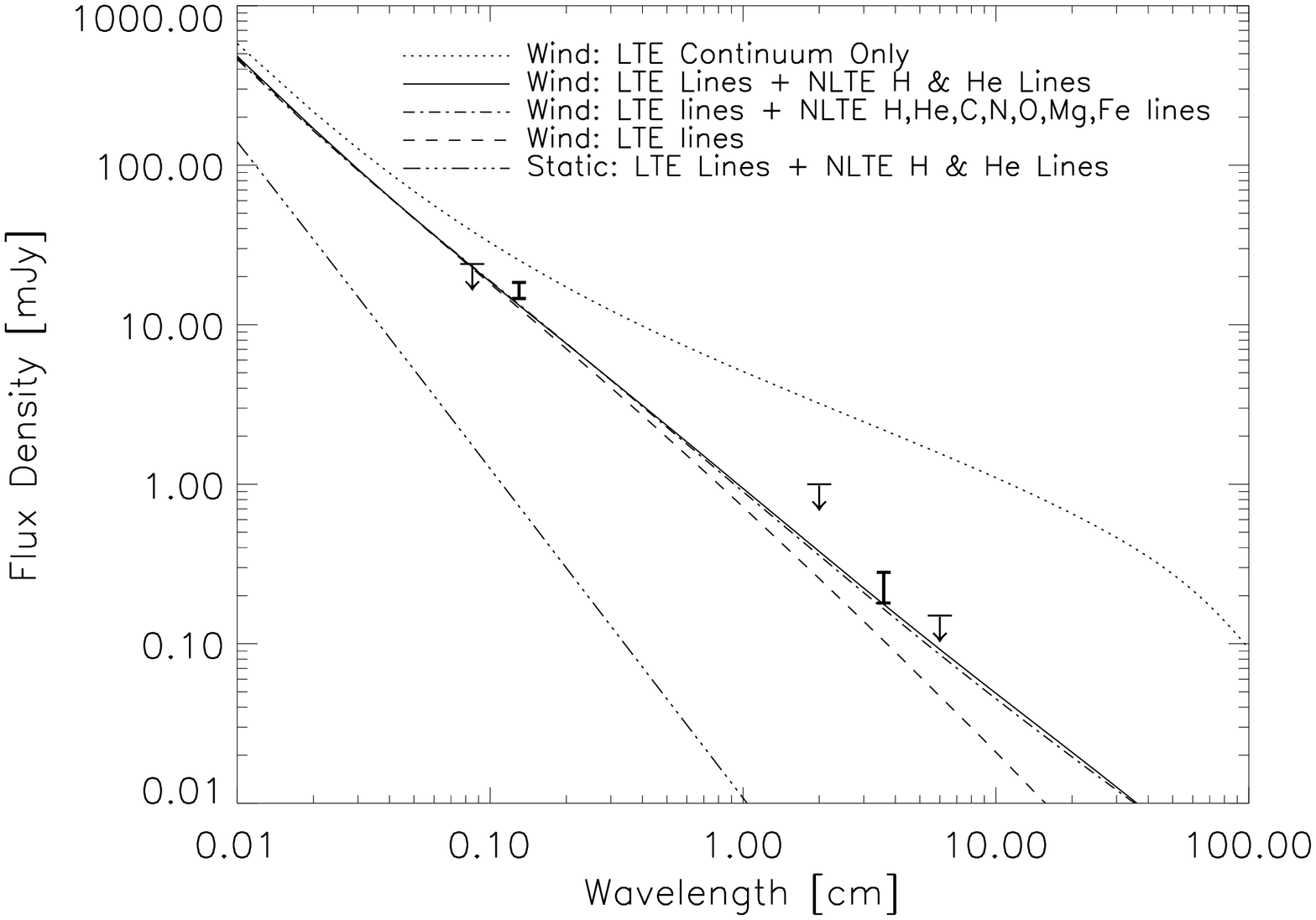}
\caption{Comparison of 4 wind models with identical parameters
($T_\star$ = 9000 K, $T_{\rm eff}^{Ross}$ = 8600 K, $\dot{M} =
10^{-6}$ \Mspyr, $R_\star$ = 1.2e13 cm, $\log g(R_\star)$ = 1.3, $\xi$
= 15 \kms, \vterm\ = 225 \kms, $\beta$ = 3.0), showing the importance
of line-blanketing on the predicted radio flux.  A hydrostatic model
($T_\star$ = 8600 K, $R_\star$ = 1.2e13 cm, $\log g(R_\star)$ = 1.3,
$\xi$ = 15 \kms) is shown for reference.  These synthetic spectra have
been scaled with $\theta_\star$ = 2.08 mas.  The HHT, SCUBA, and VLA data 
are shown for comparison.}
\label{radio_blanketing}
\end{figure}
%%%%%%%%%%%%%%%%%%%%%%%%%%%%%%%%%%%%%%%%%%%%%%%%%%%%%%%%%%%%%%%%%%%%%%%%

The synthetic continua shown in Figure \ref{radio_data} were computed
from spherical expanding atmosphere models whose temperature
structures are consistent with radiative equilibrium and where non-LTE
line-blanketing has been self-consistently included.  This is
important because at wavelengths beyond 1 mm the synthetic continua
are quite sensitive to the degree of line-blanketing employed in the
construction of the model atmosphere.  This is illustrated in Figure
\ref{radio_blanketing}, which shows the results of a test where five
wind models, which differ in the degree of line-blanketing, but have
otherwise identical parameters, were computed.  As easily seen with
reference to the hydrostatic model spectrum, the expanding atmosphere
spectra predict significantly larger fluxes.  The effects of
line-blanketing and non-LTE are no less significant.  For example, at
3 cm, the continuum opacity wind model produces $\sim$10 times more
flux than our most sophisticated non-LTE line-blanketed wind model,
and the LTE line-blanketed wind model produces $\sim$1/2 as much flux.
The LTE continuum-only model spectrum shows a turnover at $\sim30$ cm
due to truncation  at the outer boundary of the model, 200 $R_\star$.  
A similar model with an other boundary of 400 $R_\star$ does not
show a turnover.  The line-blanketed wind models, which are not fully
ionized to the outer boundary, are not sensitive to this truncation.

There are two important factors at play here.  First, the
incorporation of metal line-blanketing in the radiative equilibrium
solution cools the outermost layers of the atmosphere (classic surface
cooling), especially at depths where the radio continuum forms.  In LTE,
lower electron temperatures mean lower ionization, and therefore less
thermal radio emission is expected in the presence of line-blanketing
relative to the continuum-only opacity case.  Second, photoionization
is enhanced in non-LTE. 
Non-LTE photoionization affects the
ionization state of the outer layers because they are radiatively
coupled to the hotter radiation field at depth.  In non-LTE, the
ionization fraction in the outermost layers is enhanced relative to
the LTE line-blanketed case, which leads to more thermal radio
emission.

The sensitivity of the radio continuum to the degree of ionization in the
extended stellar envelope makes it clear that the application of a
fully ionized, uniform, spherically symmetric mass flow model commonly
employed in mass-loss rate estimates for O- and B-type supergiants
\cite[]{scud98,dl89} won't work for A-type supergiants.  Furthermore,
in such models the radio spectrum follows the power-law $S_\nu \propto
\nu^{2/3}$, but Deneb exhibits a spectral slope between 1.35 mm and
3.6 cm consistent with $S_\nu \propto \nu^{1.26\pm0.01}$, and is
therefore inconsistent with a fully ionized wind.  \cite{simon83}
studied thermal radio emission from partially ionized circumstellar
environments and winds.  They found that the slope of radio spectrum
varies from the fully ionized, optically thin case, ($S_\nu\propto
\nu^{0.6}$), to the optically thick, Rayleigh-Jeans case,
($S_\nu\propto \nu^{2}$).  Deneb is an intermediate case.  The
steepness of a radio spectrum depends on the degree of ionization and
radial extent of the ionized envelope.  We find the same result, where
the degree of ionization is controlled by line cooling and
photoionization. The optically thick case is represented by a
geometrically thin, hydrostatic model (see Figure
\ref{radio_blanketing}).

%%%%%%%%%%%%%%%%%%%%%%%%%%%%%%%%%%%%%%%%%%%%%%%%%%%%%%%%%%%%%%%%%%%%%%%%
%FIGURE Radio SED changes for beta, R, xi and logg
%
\begin{figure}
\includegraphics[scale=0.7,angle=90]{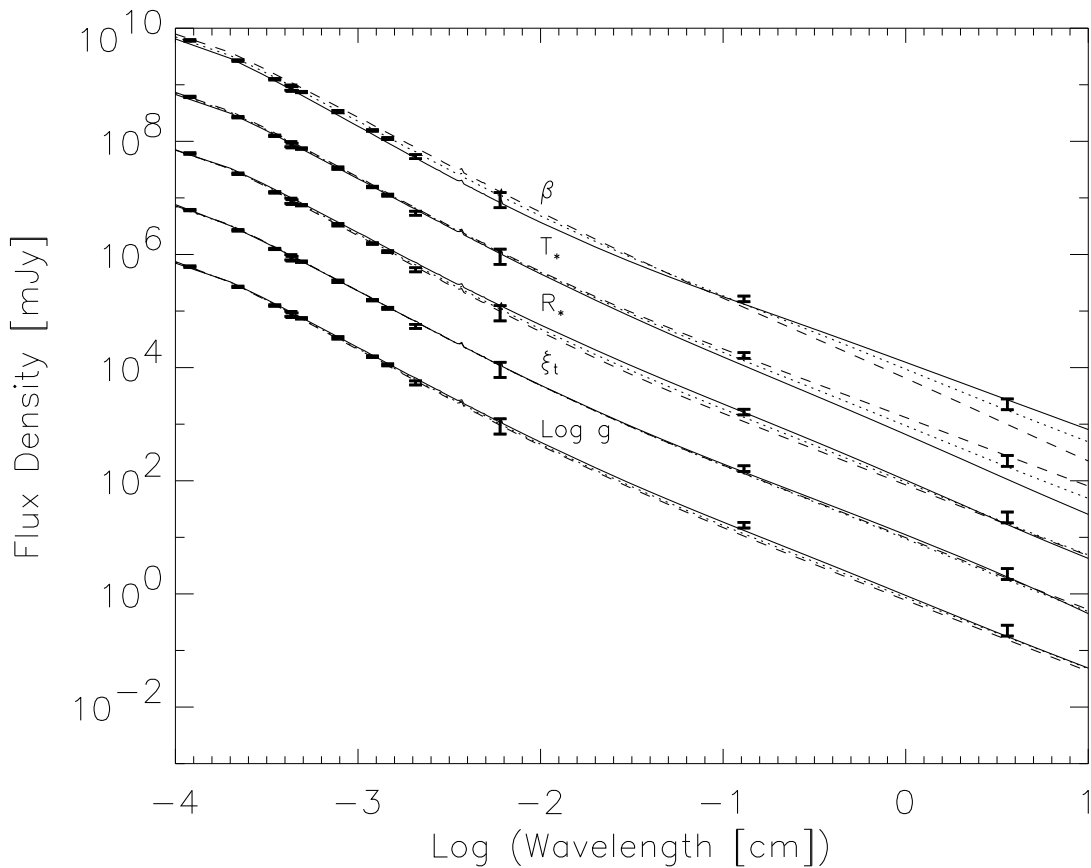}
\caption{Synthetic IR, millimeter, radio continua from five sets
(from top to bottom): the velocity law exponent,
$\beta$ = 2.0 (dotted), 3.0 (solid), 4.0 (dashed); the model
temperature, $T_\star$ = 8750 K (dotted), 9000 K (solid), 9250 K
(dashed); the model radius, $R_\star$ = 130 \Rsun\ (dotted), 172 \Rsun\
(solid), 215 \Rsun\ (dashed); microturbulence, $\xi_t$ = 5 \kms, 15
\kms, 25 \kms; and the surface gravity, \Logg\ = 1.3 (solid), 1.5
(dotted), 1.7 (dashed).  All models have the following parameters,
unless noted otherwise: $T_\star = 9000$ K , $\log g(R_\star) = 1.3$,
$\dot{M} = 10^{-6}$ \Mspyr, $\beta = 3.0$, $\xi_t$ = 15 \kms.  The
models are scaled using $\theta_\star$ = 2.08 mas.  Each set of three
models is offset by a factor of 10 for clarity.
The DIRBE, MSX, ISO, SCUBA, and VLA data are shown for comparison.}
\label{radio_beta_etal}
\end{figure}
%%%%%%%%%%%%%%%%%%%%%%%%%%%%%%%%%%%%%%%%%%%%%%%%%%%%%%%%%%%%%%%%%%%%%%%%

As demonstrated above, for a given effective temperature, the degree
of line-blanketing, photoionization, and the mass-loss rate have the
largest effects on the synthetic spectrum beyond 1 \micron.  The
synthetic spectrum is also affected to a much smaller degree by other
model parameters.  Figure
\ref{radio_beta_etal} shows the sensitivity of the synthetic spectrum,
relative to a good fitting reference model, to changes in $\beta$,
$T_\star$, $R_\star$, $\xi_t$, and \Logg.  Changing each of these
parameters has systematic effects on the synthetic radio continua.
Higher (lower) model temperatures increase (decrease) the ionization
in the wind and produce higher (lower) fluxes.  Changing $T_\star$ by
250 K has little effect below 1 mm.  Larger (smaller) values of
$R_\star$ yield lower (higher) wind densities for a fixed mass-loss
rate and as a result yields slightly lower (higher) fluxes.
Higher \Logg\ values yield slightly lower fluxes and
varying $\xi_t$ has no significant effect on the fluxes.

The velocity law parameter, $\beta$, has a very interesting effect on
the synthetic fluxes.  At wavelengths beyond 0.5 mm, larger values of
$\beta$ result in lower fluxes, while shortward of 0.5 mm the opposite
is true.  This behavior in the synthetic radio spectra is a result of
the run of electron pressure with depth in the wind as a function of
$\beta$.  At shorter wavelengths, where the opacity is lower, the
continuum forms in the inner wind ($\sim 3 R_\star$) where the gas and
electron densities are regulated primarily by the velocity law through
the continuity equation.  As a result, models with larger $\beta$
values have higher densities at this depth and thus higher fluxes.
Further out in the wind ($\sim 10 R_\star$), where the 3 cm continuum
forms, non-LTE effects are more important and the situation is
reversed.  At these depths the electron temperature and gas density
structures are essentially identical for all $\beta$ values, but the
electron densities are higher for models with the smaller $\beta$
values.  This happens because the models with smaller $\beta$ values
have warmer temperature structures at depth (see, for example, the
Balmer continuum as a function of $\beta$ in Figure \ref{beta_xi}) and
the resulting radiation field from these layers ionizes the wind more
substantially than models with larger $\beta$ values.  For example,
the $\beta$ = 2.0 model has the majority of hydrogen ionized out to a
radius of $R$ = 1640 \Rsun, while for the $\beta$ = 4.0 model, the
majority of hydrogren is ionized out to only $R$ = 610 \Rsun.

In summary, of all the model parameters, \mdot\ has the largest effect
on the synthetic continuum beyond 1 \micron.  Non-LTE photoionization,
high $T_\star$ values, and small $\beta$ values all increase model
fluxes at centimeter wavelengths for a fixed \mdot. 

\section{Deneb's Line Spectrum}

\subsection{The Ultraviolet Line Spectrum}
Deneb's rich UV line spectrum, with its multitude of iron peak
absorption lines, is a good diagnostic for the column density
in the expanding atmosphere and therefore should be a good diagnostic for the
mass-loss rate.  Modeling the detailed UV spectrum provides an
independent mass-loss rate estimate from that of the 
IR, mm, radio continuum because the latter is based on the integrated
emissivity of the extended atmosphere, while the former is based on the
line-of-sight column density toward the stellar photosphere.

\begin{figure}
\includegraphics[scale=0.7,angle=90]{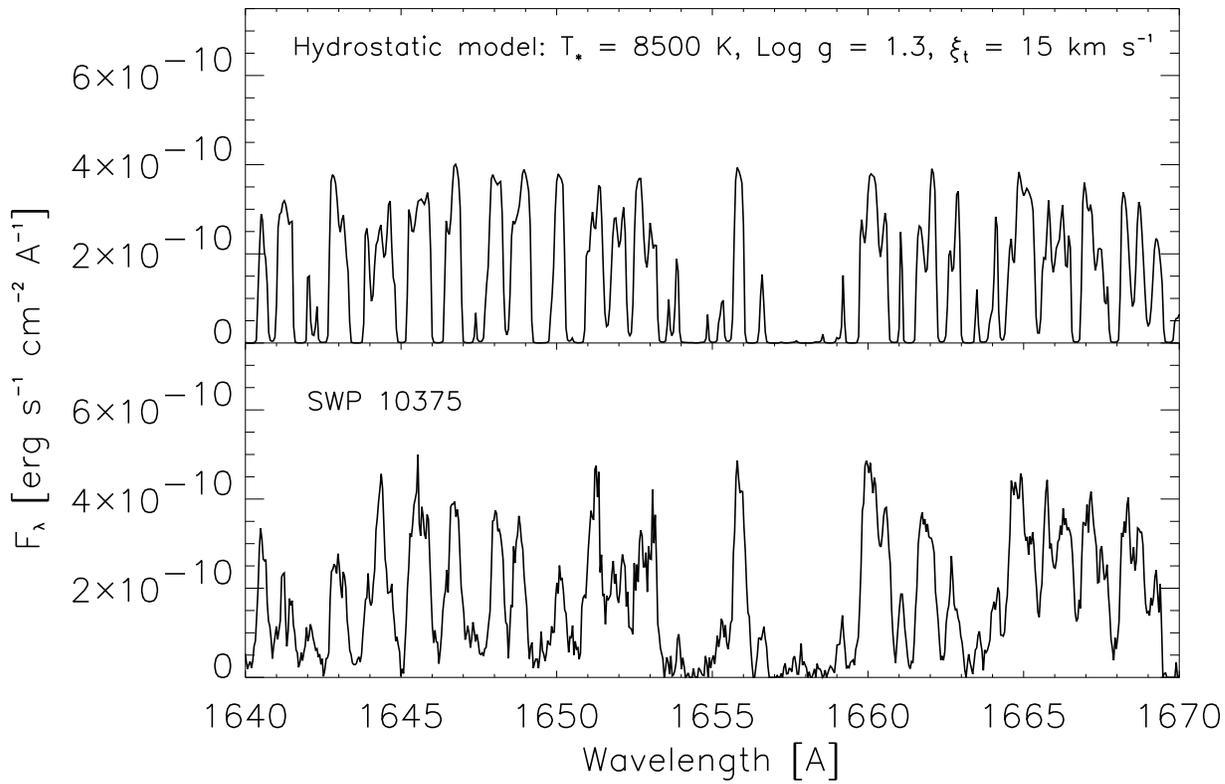}
\caption{Section of the high-dispersion \iue\ spectrum SWP 10375
(bottom panel) compared with a synthetic spectrum (top panel) from a
hydrostatic model.  There is good correspondance between the strong
absorption features in the model and the data, however the synthetic
line profiles are much too saturated relative to the observed
spectrum. The velocity field present in Deneb's atmosphere desaturates
and blends the strong absorption lines.}
\label{iue_static}
\end{figure}

Let us examine a portion of an \iue\ spectrum relative to a
hydrostatic model synthetic spectrum.  Figure \ref{iue_static} shows
that the synthetic spectrum reproduces the overall appearance of the
observed spectrum quite well.  However, in the observed spectrum the
stronger lines are considerably less saturated, broader, and blended
relative to the model spectrum.  This is almost certainly due to these
lines forming in the wind and sampling its velocity field,
as has been noted by others \cite[e.g.][]{praderie80}.
The 1640 \AA\ to 1670 \AA\ region is
chosen because here the synthetic spectra  are quite
sensitive to the value of $\dot{M}$ and this spectral region is
dominated by \ion{Fe}{2} absorption lines.
In all comparisons to the {\it IUE} data,
the models are scaled with an angular diameter
of $\theta_\star$ = 2.08 mas and the data have been corrected for
interstellar extinction (curve for HD 199579, $E(B-V)$ = 0.06, $R_V$ = 2.9).

%%%%%%%%%%%%%%%%%%%%%%%%%%%%%%
%
% IUE SWP vs. static model
%
%%%%%%%%%%%%%%%%%%%%%%%%%%%%%%%%%%%%%%%%%%%%%%%%%%%

%%%%%%%%%%%%%%%%%%%%%%%%%%%%%%
%
% IUE SWP vs. LTE wind models
%
\begin{figure}
\includegraphics[scale=0.7,angle=90]{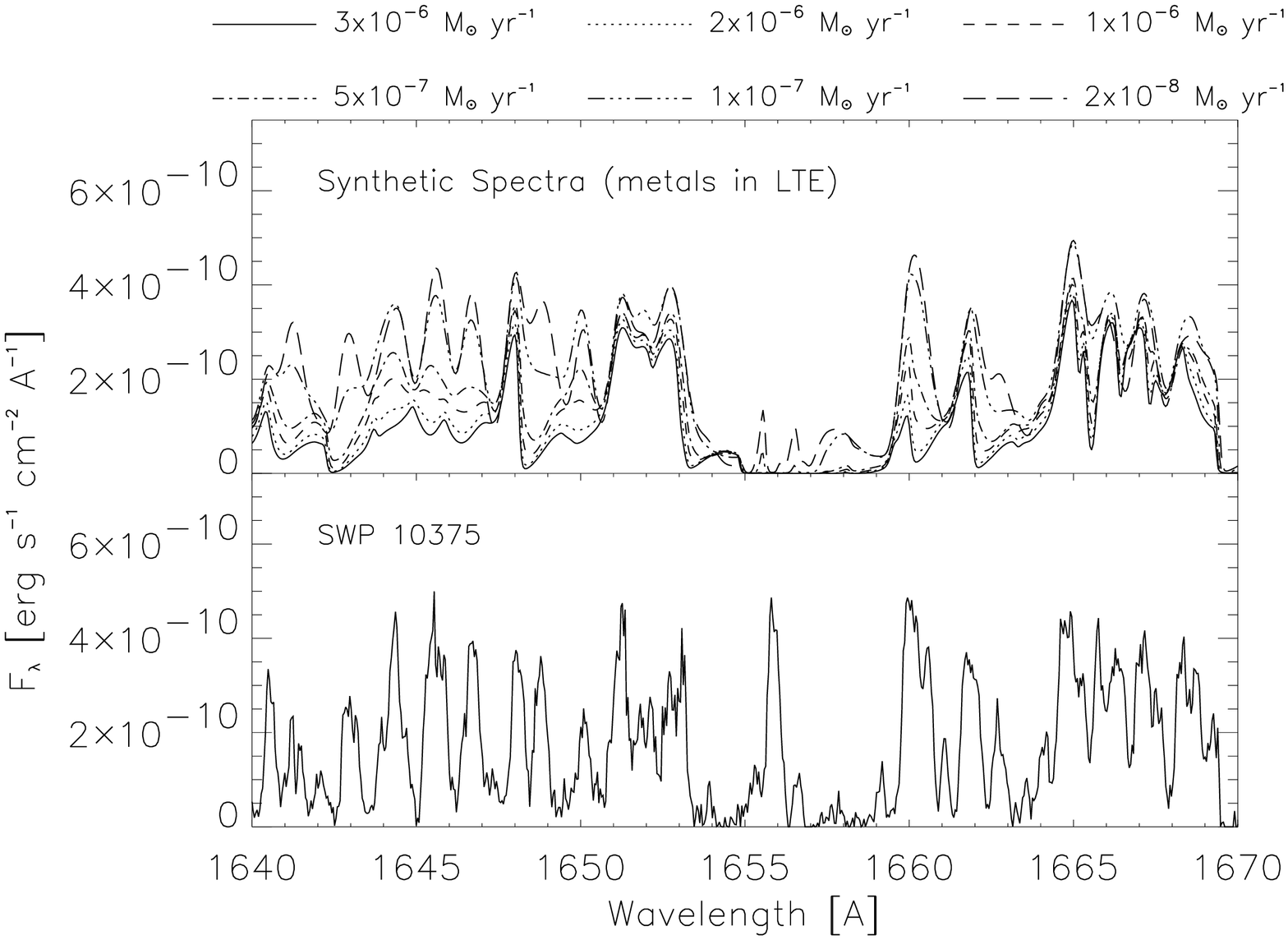}
\caption{Section of the high-dispersion \iue\ spectrum SWP 10375
(bottom panel) compared to a set of synthetic spectra (top panel) from
models with different mass-loss rates.  This spectral region is
dominated by \myion{Fe}{II} lines, however in these models all species
except \myion{H}{1} and \myion{He}{I-II}
are treated in LTE.}
\label{iue_lte_wind}
\end{figure}
%%%%%%%%%%%%%%%%%%%%%%%%%%%%%%%%%%%%%%%%%%%%%%%%%%%

Now we examine the synthetic UV spectra from the same models we
compared to the data beyond 1 \micron\ (see Figure \ref{radio_data}).
Figure \ref{iue_lte_wind} shows a comparison of six synthetic spectra
with LTE metal-line blanketing and mass-loss rates from $2\times
10^{-8}$ \Mspyr\ to $3 \times 10^{-6}$ \Mspyr\ to the \iue\ spectrum
SWP 10375.  The lowest mass-loss rate model, \mdot\ = $2 \times
10^{-8}$ \Mspyr, appears to be most consistent with the windows of low
opacity in the observed spectrum: 1645 \AA, 1656 \AA, 1660 \AA, and
1662 \AA.  Spectra from these LTE models with higher mass-loss rates (higher
wind column densities) show a very distorted spectrum, inconsistent
with the observations.  In these higher $\dot{M}$ models, the line
spectrum forms at larger radii, where the velocity field distorts and
washes out the spectrum.  However, even the \mdot\ = $2 \times
10^{-8}$ \Mspyr\ model cannot provide as good a match to the observed
spectrum as the hydrostatic model (see Figure \ref{iue_static}).
These results are very much at odds with the IR, mm, radio continuum
anaylsis which indicates $\dot{M} > 10^{-7}$ \Mspyr.

\begin{figure}
\includegraphics[scale=0.7,angle=90]{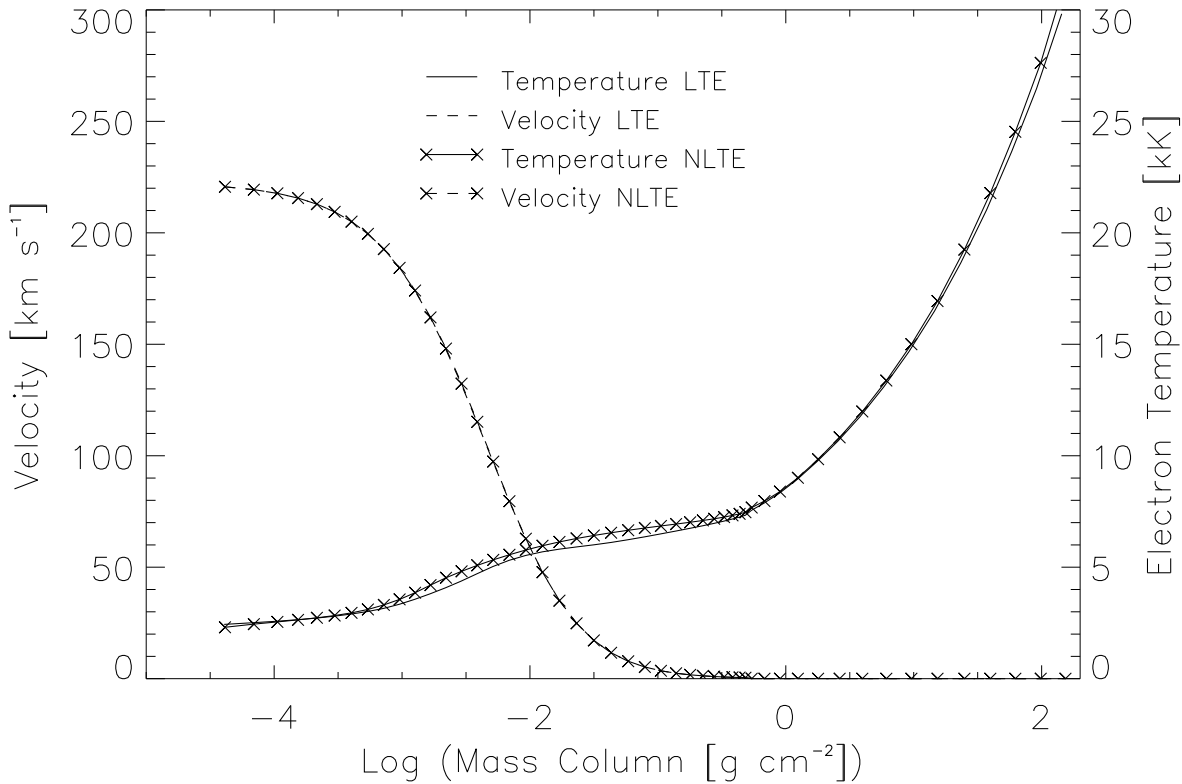}
\caption{Comparison of the velocity (left ordinate) and electron
temperature (right ordinate) structures for two models with and
without non-LTE metal line-blanketing of species
\myion{C}{i-iii}, \myion{N}{i-iii}, \myion{O}{i-iii}, \myion{Mg}{ii},
\myion{Fe}{i-iii}.  Both models have full LTE line-blanketing for
those species not treated in non-LTE.  The models have the following
parameters: $T_\star = 9000$ K , $\log g(R_\star) = 1.3$, $\dot{M} =
10^{-6}$ \Mspyr, $\beta = 3.0$, $\xi_t$ = 15 \kms.  The non-LTE model
has a slightly warmer temperature structure from the base of the wind
at $10^0$ g cm$^{-2}$ to $10^{-3}$ g cm$^{-2}$ due to photoelectric
heating.}
\label{temp_struc}
\end{figure}

\begin{figure}
\includegraphics[scale=0.7,angle=90]{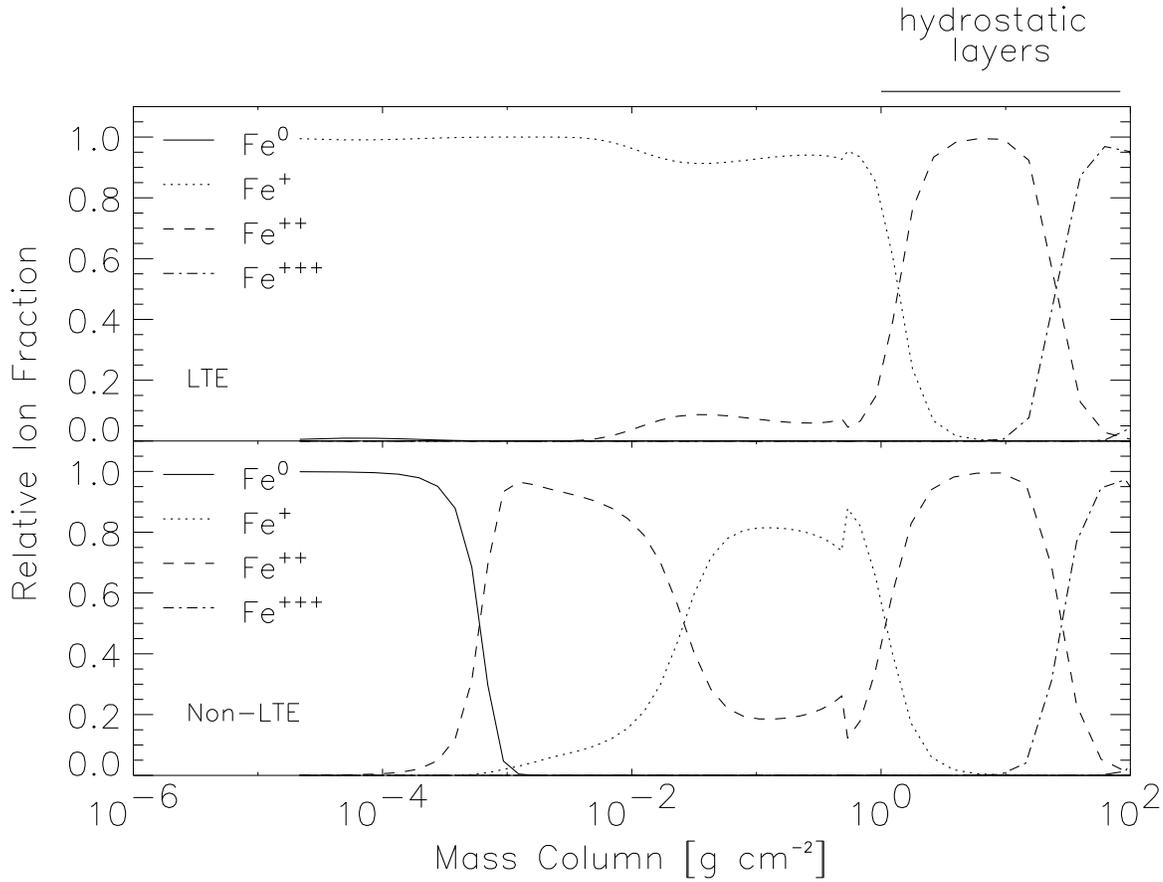}
\caption{Ionization structure of Fe from a model with LTE metal
line-blanketing (top) and a model with partial non-LTE metal
line-blanketing, including \ion{Fe}{1}, \ion{Fe}{2}, and \ion{Fe}{3}
(bottom).  A comparison of the two structures shows they are in good
agreement in the hydrostatic layers where LTE is a good assumption.
Further out in the wind, the two structures differ considerably.  The
dominant stage of Fe in the wind is Fe$^{+}$, when treated in LTE.  In
non-LTE, Fe$^{++}$ is the dominant stage of Fe throughout most of the
wind, rapidly recombining to Fe$^{0}$ in the outermost zones.  The
different ionization structures in the LTE and non-LTE cases lead to
significantly different synthetic line spectra in the UV.}
\label{iron_ion_struc}
\end{figure}

\begin{figure}
\includegraphics[scale=0.7,angle=90]{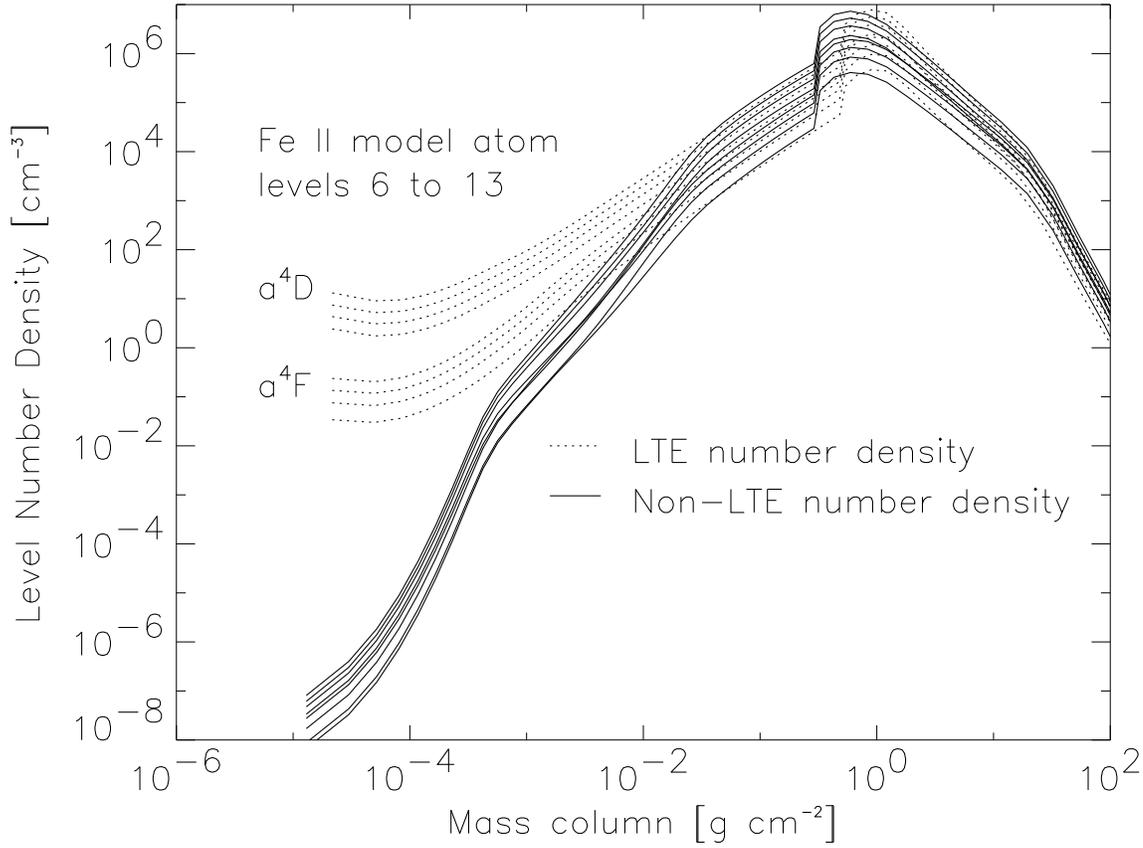}
\caption{Computed number densities for specific \ion{Fe}{2} levels
from terms a$^4F$ and a$^4D$ as a function of depth in the atmosphere.
These levels are the lower levels of transitions which are found in
the 1640 \AA\ to 1670 \AA\ spectral region shown in Figures
\ref{iue_lte_wind} and \ref{iue_nlte_wind}.  Level number densities
are shown for a model with LTE \myion{Fe}{i-iii} (dotted-lines) and a
model with non-LTE \myion{Fe}{i-iii} model (solid-lines).  At depths
greater than than 1 g cm$^{-2}$, the hydrostatic portion of the
atmosphere, the LTE and non-LTE level populuations are in close
agreement.  In the wind, non-LTE level densities are strongly
depleted relative to LTE level densities by the effects of the
ionizing radiation field.}
\label{iron_ni_struc}
\end{figure}

\begin{figure}
\includegraphics[scale=0.7,angle=90]{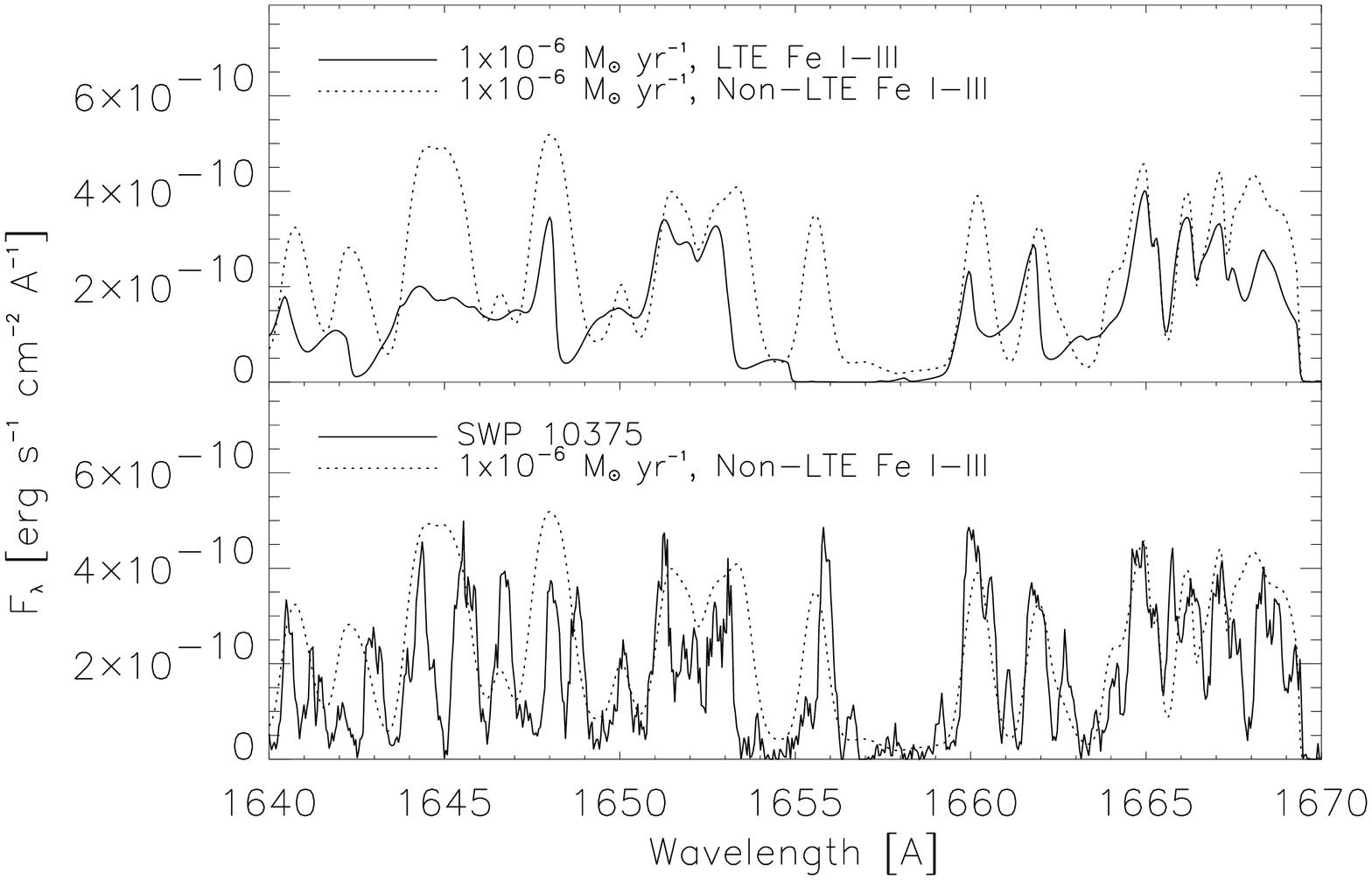}
\caption{(Top) Synthetic spectra from the two models shown in Figure
\ref{temp_struc}.  One model has non-LTE \myion{Fe}{i-iii}
line-blanketing, the other does not.  The two models have otherwise
identical parameters. The reduced \ion{Fe}{2} column density in the
non-LTE case means the strong \ion{Fe}{2} lines probe lower velocities
in the wind compared to the LTE case.  For example, the two strong
aborption features between 1640 \AA\ and 1645 \AA\ are shifted by --93 \kms\
in the LTE case relative to the non-LTE case. Most strikingly, the
opacity window at 1656 \AA\ is completely missing in the LTE case.
(Bottom) Non-LTE synthetic spectrum from the top panel is compared with the
\iue\ spectrum SWP 10375.}
\label{iue_nlte_wind}
\end{figure}

A solution to this inconsistency between the radio and UV mass-loss
rate estimates is to use a model which treats {Fe}~{\sc{I-III}}
in non-LTE (see \S A.1 for details on the construction of the model
ions and the atomic data). This is critically important.  Figure
\ref{temp_struc} shows that the temperature structures for the LTE and
non-LTE {Fe}~{\sc{i-iii}} models are quite similiar, as expected since
the synthetic radio spectra from the two models are in close agreement
(see Figure \ref{radio_blanketing}).  However, Figure
\ref{iron_ion_struc} shows the dramatic non-LTE effect of the
radiation field on the ionization structure of Fe in the wind:
Fe$^{+}$ is ionized to Fe$^{++}$, significantly dropping the Fe$^{+}$
column density in the wind relative to the LTE case.  Most of the
strong \ion{Fe}{2} transitions between 1640 \AA\ and 1670 \AA\ have
lower levels between 0.2 eV and 1 eV above the ground state, in terms
a$^4$F and a$^4$D.  These levels are seriously under-populated
relative to LTE as shown in Figure \ref{iron_ni_struc}.  The resultant
non-LTE synthetic spectrum (Figure \ref{iue_nlte_wind}) is a much
improved match to the observed spectrum with a mass-loss rate
consistent with the radio continuum result.

The bottom panel of Figure \ref{iue_nlte_wind} shows the comparison of
the non-LTE model spectrum and the \iue\ data. The overall match to
the 1640 \AA\ to 1670 \AA\ region is generally good, but several
strong lines do not appear in the model spectrum: e.g., 1645.015 \AA,
1648.403 \AA, and 1667.913 \AA.  These lines do appear in the
hydrostatic model (see Figure \ref{iue_static}) and are therefore in
the LTE line list \cite[]{jcdrom23}, but are missing or very weak in the non-LTE
spectrum.  Specifically, consulting \cite{jcdrom22} we find that the
1645.015 \AA\ line (d$^7$ a$^4$F$_{9/2}$ -- (3H)4p z$^2$I$_{11/2}$) is
missing from the synthetic spectrum because the upper level (62662
cm$^{-1}$) is missing from the model atom.  The 1648.403 \AA\ line
((5D)4s a$^6$D$_{5/2}$ -- (3P)4p y$^4$P$_{3/2}$) is an
intercombination line in LS-coupling, and its predicted $gf$-value
falls below the $\log(gf) > -3.0$ cutoff for our \ion{Fe}{2} model atom.
The data suggest the true $gf$-value is larger, which is not
surprising given the large uncertaintly likely for a theoretical
$gf$-value from a spin-forbidden transition.  The 1667.913 \AA\ line
(d$^7$ a$^2$P$_{1/2}$ -- (1D)4p w$^2$P$_{1/2}$) by comparison has a
more highly excited lower level (18882.702 cm$^{-1}$ versus 1872
cm$^{-1}$ to 8846 cm$^{-1}$ for the a$^4$F and a$^4$D terms) and may
be quenched by the Boltzmann factor, and possibly by inaccurate
radiative and collisional rates.  In constrast, the strong lines which
are well matched by the synthetic spectrum (e.g., 1643.578 \AA,
1650.704 \AA, 1663.222 \AA) have upper levels with highly pure-LS
eigenvectors (weak intermediate coupling) which are expected to yield
the most reliable transition strengths.  
The complexity of the high-dispersion UV spectrum and the limitations
of the atomic data do not allow to us to better constrain
other model parameters, such as $\beta$, from this comparison.

%%%%%%%%%%%%%%%%%%%%%%%%%%%%%%%%%%%%%%%%%%%%%%%%%%%%%%%
%
% LTE vs. NLTE Temperature structures
%
%%%%%%%%%%%%%%%%%%%%%%%%%%%%%%%%%%%%%%%%%%%%%%%%%%%%%%%%

%%%%%%%%%%%%%%%%%%%%%%%%%%%%%%%%%%%%%%%%%%%%%%%%%%%%%%%
%
% LTE vs. NLTE Fe ionization structures
%
%%%%%%%%%%%%%%%%%%%%%%%%%%%%%%%%%%%%%%%%%%%%%%%%%%%%%%%%

%%%%%%%%%%%%%%%%%%%%%%%%%%%%%%%%%%%%%%%%%%%%%%%%%%%%%%%%
%
% LTE vs. NLTE Fe ionization structures
%
%%%%%%%%%%%%%%%%%%%%%%%%%%%%%%%%%%%%%%%%%%%%%%%%%%%%%%%%

%%%%%%%%%%%%%%%%%%%%%%%%%%%%%%%%%%%%%%%%%%%%%%%%%%%%%%%%%
%
% IUE SWP vs. NLTE wind model
%
%%%%%%%%%%%%%%%%%%%%%%%%%%%%%%%%%%%%%%%%%%%%%%%%%%%%%%%%%%%

\begin{figure}
\includegraphics[scale=0.7,angle=90]{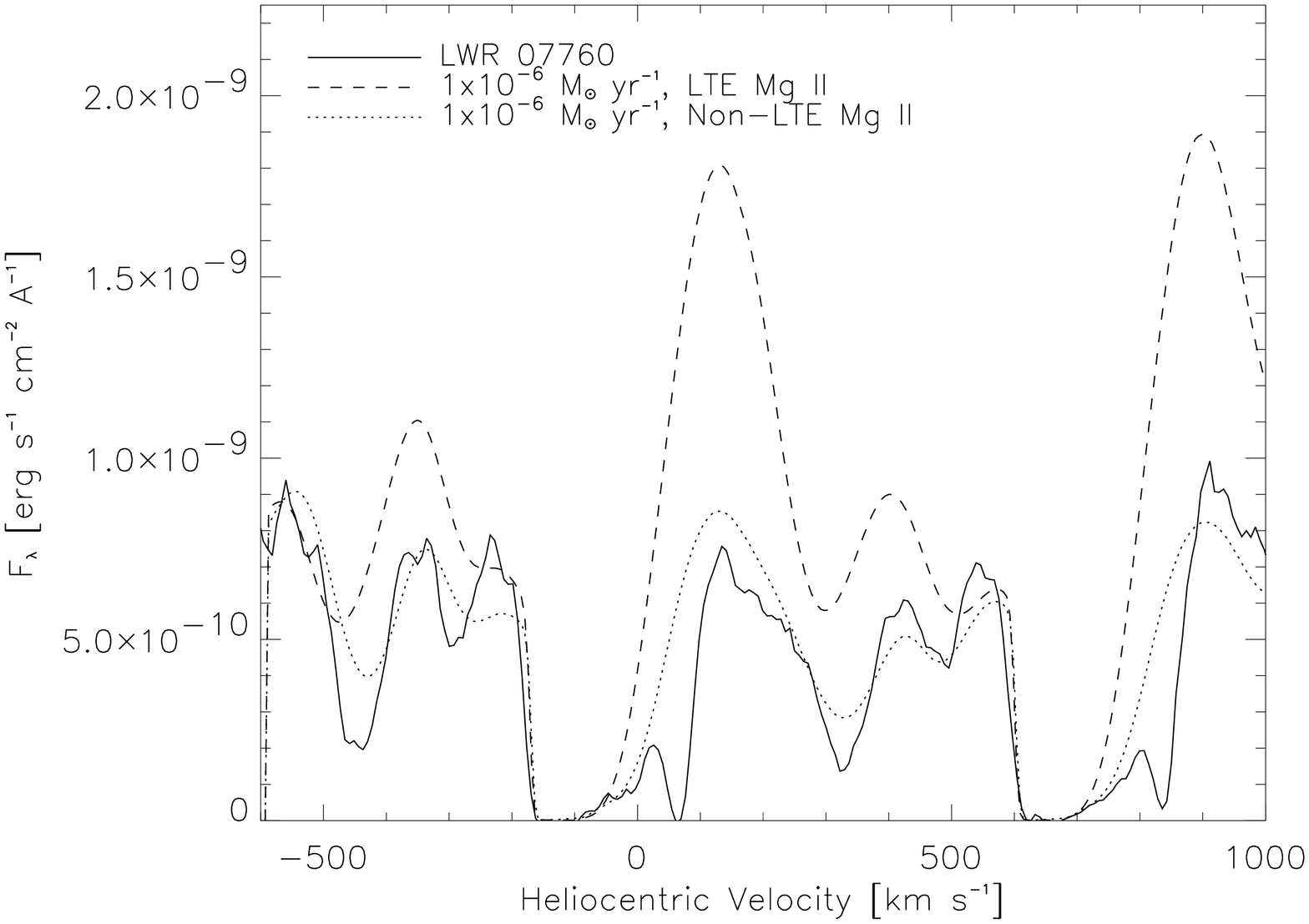}
\caption{{\it IUE} spectrum LWR 07760 in a region showing the
\ion{Mg}{2} $h$ and $k$ resonance lines.  Shown for comparison are
synthetic spectra from two models: LTE \ion{Mg}{2} and non-LTE
\ion{Mg}{2}.  With the exception of the non-LTE metal line-blanketing,
these models have identical parameters.  The zero-point of the
velocity scale is the $h$ line, $\lambda_{\rm air}$ = 2795.523 \AA\
shifted to radial velocity of Deneb, RV = $-4.5$ \kms\
\cite[]{deneb_radvel}. The interstellar components at $\sim +70$ \kms, 
have not been modeled.}
\label{mg2_vterm}
\end{figure}

A non-LTE treatment also provides a better fit to the \ion{Mg}{2} $h$
and $k$ resonance lines and their satellites as shown in Figure
\ref{mg2_vterm}.  The non-LTE synthetic spectrum more accurately
reproduces the observed spectrum relative to the LTE case in two
respects: (1) the reduced strength of the $h$ and $k$ P-Cygni emission
components and, (2) the lower velocity shifts of the \ion{Mg}{2}
satellite lines at -450 \kms\ and +325 \kms.  As in the case of
\ion{Fe}{2}, enhanced ionization in the non-LTE case reduces the
\ion{Mg}{2} column density such that the unsaturated satellite lines
form deeper in the wind at lower velocities.  The models in Figure
\ref{mg2_vterm} are parameterized with \vterm\ = 225 \kms\ (maximum
model velocity, $v_{\rm max}$ = 222 \kms), and provide a good match to
the blue edge of the $h$ and $k$ P-Cygni absorption troughs.  Because
of their strength, the resonance lines probe the highest velocities in
the wind and thus fix the \vterm\ parameter.

%%%%%%%%%%%%%%%%%%%%%%%%%%%%%%%%%%%%%%%%%%%%%%%%%%%%%%%%%%%%%%%%%%%%%%%%
%FIGURE Mg II fit to constrain vterm
%
%%%%%%%%%%%%%%%%%%%%%%%%%%%%%%%%%%%%%%%%%%%%%%%%%%%%%%%%%%%%%%%%%%%%%%%%

\subsection{Visual-IR Line Spectrum}
\subsubsection{The H$\alpha$ line}
At optical wavelengths, the most conspicuous indicator of mass-loss in
early-type supergiants is the P-Cygni character of the \ha\ line.
Like all BA-type supergiants, Deneb's \ha\ profile is variable
\cite[]{kaufer96}.  Deneb's \ha\ profile distinguishes itself from
most supergiants in that it lacks broad emission wings normally
attributed with electron scattering.  Despite being the protype of the
class, Deneb's \ha\ profile has only been modeled in detail twice to
our knowledge.

Reasonable fits to several of Deneb's \ha\ profiles have been achieved
by \cite{km82} and \cite{scuderi92}.  Unfortunately, the model
parameters and assumptions used in these works are inconsistent with
the present observational data.  In the former work, the best fits
were achieved by adopting terminal velocities of 120 \kms\ and 140
\kms\ which are inconsistant with the blue edges of the \ion{Mg}{2}
$h$ \& $k$ profiles (see Figure \ref{mg2_vterm}).  In the latter work,
the critical assumption of a fully ionized wind is made in addition to
the assumption of LTE and the Sobolev approximation.  As shown above
(\S 5.3, Figure 11), a fully ionized wind is inconsistent with the
observed slope of the millimeter-radio continuum.

\begin{figure}
\includegraphics[scale=0.7,angle=90]{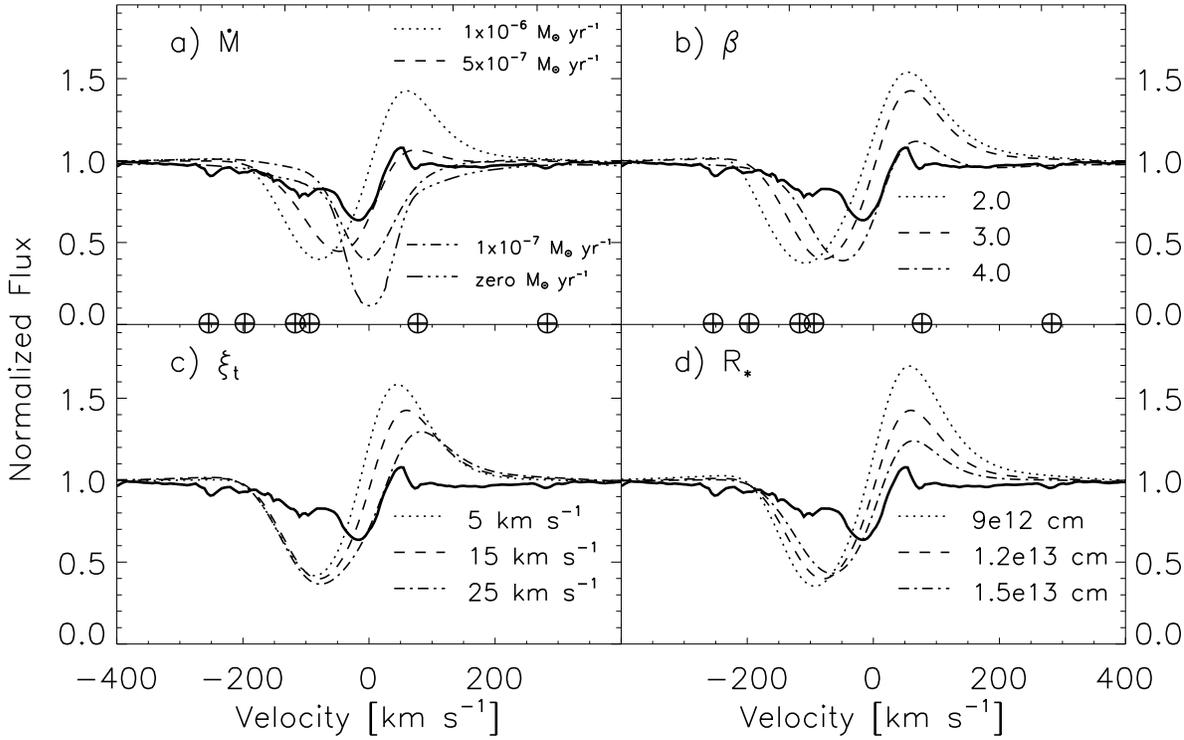}
\caption{HEROS red spectrum (thick solid line) at the \ha\ line
shown with synthetic profiles from a grid of model atmospheres with
different parameterizations about the base model ($T_\star = 9000$ K ,
$\log g(R_\star) = 1.3$, $\dot{M} = 10^{-6}$ \Mspyr, $\beta = 3.0$,
$\xi_t$ = 15 \kms). The synthetic profile for zero mass-loss is from
the hydrostatic model $T_\star = 8500$ K , $\log g(R_\star) = 1.3$,
$\xi_t$ = 15 \kms.  Shown are the effects on the synthetic profile due
to the (a) mass-loss rate, (b) velocity-law exponent, (c)
microturbulence parameter, and (d) the input radius. Positions of
telluric lines are mark by $\oplus$ symbols.  All synthetic profiles
have been shifted from vacuum to air wavelengths and rotationally
broadened for $v \sin i$ = 25 \kms.}
\label{halpha}
\end{figure}

We have attempted to fit the \ha\ profile from the HEROS red spectrum
by fixing \vterm\ at 225 \kms\ and computing the synthetic profiles in
the co-moving frame from the set of radiative equilibrium expanding
atmosphere models computed for the SED analysis.  We are unable to
improve upon earlier attempts to fit the observed \ha\ profile by
adjusting {\it any} of the model parameters (see Figure \ref{halpha}).
The most significant and persistent discrepancy between the synthetic
and observed profiles is the depth of the absorption component, which is 
significantly weaker
in the observed spectrum (residual intensity $\simeq 0.6$). 
Observed \ha\ profiles from 1991 \cite[]{kaufer96} have a broader absorption
component than shown in Figure \ref{halpha}, but they are not deeper.
Furthermore, while the velocity of absorption component minimum is
quite variable, it is rarely, if ever, shifted blueward of --50 \kms,
$\sim$20\% of the terminal velocity. Models with mass-loss rates
\mdot\ $> 10^{-7}$ \Mspyr\ produce synthetic profiles
with absorption minima shifted to --80 \kms\ and beyond.  Another problem
is that in this same mass-loss range the synthetic profiles show the
broad electron scattering wings prominent in other hot supergiants,
but lacking in Deneb.  Since we are unable to achieve a reasonable fit
to the observed \ha\ profile, it is near impossible to 
reliably estimate a mass-loss
rate from this line.  The position of the absorption component
relative to the models (Figure \ref{halpha}a) suggests
a mass-loss rate in the range: 
$5\times 10^{-7}$ \Mspyr\ $>$ \mdot\ $> 10^{-7}$ \Mspyr.

%%%%%%%%%%%%%%%%%%%%%%%%%%%%%%%%%%%%%%%%%%%%%%%%%%%%%%%%%%%%%%%%%%%%%%%%
%FIGURE halpha profiles
%
%%%%%%%%%%%%%%%%%%%%%%%%%%%%%%%%%%%%%%%%%%%%%%%%%%%%%%%%%%%%%%%%%%%%%%%%

\begin{figure}
\includegraphics[scale=0.7,angle=90]{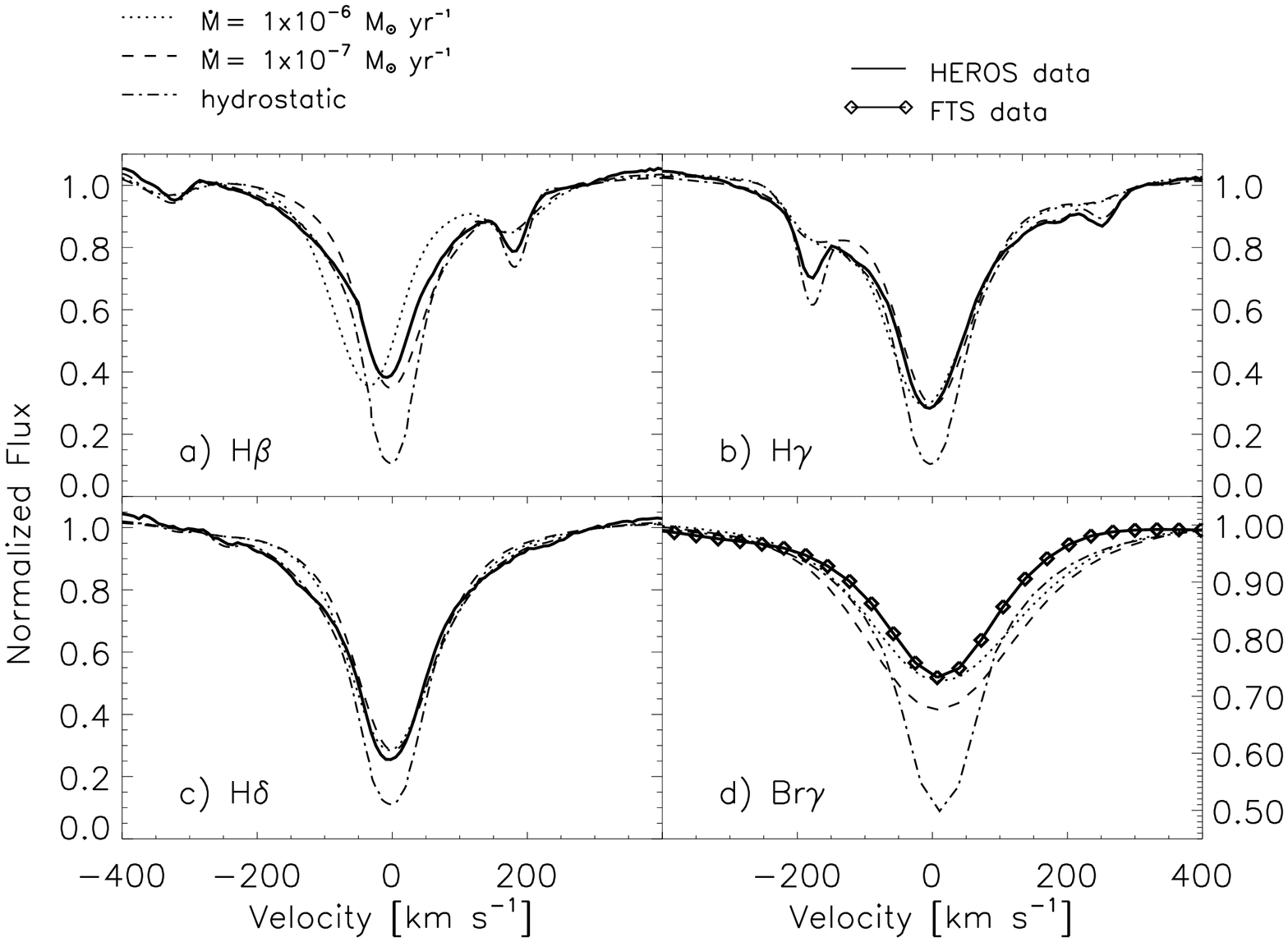}
\caption{HEROS blue spectrum (thick solid line) at (a) \hb, (b)
\hg, (c) \hd, and the FTS spectrum (thick solid line with diamonds) at
(d) \brg.  Shown for comparison are synthetic profiles from three model
atmospheres: two wind models, $T_\star = 9000$ K, $\log g(R_\star) =
1.3$, $\beta = 3.0$, $\xi_t$ = 15 \kms, $\dot{M} = 10^{-6}$ \Mspyr\
(dotted line) and $\dot{M} = 10^{-7}$ \Mspyr (dashed line), and the
hydrostatic model (dash-dot line) $T_\star = 8500$ K , $\log
g(R_\star) = 1.3$, $\xi_t$ = 15 \kms.  The synthetic profiles for
comparison with the HEROS data have been shifted from vacuum to air
wavelengths.  All synthetic profiles have been rotationally broadened
for $v \sin i$ = 25 \kms.}
\label{hbeta_etal}
\end{figure}

Higher members of the Balmer series (\hb, \hg, \hd) are
a better match to our synthetic profiles than \ha\ (see Figure
\ref{hbeta_etal}).  The depth of the observed line cores are well matched by our
wind model synthetic profiles which have weaker cores than 
hydrostatic model synthetic profiles.  However, for these same lines
the wings are generally fit better by the hydrostatic profiles.  The
wind profiles also fail to match the weak metal lines in the wings of
\hb\ and \hg.  In the wind models the column density through the wind
is clearly too large and metal lines which should form deeper in the
atmosphere are washed out by the velocity field.  The data do show the
presence of a wind in the \hb\ and \brg\ lines.  The \hb\ line core is
shifted slightly blueward from zero velocity and the \brg\ profile
(see Figure \ref{hbeta_etal}d) is noticeably asymmetric in the red
wing, suggesting a very weak P-Cygni emission component to this line.

%%%%%%%%%%%%%%%%%%%%%%%%%%%%%%%%%%%%%%%%%%%%%%%%%%%%%%%%%%%%%%%%%%%%%%%%
%FIGURE h-beta, h-gamma, h-delta, and Brackett gamma
%
%%%%%%%%%%%%%%%%%%%%%%%%%%%%%%%%%%%%%%%%%%%%%%%%%%%%%%%%%%%%%%%%%%%%%%%%

\begin{figure}
\includegraphics[scale=0.7,angle=90]{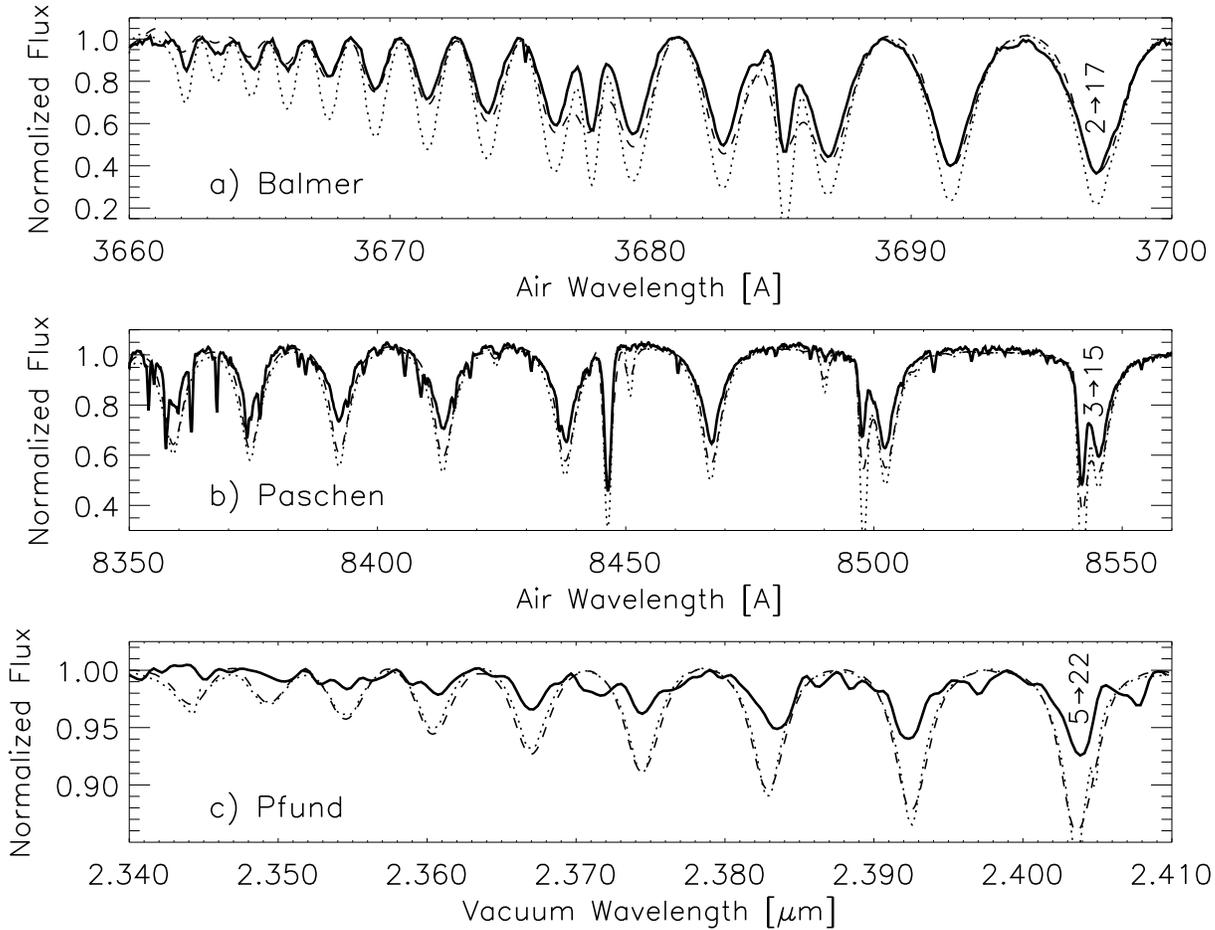}
\caption{Observed hydrogen lines (thick solid lines), from three
series: (a) Balmer, (b) Paschen, and (c) Pfund.  These lines are all
high members of there respective series: (a) 2$\rightarrow$17 to
2$\rightarrow$30, (b) 3$\rightarrow$15 to 3$\rightarrow$22, and (c)
5$\rightarrow$22 to $\sim$5$\rightarrow$26.  Other strong features
in the spectra include \ion{Cr}{2} $\lambda$3678, \ion{Ti}{2}
$\lambda$3685, \ion{O}{1} $\lambda$8446, and \ion{Ca}{2}
$\lambda\lambda$8498,8542. The Paschen series is contaminated by many
weak telluric features.  The observed Balmer and Paschen series are
from the HEROS blue and red spectra respectively and the Pfund series
is from the FTS spectrum.  Shown for comparison are synthetic profiles
from two model atmospheres: a wind model (dashed line), $T_\star =
9000$ K, $\log g(R_\star) = 1.3$, $\beta = 3.0$, $\xi_t$ = 15 \kms,
$\dot{M} = 10^{-6}$ \Mspyr\ and a hydrostatic model (dotted line)
$T_\star = 8500$ K , $\log g(R_\star) = 1.3$, $\xi_t$ = 15 \kms.  The
synthetic profiles for comparison with the HEROS data have been
shifted from vacuum to air wavelengths.  All synthetic profiles have
been rotationally broadened for $v \sin i$ = 25 \kms.}
\label{h_lines}
\end{figure}

For still higher members of the Balmer series, 2$\rightarrow$17 to
2$\rightarrow$30, the observed lines are closely matched by the wind
model synthetic profiles (see Figure \ref{h_lines}a). The hydrostatic
model line cores are again too strong.  For high members of the
Paschen series, 3$\rightarrow$15 to 3$\rightarrow$22, the difference
between the wind and hydrostatic synthetic profiles is less distinct
and both predict profiles which are a bit too strong (see Figure
\ref{h_lines}b).  For the last few members of the Pfund series,
5$\rightarrow$ 22 to roughly 5$\rightarrow$26, the wind and
hydrostatic synthetic profiles are nearly indentical and are much
stronger than the observed profiles (see Figure \ref{h_lines}c).  It
is expected that for progressively weaker hydrogen series the
differences between the wind and hydrostatic models should disappear
since these weak lines form deep in the photosphere.  It also appears
clear that the $n=5$ level of hydrogen in the models is overpopulated
relative to Deneb's photosphere.

%%%%%%%%%%%%%%%%%%%%%%%%%%%%%%%%%%%%%%%%%%%%%%%%%%%%%%%%%%%%%%%%%%%%%%%%
%FIGURE high Balmer lines
%
%%%%%%%%%%%%%%%%%%%%%%%%%%%%%%%%%%%%%%%%%%%%%%%%%%%%%%%%%%%%%%%%%%%%%%%%

\begin{figure}
\includegraphics[scale=0.7,angle=90]{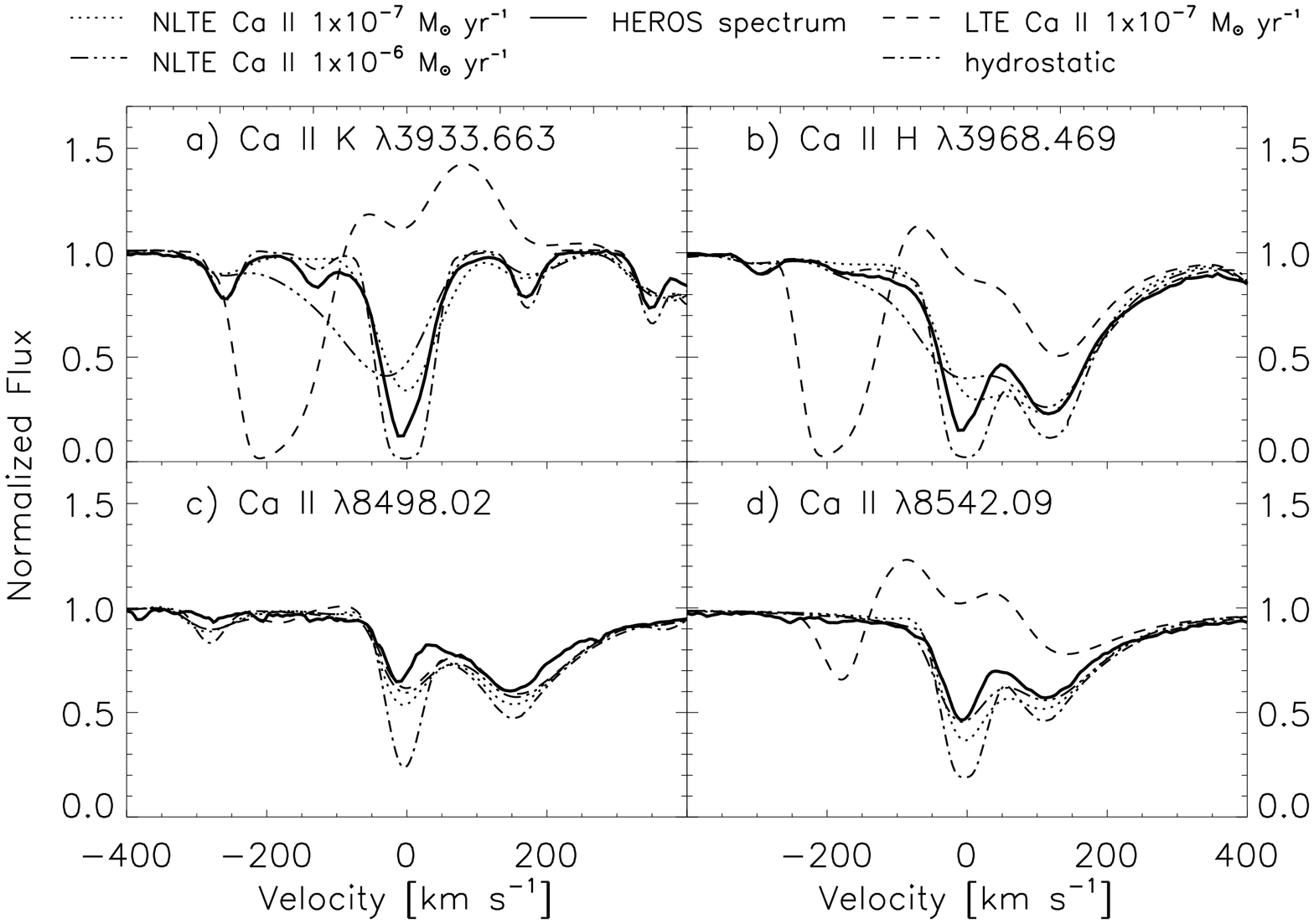}
\caption{HEROS spectra (thick solid lines) at the \ion{Ca}{2}
lines (a) $K$, (b) $H$, (c) $\lambda$8498, and (d) $\lambda$8542.  Shown
for comparison are synthetic profiles from four model atmospheres: two
wind models with non-LTE \myion{Ca}{i-iii}, plus one wind model and
one hydrostatic model with LTE \myion{Ca}{i-iii}.  The wind models:
$T_\star = 9000$ K, $\log g(R_\star) = 1.3$, $\beta = 3.0$, $\xi_t$ =
15 \kms, $\dot{M} = 10^{-6}$ \Mspyr\ (long dash-dot line) and $\dot{M}
= 10^{-7}$ \Mspyr (LTE dashed line, non-LTE dotted line) and the
hydrostatic model: $T_\star = 8500$ K , $\log g(R_\star) = 1.3$,
$\xi_t$ = 15 \kms\ (short dash-dot line).  The synthetic profiles have
been shifted from vacuum to air wavelengths.  All synthetic profiles
have been rotationally broadened for $v \sin i$ = 25 \kms.}
\label{ca2_4box}
\end{figure}

The last set of line profiles we show are those from \ion{Ca}{2}: the
$H$ \& $K$ lines and two members of the infrared triplet (IRT) (see Figure
\ref{ca2_4box}).  The most striking characteristic of the synthetic
profiles for the $H$, $K$, and $\lambda$8542 lines is the importance of
non-LTE.  In LTE, Ca$^+$ is the dominant stage of Ca in the wind and
the strong synthetic P-Cygni profiles of the $H$ \& $K$ are evidence for
this. In the non-LTE models, and apparently in Deneb's expanding
atmosphere, most of the Ca is in the form of Ca$^{++}$, which has its
first excited level 25 eV above the ground state, with no strong
lines.  For the $H$ \& $K$ lines, the hydrostatic model synthetic profiles
are a bit too strong and the non-LTE wind profiles are a bit too
weak. For the model with \mdot\ = $10^{-6}$ \Mspyr, the synthetic
profiles are also significantly blueshifted, suggesting \mdot $< 10^{-7}$
\Mspyr.  For the two IRT lines, only the stronger $\lambda$8542 line
shows a significant departure from LTE.  The non-LTE wind model lines
are a bit too strong and the hydrostatic profiles are stronger still.

%%%%%%%%%%%%%%%%%%%%%%%%%%%%%%%%%%%%%%%%%%%%%%%%%%%%%%%%%%%%%%%%%%%%%%%%
%FIGURE Ca II lines
%
%%%%%%%%%%%%%%%%%%%%%%%%%%%%%%%%%%%%%%%%%%%%%%%%%%%%%%%%%%%%%%%%%%%%%%%%

\section{Summary and Conclusions}
A stellar wind module has been developed for the \phoenix\ stellar
atmosphere code for the purpose of computing atmospheric structures
and detailed synthetic spectra of hot luminous stars.  These models
incorporate a solution of the expanding spherically symmetric
radiative transfer equation in the co-moving frame, a treatment of
detailed metal line-blanketing in non-local thermodynamic equilibrium,
and a convergence of the model temperature structure under the
constraint of radiative equilibrium.  We have used the model
atmospheres and synthetic spectra from this code to better illuminate
what we do not understand about the spectral energy distribution, the
mass-loss rate, and the fundamental stellar parameters of the A-type
supergiant Deneb.

\subsection{Millimeter and Centimeter Radiation Detections}
We have reported the first detections of Deneb at millimeter and
centimeter wavelengths with the SCUBA and the VLA.  These new data, in
conjunction with IR photometery from 2 \micron\ to 60 \micron, provide
a better glimpse at the thermal continuum from Deneb's extended
atmosphere.  The slope of the radio spectrum implies a stellar wind
where hydrogen is partially ionized, a result reproduced by our
models.  We find that the effects of line-blanketing on the synthetic
millimeter and radio continua are significant and must be taken into
when analyzing the continua of partially ionized winds.

\subsection{Angular Diameter Measurement and Limb-darkening}
Visibility measurements from the NPOI indicate a uniform-disk angular
diameter for Deneb of $\overline{\theta_{\rm UD}} = 2.40 \pm 0.06\
{\rm mas}$.  Our models predict that the center-to-limb profile for
Deneb in the $I$-band is significantly extended relative to hydrostatic
model predictions.  We believe this limb profile causes the uniform
disk angular diameter to be larger than the angular diameter
$\theta_{\rm Ross} = 2.32 \pm 0.06 {\rm\ mas}$ corresponding to
$R_{\rm Ross}$, the radius at the Rosseland mean optical depth
$\tau_{\rm Ross}$ = 2/3.  The result is a correction to the uniform
disk diameter which is less than unity, contrary to the limb-darkening
correction made in the calculation of the effective temperature for
more compact stars.  This limb profile prediction 
will hopefully be
testable in the coming years as longer base-line optical/IR
interferometers come on-line.

\subsection{Interferometic Effective Temperature}
Considering the uncertainites in the spectrophotometry, the
interstellar extinction, and the corrected NPOI angular diameter, we
find an inteferometric fundamental effective temperature of $T_{\rm
eff}^{\rm Ross} = 8600 \pm 500\ {\rm K}$ for Deneb. The biggest
uncertainty in the value of the effective temperature is the
uncertainty in the color excess, $E(B-V) = 0.09^{+0.04}_{-0.03}$. 
This
effective temperature together with the distance estimates from {\it
Hipparcos} and Deneb's Cyg OB7 association membership, and theoretical
evolutionary tracks yield: $R_{\rm Ross} \simeq$ 180 \Rsun, $L \simeq
1.6 \pm 0.4 \times 10^4$ \Lsun, $M \sim 20 -25$ \Msun, and \Logg\
$\simeq 1.3$.

\subsection{Least-squares Spectrum Analysis}
A least-squares comparison of 55 synthetic spectral energy
distributions from 1220 \AA\ to 3.6 cm with the spectral energy
distibution of Deneb provide estimates (1$\sigma$) for the star's
effective temperature and mass-loss rate of: $T_{\rm eff}^{\rm Ross}
\simeq 8420 \pm 100$ K and $\dot{M} \simeq 8 \pm 3 \times 10^{-7}$
\Mspyr.  The best fitting models to the SED indicate that the color
excess toward Deneb is $E(B-V) = 0.06 \pm 0.01$.  This effective
temperature is consistent with the inteferometric fundamental
effective temperature.  The smaller 1$\sigma$ error in the
least-squares value of $T_{\rm eff}^{\rm Ross}$ relative to
interferometric value reflects the tighter contraint on the color
excess from the least-squares fit.  However, both values of
$T_{\rm eff}^{\rm Ross}$ are subject to systematic errors due to the
poorly established interstellar extinction curve toward Deneb.

The least-squares analysis shows that, in addition to the effective
temperature and mass-loss rate, the steepness of $\beta$-law velocity
field, the microturbulence, and the surface gravity all have a
significant effect on the model temperature structure and ultraviolet
spectral energy distribution.  It is interesting to point out that
models with $\beta$ outside the range $3.0\pm1.0$ provide very poor
fits to the UV spectrophotometry and that $\beta$ values near 3.0 are
not consistent with the radiation driven wind models of \cite{alp97},
which predict $\beta < 1$ for A-type supergiant winds.

\subsection{Non-LTE Effects on the Line Spectrum}
The mass-loss rate range derived from the least-squares analysis is
reasonably consistent with that derived from high dispersion spectra of Deneb
when non-LTE metal-line blanketing is treated, in particular for
species \myion{Fe}{i-iii}, \myion{Mg}{ii}, and \myion{Ca}{i-iii}.
The primary effect of the non-LTE treatment is to enhance the ionization
of these elements in the wind relative to LTE.  The non-LTE model
ionization structure produces much lower column densities of
\ion{Fe}{2}, \ion{Mg}{2}, and \ion{Ca}{2} relative to LTE and causes
strong lines to form deeper in the wind at lower velocities, in better
agreement with the observations.  For the \ion{Fe}{2} UV lines and
the \ion{Ca}{2} $H$ \& $K$ lines, the non-LTE effects are striking.  The
mass-loss rate inferred from the \ion{Fe}{2} UV lines is 50 to 100
times larger in non-LTE than in LTE.  The synthetic $H$ \& $K$ profiles
show strong synthetic P-Cygni profiles in LTE which are not present in
the data or in the non-LTE synthetic profiles.

\subsection{Problems with \ha\ and Other Hydrogen Lines}
We are unable achieve a reasonable fit to a typical \ha\ P-Cygni
profile for Deneb with any model parameters over a reasonable range.
The most prominent problem is that the model absorption component is
always too strong.  While Deneb's \ha\ profile is variable, it
is not hugely variable, and the model \ha\ profiles don't look
like the observed profile at anytime. 
The wind model line profiles generally provide a
better match to the higher Balmer series lines than hydrostatic model
line profiles.  Both wind models and hydrostatic models match the
higher Paschen series lines equally well.  Both types of models fail
badly for the higher Pfund series lines.  These failures indicate that
a spherically symmetric, expanding, steady state, line-blanketed,
radiative equilibrium structure is not consistent with the conditions
under which a typical \ha\ and the higher Pfund lines form.  The
strong lines of \ion{Fe}{2}, \ion{Mg}{2}, and \ion{Ca}{2} that we can
reproduce with our models originate from the ground state or
low-exitation levels and apparently are not sensitive to the
conditions which affect the 2$\rightarrow$3 transition of hydrogen. 
Our inability to fit a typical \ha\ profile from Deneb is very bothersome
because \ha\ is the most commonly used mass-loss rate diagnositic
for A-type supergiants.

\subsection{Mass-loss Rate}
The thermal millimeter and radio continua of Deneb when compared with
our models provide both lower and upper limits to its mass-loss rate.
The detection of radiation at 3.6 cm, far in excess of what is expected
for no mass-loss, provides the lower limit at about $10^{-7}$ \Mspyr.
The $3\sigma$ upper limit to the 870 \micron\ flux provides the upper
limit at about $10^{-6}$ \Mspyr.  If Deneb's extended atmosphere
departs significantly from spherical symmetry, this would
systematically bias our estimates for the mass-loss rate.  However,
there is no evidence for deviations from spherical symmetry from
instrinsic polarization studies, either in the continuum or in the
spectral lines \hb, \hg, \ion{Ca}{2} $H$ \cite[][and references
therein]{notpolarized}.

The line spectrum of Deneb provides mainly upper limits to the
mass-loss rate.  While the observed P-Cygni character of the \ha\ and
\ion{Mg}{2} $h$ and $k$ lines are clear indicators of the stellar
outflow, a hydrostatic model with no mass loss is really not that bad a
fit to most of the UV line spectrum.  It is the more subtle
characteristics of the UV lines, their widths and desaturated profiles
which show evidence of the velocity field.  Many weak metal lines in
the optical spectrum are also very well matched by models with no
mass-loss and all our models with \mdot $> 10^{-8}$ \Mspyr\ predict
profiles for these lines which on average tend to be washed out and
too shallow.  Certainly the column densities in many of these
transitions are too high.  While non-LTE ionization effects solve this
problem for some lines, others are still problematic.  A line-by-line
study with a complete grid of models which include \myion{Ti}{i-iii},
\myion{Cr}{i-iii}, and other metals in non-LTE is needed to 
analyze this issue fully.  We have model atoms for these species and
will investigate this in future work.

While previously published estimates for Deneb's mass-loss rate range
over more than three orders of magnitude, we can with confidence
reduce this uncertainty to one order of magnitude, but not much
better.  Reducing this uncertainty further requires a better match
simultaneously to the spectral energy distribution and the spectral lines.
This may require removing several of our simplifying assumptions such
as radial symmetry, homogeniety (filling-factor equal to unity),
time-independence, and no mechanical dissipation.

We suggest that before the Wind Momentum-Luminosity Relationship, thus
far based solely on Balmer line fits, can be firmly established for
A-type supergiants, researchers should work to toward firmly
establishing fundamental parameters and stellar wind properties of
Galactic A-type supergiants by checking that models are consistant with
both the spectral energy distributions and the line spectra.

\acknowledgments

Thanks to S. Shore, N. Morrison, G. Schwarz, S. Starrfield,
J. Monnier, D. Sasselov, and R. Kurucz for valuable
discussions. Thanks to A. Kaufer for kindly providing the HEROS
spectra.  We thank Michael Dumke for his invaluable help in teaching
two of us (JAS and KDG) how to do the submillimeter observations of
point sources at the Heinrich Hertz Telescope.  Thanks to H. Dole for
help with the ISOPHOT data reduction.  The completion of the paper was
supported by a Harvard-Smithsonian Center for Astrophysics
Postdoctoral Fellowship to JPA. AWB is supported by PPARC, and IDH is
a Fresia Jolligoode Fellow.  This work was also supported by NSF grant
AST-9819795 to Arizona State University.  This work was supported in
part by NSF grants AST-9720704 and AST-0086246, NASA grants NAG5-8425,
NAG5-9222, as well as NASA/JPL grant 961582 to the University of
Georgia and in part by NSF grants AST-97314508, by NASA grant
NAG5-3505 and an IBM SUR grant to the University of Oklahoma.  This
work was supported in part by the P\^ole Scientifique de
Mod\'elisation Num\'erique at ENS-Lyon.  Some of the calculations
presented in this paper were performed on the IBM SP2 of the UGA UCNS,
on the IBM SP ``Blue Horizon'' of the San Diego Supercomputer Center
(SDSC), with support from the National Science Foundation, and on the
IBM SP of the NERSC with support from the DOE.  We thank all these
institutions for a generous allocation of computer time.  This work
made use of the SIMBAD database, Strasbourg, France.  The ISOPHOT data
are based on observations with ISO, an ESA project with instruments
funded by ESA Member States (especially the PI countries: France,
Germany, the Netherlands and the United Kingdom) and with the
participation of ISAS and NASA.  The ISOPHOT data presented in this
paper were reduced using PIA, which is a joint development by the ESA
Astrophysics Division and the ISOPHOT Consortium with the
collaboration of the Infrared Processing and Analysis Center
(IPAC). Contributing ISOPHOT Consortium institutes are DIAS, RAL, AIP,
MPIK, and MPIA.  This research used the DIRBE Point Source Photometry
Research Tool, a service provided by the Astrophysics Data Facility at
NASA's Goddard Space Flight Center.

\appendix

\section{Construction of Model Atmospheres}
The stellar wind module developed by \cite{jpaphd} for the \phoenix\
code links a unified model atmosphere structure to the non-LTE
line-blanketing and expanding atmosphere radiative transfer modules
already in the code.  Important improvements over previous models lie
in the solution of the radiative transfer equation, the degree of
consistent LTE and non-LTE metal line-blanketing, and the solution of
the temperature structure from the condition of energy conservation.

\subsection{Line-blanketing Specifics and Non-LTE Model Atoms}
All models presented in this paper include in LTE approximately
$7\times10^5$ of the most important blanketing lines dynamically
selected from a list of $42\times10^{6}$ lines \cite[]{jcdrom23} from
39 elements with up to 26 ionization stages in addition to the
strongest 517 lines for the spectra \ion{H}{1}, \ion{He}{1},
\ion{He}{2}, which are treated in non-LTE.  More sophisticated models
include an additional $2.3\times10^4$ lines in non-LTE for these
spectra: \ion{C}{1}, \ion{C}{2}, \ion{C}{3}, \ion{N}{1}, \ion{N}{2},
\ion{N}{3}, \ion{O}{1}, \ion{O}{2},\ion{O}{3},
\ion{Mg}{1},\ion{Mg}{2},\ion{Mg}{3}, \ion{Ca}{1}, \ion{Ca}{2},
\ion{Ca}{3}, \ion{Fe}{1}, \ion{Fe}{2}, \ion{Fe}{3}.  The number of
levels and lines from each species are shown in Table
\ref{tab:NLTE_levels_lines}.  For the complete set of non-LTE model
ions in the \phoenix\ code see \cite{massive}.  The references to the
atomic data used to construct the model atoms and ions are given in
Table \ref{tab:atomic_sources}.

The non-LTE lines from the model atoms and ions replace the LTE lines
and all the lines that are not in non-LTE are handled in LTE as an
approximation.  Details on non-LTE blanketing techniques can be found
in \cite[][and references therein]{massive}.  Even in our most
sophisticated models the vast majority of lines are treated in LTE.
The bulk of the {\em total opacity} is, however, computed in detailed
non-LTE.  The line-blanketing is {\it true} line-blanketing in the
sense that the effect of these millions of lines on the temperature
structure is computed through the condition of radiative equilibrium
(see below).  Therefore, the line-blanketing is consistently included
in both the model atmosphere and final synthetic spectrum
calculations.

\begin{deluxetable}{lccc}
\tablecolumns{4} \tabletypesize{\small} \tablecaption{Number of Levels and Primary Transitions\tablenotemark{a}\ \ Treated in Non-LTE}
\tablewidth{0pt} \tablehead{ 
&\multicolumn{3}{c}{Stage of Ionization} \\
\cline{2-4} \\
\colhead{Element} & \colhead{\sc{i}} &\colhead{\sc{ii}}    & \colhead{\sc{iii}}}
\startdata 
H                          &30/435       &...       &...    \\
He                         &19/37        &10/45     &...    \\
C                          &228/1387     &85/336    &79/365 \\ 
N                          &252/2313     &152/1110  &87/226 \\
O                          &36/66        &171/1304  &137/765 \\ 
Mg                         &273/835      &72/340    &91/656   \\
Ca                         &194/1029     &87/455    &150/1661 \\
Fe\tablenotemark{b}        &494/6903     &617/13675 &566/9721 \\   
\enddata
\tablenotetext{a}{Primary transitions are those that connect observed levels and have $\log(gf)$ values greater than $-3$.} 
\tablenotetext{b}{Values for the complete models of stages {\sc{i-iii}} of Fe are shown.  
In our calculations, only the first 200 levels of each stage are used.}
\label{tab:NLTE_levels_lines}
\end{deluxetable}

\begin{deluxetable}{lcccccc}
\tablecolumns{7} \tabletypesize{\scriptsize} \tablecaption{Atomic Data Sources for Model Atoms and Ions\tablenotemark{a}}
\tablewidth{0pt} \tablehead{ 
Model Species &Levels &\multicolumn{2}{c}{$b-b$ cross-sections} & &\multicolumn{2}{c}{$b-f$ cross-sections} \\
\cline{3-4} \cline{6-7} \\ 
\colhead{ } & \colhead{ } &\colhead{Radiative}    & \colhead{Collisional\tablenotemark{b}} 
& \colhead{ } &\colhead{Photoionization\tablenotemark{c}} & \colhead{Collisional}} 
\startdata 
{H}~{\sc{i}}         &  Johnson       & Johnson     &   Johnson        &   &   M84          &     Drawin \\
{He}~{\sc{i}}        &   Martin       &   WSG       &   BK, Lanzafame  &   &   M84          &     MS \\
{He}~{\sc{ii}}       &  Johnson       & Johnson     &   Johnson        &   &   M84          &     MS    \\                               
{C}~{\sc{i-iii}}     &   KB        &   KB     &   Allen, VR      &   & BRP, RM        &     Drawin \\
{N}~{\sc{i-iii}}     &   KB        &   KB     &   Allen, VR      &   & BRP, RM        &     Darwin \\
{O}~{\sc{i}}         &   KB        &   KB     &   Allen, VR      &   & BRP, RM        &     CLL   \\
{O}~{\sc{ii-iii}}    &   KB        &   KB     &   Allen, VR      &   & BRP, RM        &     Drawin \\
{Mg}~{\sc{i-iii}}    &   KB        &   KB     &   Allen, VR      &   &      RM        &     Drawin \\
{Ca}~{\sc{i-iii}}    &   KB        &   KB     &   Allen, VR      &   & BRP, RM        &     Drawin \\
{Fe}~{\sc{i-iii}}    &   KB        &   KB     &   Allen, VR      &   & BRP, RM        &     Drawin \\
\enddata
\tablenotetext{a}{References - Allen = \cite{general_coll_bf}; BK =
\cite{he1_bb_coll_1}; BRP = \cite{general_ground_photo_bf}; KB =
\cite{jcdrom23}; CLL=\cite{o1_bf_coll_1,o1_bf_coll_2}; Drawin =
\cite{drawin61}; Lanzafame = \cite{he1_bb_coll_2}; M84 =
\cite{mathisen}; MS = \cite{he1_bf_coll}; Johnson = \cite{hi_atom}; RM
= \cite{general_excited_photo_bf}; VR = \cite{vr62}; WSG =
\cite{wgs66}} 
\tablenotetext{b}{Reference to Allen is only for
cross-sections between bound levels not coupled by radiative
transitions} 
\tablenotetext{c}{Reference to BRP is only for
ground-state cross-sections}
\label{tab:atomic_sources}
\end{deluxetable}

\subsection{The Wind Module} 
We treat the problem of radiation-driven winds with the following
simplifying assumptions: (1) time-independent, steady-state; (2) no
magnetic fields or rotation; (3) radial symmetry; and (4) a smooth
single-phase flow with no mechanical dissipation, heat conduction, 
viscosity, or advection.

\subsubsection{Atmospheric Structure and Boundary Conditions}
The gas escaping from the model star is treated as a hydrodynamic flow,
satisfying the conservation of mass, momentum, and energy.  
The mass conservation is specified by the equation of
continuity,
\begin{equation}
\dot{M} = 4\pi r^2\rho(r)v(r)
\label{cont}
\end{equation}
where $\dot{M}$ is the steady state mass-loss rate, 
$\rho(r)$ is the gas density, and $v(r)$ is the velocity field.
The velocity field is specified by a
$\beta$-law of the form
\begin{equation}
\label{betalaw}
v(r) = v_{\infty}(1 - R_\star/r)^{\beta}.
\end{equation}
The layers of the atmosphere below $R_\star$ are modeled
as hydrostatic and in these layers the velocity is set to zero.

The structure of the outer wind layers is computed on a fixed radial
grid. The structure of the inner hydrostatic layers of the atmosphere
is computed on an optical depth grid. Typically 50 layers are used
consisting of 35 layers for the wind region and 15 layers for the
hydrostatic region.  The radial grid points are prescribed by a
power-law in order to finely sample the inner region of the wind
\cite[]{steffen97}, where the velocity gradient is steepest, and to
coarsely sample the outer portion of the wind where the velocity
gradient is small.

\subsubsection{Wind region}
The radii $r(l)$ in layers $l=1,...,l_{\star}$ are specified by,
\begin{equation}
r(l) = r(1) + \bigl\{r(l_{\star})-r(1)\bigr\}\cdot
\Biggl\{\frac{\alpha^{(l-1)} -1 }{\alpha^{(l_{\star}-1)}-1}\Biggr\}
\end{equation}

\noindent where $l_{\star}$ is the layer (typically layer 35)
immediately above the dynamic-static transition.  Layer 1 is the
outermost layer and typically we choose the radial extension in the
wind to be a factor of 200, $r(1) = 200R_\star$.  The value of
$r(l_{\star})$ is iteratively selected to provide a predetermined
degree of smoothness in the density profile across the dynamic-static
transition.  Generally the jump in density must be less than a factor
of 5 across the transition.  The input parameter $\alpha$ 
may be adjusted to optimize the distribution of radial points for a specific
velocity field (typically, $\alpha \simeq 1.5$).

The density $\rho(r)$ in the expanding layers is specified from the
mass flux continuity equation (\ref{cont}),
where the velocity field has the form of a $\beta$-law 
from equation (\ref{betalaw}).
The input radius $R_\star$ and the model effective temperature $T_\star$
set the luminosity L of the star,
\begin{equation}
L = 4{\pi}R_{\star}^2{\sigma}T_\star^4.
\label{eqn:lum}
\end{equation}

Since we use a form of the $\beta$-law lacking a macroturbulent
velocity, the switch in the structure setup from the dynamic layers to
the static layers occurs near the point in the atmosphere where the
wind velocity, $v(r)$, drops below the microturbulent velocity,
$\xi_t$.  An initial guess at this radius, $r(l_{\star}) \gtrsim
R_\star$, is calculated by finding the radius at which $v(r) = \xi_t$
from equation (\ref{betalaw}):
\begin{equation}
r(l_{\star}) = \frac{R_\star}{ \biggl\{ {1- \bigl(\xi_t/
v_{\infty}\bigr)^{1/\beta}} \biggr\}}.
\end{equation}
This radius is then adjusted inward slightly to improve the degree of
smoothness of the density profile across the dynamic-static
transition.  In practice, this results in a discontinuity in the
velocity field across dynamic-static boundary which is very small
($\sim$0.2 \kms), much smaller than the microturbulent width of the
lines, $\xi_t \ge 5$ \kms.

\subsubsection{Hydrostatic region}

In the hydrostatic layers ($l=l_{\star}+1,...,N$), the structure is
computed on a fixed logarithmically spaced optical depth grid,

\begin{equation}
\log\{\tau(l)\} = \log\{\tau(l_{\star})\} + (l-l{_\star}) \Biggl\{
\frac{\log\{\tau(N)\}-\log\{\tau{(\l_{\star})}\}} {N-l_{\star}}
\Biggr\} 
%\qquad\qquad l=l_{\star}+1,...,N
\end{equation}

\noindent The inner continuum extinction optical depth boundary,
$\tau(N)=\tau({\rm 2000 \rm{\AA}})$ = $10^{2}$, is set well below the
thermalization depth.  The pressure structure is computed by
numerically integrating the hydrostatic equation,
\begin{equation}
\frac{dP}{d\tau} = \frac{g_{\rm eff}}{\kappa}
\label{eqn:hydro}
\end{equation}
\noindent where $g_{\rm eff}$ is the effective gravitational acceleration
\begin{equation}
g_{\rm eff} = g(R_{\star})\frac{R_{\star}^2}{r^2} - g_{\rm rad}(r)
\label{geff}
\end{equation}
the difference of the Newtonian acceleration and
the radiative acceleration $g_{\rm rad}$ computed from
\begin{equation}
\label{arad}
g_{\rm rad}(r) = \frac{1}{c\rho(r)}
4\pi\int_{0}^{\infty}\chi_{\lambda}(r)H_{\lambda}(r)\,d\lambda,
\end{equation}
where $\chi_{\lambda}(r)H_{\lambda}(r)$ is the product of the total line and 
continuum opacity with the first moment of the radiation field
provided by solution of the radiative transfer equation (see below).

The radii in the hydrostatic layers are computed by integrating
\begin{equation}
dr = \frac{1}{\kappa\rho}\, d\tau.
\end{equation}
\noindent Once the velocity field and radii are computed for
all layers, the non-grey, non-LTE radiative transfer problem is solved and the
temperature structure is adjusted iteratively to achieve energy
conservation.  

\subsection{Radiative Transfer and Radiative Equilibrium}
The spherically symmetric radiative transfer equation (SSRTE) 
in the co-moving (Lagrangian) frame \cite[]{FRH}
\begin{equation}
\label{SSRTE}
\begin{array}{cc}
  \g (\mu + \b) \pder{I}{r}
   &  + \pder{}{\mu}\left\{ \g (1-\mu^2)
      \left[ {\frac{(1+\b\mu)}{r}}
                - \g^2(\mu+\b) \pder{\b}{r} \right] I \right\} \\
  &  - \pder{}{\lambda} \left\{ \g
       \left[ \frac{\b(1-\mu^2)}{r} + \g^2\mu(\mu+\b)\pder{\b}{r} \right]
         \lambda I \right\} \\
  &  + \g\left\{
       \frac{2\mu+\b(3-\mu^2)}{r} + \g^2(1+\mu^2+2\b\mu)\pder{\b}{r}\right\}
      I \\
  & \qquad = \eta - \chi I. \\
\end{array}
\end{equation}
is solved using an Operator Splitting/Approximate $\Lambda$-operator
Iteration (OS/ALI) method using short-characteristics \cite[]{s3pap}.
Improvements since this work include a more accurate discretization of
the $\partial(\lambda I)/\partial\lambda$ term needed to improve the numerical
accuracy necessary for $\beta$-law winds.

In equation (\ref{SSRTE}), $\mu$ is the direction-cosine, $I$ is the
specific intensity, the velocity is measured in units of the speed of
light c, $\beta(r)=v(r)/c$, and $\gamma= 1/(1-\beta^2)^{1/2}$.  The
principal reason for choosing to solve the radiative transfer problem
in the co-moving, fluid rest frame is that although the frequency
spectrum may be complex, with many overlapping lines, the opacity is
isotropic. This leads to great simplification in the radiation-matter
interaction terms relative to the observer's frame equation.  For
example, the flow velocity is not an argument of the line profile
function in the co-moving frame. Frequencies are measured in the frame
of the moving fluid, therefore we can use a static line profile.  In
this frame one can directly integrate over angle and use
co-moving-frame moments of the radiation field in the computation of
the radiation pressure and radiative equilibrium equations.
The angular dependence of the opacity in the observer's frame leads to
computational difficulties, and in flows with small velocity gradients
and many overlapping spectral lines, escape probability methods, like
the Sobolev approximation, are not valid.  

We iteratively compute the temperature structure from the condition of energy
conservation in the co-moving frame. 
The conservation of energy, assuming radiative equilibrium, can be
specified in the Lagrangian frame \cite[]{REpap} by
\begin{equation}
\label{RE}
\int_0^\infty (\eta_{\lambda}-\chi_{\lambda} J_{\lambda})\,d\lambda 
+ \dot{S}=0
\end{equation}
where $\eta_{\lambda}$ and $J_{\lambda}$
describe the emissivity and mean intensity 
respectively. 
We include the kinetic and gravitational energy of the wind,
$\dot{S}$, in the energy equation, although they are less than
$0.1$\% of the total nuclear energy generated by the star for the
models presented here and thus have negligible effect on the
temperature structure.  For very high wind momenta and small
luminosities, the energy required to drive the wind might be large
enough to affect the temperature structure.  Our numerical solution of
the special relativistic radiative energy equation is 
described in \cite{hb99}.

Energy conservation errors are typically $< 1\%$ in both the
luminosity and luminosity derivative in each layer.  Models are
started from a grey temperature structure and iterations are performed
with continuum opacity only.  Next, more iterations are performed with
the addition of full LTE metal-line blanketing opacity.  This yields a
converged LTE line-blanketed model.  Non-LTE models are converged
starting from converged LTE line-blanketed models. Non-LTE species are
added and the model structure is converged in a series of steps: first
H and He; next C, N, O; next Mg, Ca, Fe, etc.

\bibliography{jaufdenb,newyeti,supernovae,opacity,opacity-fa,novae,mdwarf,radtran,winds,hotstars,general,lbv,thesis,alfcyg}

\end{document}